\documentclass[longauth]{aa1}

\usepackage{siunitx}
\usepackage{graphicx}
\usepackage{natbib}
\usepackage{scalerel}

\usepackage[table]{xcolor}

\usepackage{txfonts}
\usepackage[pdfencoding=auto,psdextra]{hyperref}
\hypersetup{
    colorlinks=true,
    linkcolor=blue,
    filecolor=magenta,      
    urlcolor=blue,
    citecolor=blue
}
\makeatletter
\renewcommand*\aa@pageof{, page \thepage{} of \pageref*{LastPage}}
\makeatother

\usepackage[utf8]{inputenc}
\usepackage{supertabular}
\usepackage[switch, modulo]{lineno}

\defcitealias{Q1-SP031}{E25}
\defcitealias{Q1-SP040}{Q25}

\usepackage{euclid}
\newcommand{\msun}{\text{M}_{\odot}}

\begin{document}
%
%

\title{Euclid Quick Data Release (Q1)} \subtitle{The evolution of the passive-density and morphology-density relations between $z=0.25$ and $z=1$}

   


\newcommand{\orcid}[1]{} 
\author{Euclid Collaboration: C.~Cleland\orcid{0009-0002-1769-1437}\thanks{\email{cleland@apc.in2p3.fr}}\inst{\ref{aff1}}
\and S.~Mei\orcid{0000-0002-2849-559X}\inst{\ref{aff1},\ref{aff2}}
\and G.~De~Lucia\orcid{0000-0002-6220-9104}\inst{\ref{aff3}}
\and F.~Fontanot\orcid{0000-0003-4744-0188}\inst{\ref{aff3},\ref{aff4}}
\and H.~Fu\orcid{0009-0002-8051-1056}\inst{\ref{aff5},\ref{aff6}}
\and C.~C.~Lovell\orcid{0000-0001-7964-5933}\inst{\ref{aff7}}
\and M.~Magliocchetti\orcid{0000-0001-9158-4838}\inst{\ref{aff8}}
\and N.~Mai\inst{\ref{aff1},\ref{aff2}}
\and D.~Roberts\orcid{0009-0009-7662-0445}\inst{\ref{aff6}}
\and F.~Shankar\orcid{0000-0001-8973-5051}\inst{\ref{aff6}}
\and J.~G.~Sorce\orcid{0000-0002-2307-2432}\inst{\ref{aff9},\ref{aff10}}
\and M.~Baes\orcid{0000-0002-3930-2757}\inst{\ref{aff11}}
\and P.~Corcho-Caballero\orcid{0000-0001-6327-7080}\inst{\ref{aff12}}
\and S.~Eales\inst{\ref{aff13}}
\and C.~Tortora\orcid{0000-0001-7958-6531}\inst{\ref{aff14}}
\and N.~Aghanim\orcid{0000-0002-6688-8992}\inst{\ref{aff10}}
\and B.~Altieri\orcid{0000-0003-3936-0284}\inst{\ref{aff15}}
\and A.~Amara\inst{\ref{aff16}}
\and S.~Andreon\orcid{0000-0002-2041-8784}\inst{\ref{aff17}}
\and N.~Auricchio\orcid{0000-0003-4444-8651}\inst{\ref{aff18}}
\and H.~Aussel\orcid{0000-0002-1371-5705}\inst{\ref{aff19}}
\and C.~Baccigalupi\orcid{0000-0002-8211-1630}\inst{\ref{aff4},\ref{aff3},\ref{aff20},\ref{aff21}}
\and M.~Baldi\orcid{0000-0003-4145-1943}\inst{\ref{aff22},\ref{aff18},\ref{aff23}}
\and A.~Balestra\orcid{0000-0002-6967-261X}\inst{\ref{aff24}}
\and S.~Bardelli\orcid{0000-0002-8900-0298}\inst{\ref{aff18}}
\and P.~Battaglia\orcid{0000-0002-7337-5909}\inst{\ref{aff18}}
\and A.~Biviano\orcid{0000-0002-0857-0732}\inst{\ref{aff3},\ref{aff4}}
\and A.~Bonchi\orcid{0000-0002-2667-5482}\inst{\ref{aff25}}
\and D.~Bonino\orcid{0000-0002-3336-9977}\inst{\ref{aff26}}
\and E.~Branchini\orcid{0000-0002-0808-6908}\inst{\ref{aff27},\ref{aff28},\ref{aff17}}
\and M.~Brescia\orcid{0000-0001-9506-5680}\inst{\ref{aff29},\ref{aff14}}
\and J.~Brinchmann\orcid{0000-0003-4359-8797}\inst{\ref{aff30},\ref{aff31}}
\and S.~Camera\orcid{0000-0003-3399-3574}\inst{\ref{aff32},\ref{aff33},\ref{aff26}}
\and G.~Ca\~nas-Herrera\orcid{0000-0003-2796-2149}\inst{\ref{aff34},\ref{aff35},\ref{aff36}}
\and V.~Capobianco\orcid{0000-0002-3309-7692}\inst{\ref{aff26}}
\and C.~Carbone\orcid{0000-0003-0125-3563}\inst{\ref{aff37}}
\and J.~Carretero\orcid{0000-0002-3130-0204}\inst{\ref{aff38},\ref{aff39}}
\and S.~Casas\orcid{0000-0002-4751-5138}\inst{\ref{aff40}}
\and F.~J.~Castander\orcid{0000-0001-7316-4573}\inst{\ref{aff41},\ref{aff42}}
\and M.~Castellano\orcid{0000-0001-9875-8263}\inst{\ref{aff43}}
\and G.~Castignani\orcid{0000-0001-6831-0687}\inst{\ref{aff18}}
\and S.~Cavuoti\orcid{0000-0002-3787-4196}\inst{\ref{aff14},\ref{aff44}}
\and K.~C.~Chambers\orcid{0000-0001-6965-7789}\inst{\ref{aff45}}
\and A.~Cimatti\inst{\ref{aff46}}
\and C.~Colodro-Conde\inst{\ref{aff47}}
\and G.~Congedo\orcid{0000-0003-2508-0046}\inst{\ref{aff48}}
\and C.~J.~Conselice\orcid{0000-0003-1949-7638}\inst{\ref{aff49}}
\and L.~Conversi\orcid{0000-0002-6710-8476}\inst{\ref{aff50},\ref{aff15}}
\and Y.~Copin\orcid{0000-0002-5317-7518}\inst{\ref{aff51}}
\and F.~Courbin\orcid{0000-0003-0758-6510}\inst{\ref{aff52},\ref{aff53}}
\and H.~M.~Courtois\orcid{0000-0003-0509-1776}\inst{\ref{aff54}}
\and M.~Cropper\orcid{0000-0003-4571-9468}\inst{\ref{aff55}}
\and A.~Da~Silva\orcid{0000-0002-6385-1609}\inst{\ref{aff56},\ref{aff57}}
\and H.~Degaudenzi\orcid{0000-0002-5887-6799}\inst{\ref{aff58}}
\and A.~M.~Di~Giorgio\orcid{0000-0002-4767-2360}\inst{\ref{aff8}}
\and C.~Dolding\orcid{0009-0003-7199-6108}\inst{\ref{aff55}}
\and H.~Dole\orcid{0000-0002-9767-3839}\inst{\ref{aff10}}
\and F.~Dubath\orcid{0000-0002-6533-2810}\inst{\ref{aff58}}
\and X.~Dupac\inst{\ref{aff15}}
\and A.~Ealet\orcid{0000-0003-3070-014X}\inst{\ref{aff51}}
\and S.~Escoffier\orcid{0000-0002-2847-7498}\inst{\ref{aff59}}
\and M.~Farina\orcid{0000-0002-3089-7846}\inst{\ref{aff8}}
\and R.~Farinelli\inst{\ref{aff18}}
\and F.~Faustini\orcid{0000-0001-6274-5145}\inst{\ref{aff43},\ref{aff25}}
\and S.~Ferriol\inst{\ref{aff51}}
\and F.~Finelli\orcid{0000-0002-6694-3269}\inst{\ref{aff18},\ref{aff60}}
\and S.~Fotopoulou\orcid{0000-0002-9686-254X}\inst{\ref{aff61}}
\and M.~Frailis\orcid{0000-0002-7400-2135}\inst{\ref{aff3}}
\and E.~Franceschi\orcid{0000-0002-0585-6591}\inst{\ref{aff18}}
\and M.~Fumana\orcid{0000-0001-6787-5950}\inst{\ref{aff37}}
\and S.~Galeotta\orcid{0000-0002-3748-5115}\inst{\ref{aff3}}
\and K.~George\orcid{0000-0002-1734-8455}\inst{\ref{aff62}}
\and B.~Gillis\orcid{0000-0002-4478-1270}\inst{\ref{aff48}}
\and C.~Giocoli\orcid{0000-0002-9590-7961}\inst{\ref{aff18},\ref{aff23}}
\and J.~Gracia-Carpio\inst{\ref{aff63}}
\and B.~R.~Granett\orcid{0000-0003-2694-9284}\inst{\ref{aff17}}
\and A.~Grazian\orcid{0000-0002-5688-0663}\inst{\ref{aff24}}
\and F.~Grupp\inst{\ref{aff63},\ref{aff62}}
\and S.~Gwyn\orcid{0000-0001-8221-8406}\inst{\ref{aff64}}
\and S.~V.~H.~Haugan\orcid{0000-0001-9648-7260}\inst{\ref{aff65}}
\and J.~Hoar\inst{\ref{aff15}}
\and W.~Holmes\inst{\ref{aff66}}
\and F.~Hormuth\inst{\ref{aff67}}
\and A.~Hornstrup\orcid{0000-0002-3363-0936}\inst{\ref{aff68},\ref{aff69}}
\and P.~Hudelot\inst{\ref{aff70}}
\and K.~Jahnke\orcid{0000-0003-3804-2137}\inst{\ref{aff71}}
\and M.~Jhabvala\inst{\ref{aff72}}
\and B.~Joachimi\orcid{0000-0001-7494-1303}\inst{\ref{aff73}}
\and E.~Keih\"anen\orcid{0000-0003-1804-7715}\inst{\ref{aff74}}
\and S.~Kermiche\orcid{0000-0002-0302-5735}\inst{\ref{aff59}}
\and A.~Kiessling\orcid{0000-0002-2590-1273}\inst{\ref{aff66}}
\and B.~Kubik\orcid{0009-0006-5823-4880}\inst{\ref{aff51}}
\and M.~K\"ummel\orcid{0000-0003-2791-2117}\inst{\ref{aff62}}
\and M.~Kunz\orcid{0000-0002-3052-7394}\inst{\ref{aff75}}
\and H.~Kurki-Suonio\orcid{0000-0002-4618-3063}\inst{\ref{aff76},\ref{aff77}}
\and O.~Lahav\orcid{0000-0002-1134-9035}\inst{\ref{aff73}}
\and Q.~Le~Boulc'h\inst{\ref{aff78}}
\and A.~M.~C.~Le~Brun\orcid{0000-0002-0936-4594}\inst{\ref{aff79}}
\and D.~Le~Mignant\orcid{0000-0002-5339-5515}\inst{\ref{aff80}}
\and S.~Ligori\orcid{0000-0003-4172-4606}\inst{\ref{aff26}}
\and P.~B.~Lilje\orcid{0000-0003-4324-7794}\inst{\ref{aff65}}
\and V.~Lindholm\orcid{0000-0003-2317-5471}\inst{\ref{aff76},\ref{aff77}}
\and I.~Lloro\orcid{0000-0001-5966-1434}\inst{\ref{aff81}}
\and G.~Mainetti\orcid{0000-0003-2384-2377}\inst{\ref{aff78}}
\and D.~Maino\inst{\ref{aff82},\ref{aff37},\ref{aff83}}
\and E.~Maiorano\orcid{0000-0003-2593-4355}\inst{\ref{aff18}}
\and O.~Mansutti\orcid{0000-0001-5758-4658}\inst{\ref{aff3}}
\and S.~Marcin\inst{\ref{aff84}}
\and O.~Marggraf\orcid{0000-0001-7242-3852}\inst{\ref{aff85}}
\and M.~Martinelli\orcid{0000-0002-6943-7732}\inst{\ref{aff43},\ref{aff86}}
\and N.~Martinet\orcid{0000-0003-2786-7790}\inst{\ref{aff80}}
\and F.~Marulli\orcid{0000-0002-8850-0303}\inst{\ref{aff87},\ref{aff18},\ref{aff23}}
\and R.~Massey\orcid{0000-0002-6085-3780}\inst{\ref{aff88}}
\and S.~Maurogordato\inst{\ref{aff89}}
\and E.~Medinaceli\orcid{0000-0002-4040-7783}\inst{\ref{aff18}}
\and Y.~Mellier\inst{\ref{aff90},\ref{aff70}}
\and M.~Meneghetti\orcid{0000-0003-1225-7084}\inst{\ref{aff18},\ref{aff23}}
\and E.~Merlin\orcid{0000-0001-6870-8900}\inst{\ref{aff43}}
\and G.~Meylan\inst{\ref{aff91}}
\and A.~Mora\orcid{0000-0002-1922-8529}\inst{\ref{aff92}}
\and M.~Moresco\orcid{0000-0002-7616-7136}\inst{\ref{aff87},\ref{aff18}}
\and L.~Moscardini\orcid{0000-0002-3473-6716}\inst{\ref{aff87},\ref{aff18},\ref{aff23}}
\and R.~Nakajima\orcid{0009-0009-1213-7040}\inst{\ref{aff85}}
\and C.~Neissner\orcid{0000-0001-8524-4968}\inst{\ref{aff93},\ref{aff39}}
\and S.-M.~Niemi\inst{\ref{aff34}}
\and J.~W.~Nightingale\orcid{0000-0002-8987-7401}\inst{\ref{aff94}}
\and C.~Padilla\orcid{0000-0001-7951-0166}\inst{\ref{aff93}}
\and S.~Paltani\orcid{0000-0002-8108-9179}\inst{\ref{aff58}}
\and F.~Pasian\orcid{0000-0002-4869-3227}\inst{\ref{aff3}}
\and K.~Pedersen\inst{\ref{aff95}}
\and W.~J.~Percival\orcid{0000-0002-0644-5727}\inst{\ref{aff96},\ref{aff97},\ref{aff98}}
\and V.~Pettorino\inst{\ref{aff34}}
\and S.~Pires\orcid{0000-0002-0249-2104}\inst{\ref{aff19}}
\and G.~Polenta\orcid{0000-0003-4067-9196}\inst{\ref{aff25}}
\and M.~Poncet\inst{\ref{aff99}}
\and L.~A.~Popa\inst{\ref{aff100}}
\and L.~Pozzetti\orcid{0000-0001-7085-0412}\inst{\ref{aff18}}
\and F.~Raison\orcid{0000-0002-7819-6918}\inst{\ref{aff63}}
\and R.~Rebolo\orcid{0000-0003-3767-7085}\inst{\ref{aff47},\ref{aff101},\ref{aff102}}
\and A.~Renzi\orcid{0000-0001-9856-1970}\inst{\ref{aff103},\ref{aff104}}
\and J.~Rhodes\orcid{0000-0002-4485-8549}\inst{\ref{aff66}}
\and G.~Riccio\inst{\ref{aff14}}
\and E.~Romelli\orcid{0000-0003-3069-9222}\inst{\ref{aff3}}
\and M.~Roncarelli\orcid{0000-0001-9587-7822}\inst{\ref{aff18}}
\and R.~Saglia\orcid{0000-0003-0378-7032}\inst{\ref{aff62},\ref{aff63}}
\and Z.~Sakr\orcid{0000-0002-4823-3757}\inst{\ref{aff105},\ref{aff106},\ref{aff107}}
\and D.~Sapone\orcid{0000-0001-7089-4503}\inst{\ref{aff108}}
\and B.~Sartoris\orcid{0000-0003-1337-5269}\inst{\ref{aff62},\ref{aff3}}
\and J.~A.~Schewtschenko\orcid{0000-0002-4913-6393}\inst{\ref{aff48}}
\and P.~Schneider\orcid{0000-0001-8561-2679}\inst{\ref{aff85}}
\and M.~Scodeggio\inst{\ref{aff37}}
\and A.~Secroun\orcid{0000-0003-0505-3710}\inst{\ref{aff59}}
\and G.~Seidel\orcid{0000-0003-2907-353X}\inst{\ref{aff71}}
\and S.~Serrano\orcid{0000-0002-0211-2861}\inst{\ref{aff42},\ref{aff109},\ref{aff41}}
\and P.~Simon\inst{\ref{aff85}}
\and C.~Sirignano\orcid{0000-0002-0995-7146}\inst{\ref{aff103},\ref{aff104}}
\and G.~Sirri\orcid{0000-0003-2626-2853}\inst{\ref{aff23}}
\and L.~Stanco\orcid{0000-0002-9706-5104}\inst{\ref{aff104}}
\and J.~Steinwagner\orcid{0000-0001-7443-1047}\inst{\ref{aff63}}
\and P.~Tallada-Cresp\'{i}\orcid{0000-0002-1336-8328}\inst{\ref{aff38},\ref{aff39}}
\and A.~N.~Taylor\inst{\ref{aff48}}
\and H.~I.~Teplitz\orcid{0000-0002-7064-5424}\inst{\ref{aff110}}
\and I.~Tereno\inst{\ref{aff56},\ref{aff111}}
\and N.~Tessore\orcid{0000-0002-9696-7931}\inst{\ref{aff73}}
\and S.~Toft\orcid{0000-0003-3631-7176}\inst{\ref{aff112},\ref{aff113}}
\and R.~Toledo-Moreo\orcid{0000-0002-2997-4859}\inst{\ref{aff114}}
\and F.~Torradeflot\orcid{0000-0003-1160-1517}\inst{\ref{aff39},\ref{aff38}}
\and I.~Tutusaus\orcid{0000-0002-3199-0399}\inst{\ref{aff106}}
\and L.~Valenziano\orcid{0000-0002-1170-0104}\inst{\ref{aff18},\ref{aff60}}
\and J.~Valiviita\orcid{0000-0001-6225-3693}\inst{\ref{aff76},\ref{aff77}}
\and T.~Vassallo\orcid{0000-0001-6512-6358}\inst{\ref{aff62},\ref{aff3}}
\and G.~Verdoes~Kleijn\orcid{0000-0001-5803-2580}\inst{\ref{aff12}}
\and A.~Veropalumbo\orcid{0000-0003-2387-1194}\inst{\ref{aff17},\ref{aff28},\ref{aff27}}
\and Y.~Wang\orcid{0000-0002-4749-2984}\inst{\ref{aff110}}
\and J.~Weller\orcid{0000-0002-8282-2010}\inst{\ref{aff62},\ref{aff63}}
\and A.~Zacchei\orcid{0000-0003-0396-1192}\inst{\ref{aff3},\ref{aff4}}
\and G.~Zamorani\orcid{0000-0002-2318-301X}\inst{\ref{aff18}}
\and F.~M.~Zerbi\inst{\ref{aff17}}
\and I.~A.~Zinchenko\orcid{0000-0002-2944-2449}\inst{\ref{aff62}}
\and E.~Zucca\orcid{0000-0002-5845-8132}\inst{\ref{aff18}}
\and V.~Allevato\orcid{0000-0001-7232-5152}\inst{\ref{aff14}}
\and M.~Ballardini\orcid{0000-0003-4481-3559}\inst{\ref{aff115},\ref{aff116},\ref{aff18}}
\and M.~Bolzonella\orcid{0000-0003-3278-4607}\inst{\ref{aff18}}
\and E.~Bozzo\orcid{0000-0002-8201-1525}\inst{\ref{aff58}}
\and C.~Burigana\orcid{0000-0002-3005-5796}\inst{\ref{aff117},\ref{aff60}}
\and R.~Cabanac\orcid{0000-0001-6679-2600}\inst{\ref{aff106}}
\and A.~Cappi\inst{\ref{aff18},\ref{aff89}}
\and D.~Di~Ferdinando\inst{\ref{aff23}}
\and J.~A.~Escartin~Vigo\inst{\ref{aff63}}
\and L.~Gabarra\orcid{0000-0002-8486-8856}\inst{\ref{aff118}}
\and J.~Mart\'{i}n-Fleitas\orcid{0000-0002-8594-569X}\inst{\ref{aff92}}
\and S.~Matthew\orcid{0000-0001-8448-1697}\inst{\ref{aff48}}
\and M.~Maturi\orcid{0000-0002-3517-2422}\inst{\ref{aff105},\ref{aff119}}
\and N.~Mauri\orcid{0000-0001-8196-1548}\inst{\ref{aff46},\ref{aff23}}
\and R.~B.~Metcalf\orcid{0000-0003-3167-2574}\inst{\ref{aff87},\ref{aff18}}
\and A.~Pezzotta\orcid{0000-0003-0726-2268}\inst{\ref{aff120},\ref{aff63}}
\and M.~P\"ontinen\orcid{0000-0001-5442-2530}\inst{\ref{aff76}}
\and C.~Porciani\orcid{0000-0002-7797-2508}\inst{\ref{aff85}}
\and I.~Risso\orcid{0000-0003-2525-7761}\inst{\ref{aff121}}
\and V.~Scottez\inst{\ref{aff90},\ref{aff122}}
\and M.~Sereno\orcid{0000-0003-0302-0325}\inst{\ref{aff18},\ref{aff23}}
\and M.~Tenti\orcid{0000-0002-4254-5901}\inst{\ref{aff23}}
\and M.~Viel\orcid{0000-0002-2642-5707}\inst{\ref{aff4},\ref{aff3},\ref{aff21},\ref{aff20},\ref{aff123}}
\and M.~Wiesmann\orcid{0009-0000-8199-5860}\inst{\ref{aff65}}
\and Y.~Akrami\orcid{0000-0002-2407-7956}\inst{\ref{aff124},\ref{aff125}}
\and S.~Alvi\orcid{0000-0001-5779-8568}\inst{\ref{aff115}}
\and I.~T.~Andika\orcid{0000-0001-6102-9526}\inst{\ref{aff126},\ref{aff127}}
\and S.~Anselmi\orcid{0000-0002-3579-9583}\inst{\ref{aff104},\ref{aff103},\ref{aff128}}
\and M.~Archidiacono\orcid{0000-0003-4952-9012}\inst{\ref{aff82},\ref{aff83}}
\and F.~Atrio-Barandela\orcid{0000-0002-2130-2513}\inst{\ref{aff129}}
\and C.~Benoist\inst{\ref{aff89}}
\and K.~Benson\inst{\ref{aff55}}
\and D.~Bertacca\orcid{0000-0002-2490-7139}\inst{\ref{aff103},\ref{aff24},\ref{aff104}}
\and M.~Bethermin\orcid{0000-0002-3915-2015}\inst{\ref{aff130}}
\and A.~Blanchard\orcid{0000-0001-8555-9003}\inst{\ref{aff106}}
\and L.~Blot\orcid{0000-0002-9622-7167}\inst{\ref{aff131},\ref{aff128}}
\and H.~B\"ohringer\orcid{0000-0001-8241-4204}\inst{\ref{aff63},\ref{aff132},\ref{aff133}}
\and S.~Borgani\orcid{0000-0001-6151-6439}\inst{\ref{aff134},\ref{aff4},\ref{aff3},\ref{aff20},\ref{aff123}}
\and M.~L.~Brown\orcid{0000-0002-0370-8077}\inst{\ref{aff49}}
\and S.~Bruton\orcid{0000-0002-6503-5218}\inst{\ref{aff135}}
\and A.~Calabro\orcid{0000-0003-2536-1614}\inst{\ref{aff43}}
\and F.~Caro\inst{\ref{aff43}}
\and C.~S.~Carvalho\inst{\ref{aff111}}
\and T.~Castro\orcid{0000-0002-6292-3228}\inst{\ref{aff3},\ref{aff20},\ref{aff4},\ref{aff123}}
\and F.~Cogato\orcid{0000-0003-4632-6113}\inst{\ref{aff87},\ref{aff18}}
\and A.~R.~Cooray\orcid{0000-0002-3892-0190}\inst{\ref{aff136}}
\and O.~Cucciati\orcid{0000-0002-9336-7551}\inst{\ref{aff18}}
\and S.~Davini\orcid{0000-0003-3269-1718}\inst{\ref{aff28}}
\and F.~De~Paolis\orcid{0000-0001-6460-7563}\inst{\ref{aff137},\ref{aff138},\ref{aff139}}
\and G.~Desprez\orcid{0000-0001-8325-1742}\inst{\ref{aff12}}
\and A.~D\'iaz-S\'anchez\orcid{0000-0003-0748-4768}\inst{\ref{aff140}}
\and J.~J.~Diaz\inst{\ref{aff141}}
\and S.~Di~Domizio\orcid{0000-0003-2863-5895}\inst{\ref{aff27},\ref{aff28}}
\and J.~M.~Diego\orcid{0000-0001-9065-3926}\inst{\ref{aff142}}
\and P.-A.~Duc\orcid{0000-0003-3343-6284}\inst{\ref{aff130}}
\and A.~Enia\orcid{0000-0002-0200-2857}\inst{\ref{aff22},\ref{aff18}}
\and Y.~Fang\inst{\ref{aff62}}
\and A.~G.~Ferrari\orcid{0009-0005-5266-4110}\inst{\ref{aff23}}
\and P.~G.~Ferreira\orcid{0000-0002-3021-2851}\inst{\ref{aff118}}
\and A.~Finoguenov\orcid{0000-0002-4606-5403}\inst{\ref{aff76}}
\and A.~Fontana\orcid{0000-0003-3820-2823}\inst{\ref{aff43}}
\and A.~Franco\orcid{0000-0002-4761-366X}\inst{\ref{aff138},\ref{aff137},\ref{aff139}}
\and K.~Ganga\orcid{0000-0001-8159-8208}\inst{\ref{aff1}}
\and J.~Garc\'ia-Bellido\orcid{0000-0002-9370-8360}\inst{\ref{aff124}}
\and T.~Gasparetto\orcid{0000-0002-7913-4866}\inst{\ref{aff3}}
\and V.~Gautard\inst{\ref{aff143}}
\and E.~Gaztanaga\orcid{0000-0001-9632-0815}\inst{\ref{aff41},\ref{aff42},\ref{aff7}}
\and F.~Giacomini\orcid{0000-0002-3129-2814}\inst{\ref{aff23}}
\and F.~Gianotti\orcid{0000-0003-4666-119X}\inst{\ref{aff18}}
\and A.~H.~Gonzalez\orcid{0000-0002-0933-8601}\inst{\ref{aff144}}
\and G.~Gozaliasl\orcid{0000-0002-0236-919X}\inst{\ref{aff145},\ref{aff76}}
\and M.~Guidi\orcid{0000-0001-9408-1101}\inst{\ref{aff22},\ref{aff18}}
\and C.~M.~Gutierrez\orcid{0000-0001-7854-783X}\inst{\ref{aff146}}
\and A.~Hall\orcid{0000-0002-3139-8651}\inst{\ref{aff48}}
\and W.~G.~Hartley\inst{\ref{aff58}}
\and C.~Hern\'andez-Monteagudo\orcid{0000-0001-5471-9166}\inst{\ref{aff102},\ref{aff47}}
\and H.~Hildebrandt\orcid{0000-0002-9814-3338}\inst{\ref{aff147}}
\and J.~Hjorth\orcid{0000-0002-4571-2306}\inst{\ref{aff95}}
\and J.~J.~E.~Kajava\orcid{0000-0002-3010-8333}\inst{\ref{aff148},\ref{aff149}}
\and Y.~Kang\orcid{0009-0000-8588-7250}\inst{\ref{aff58}}
\and V.~Kansal\orcid{0000-0002-4008-6078}\inst{\ref{aff150},\ref{aff151}}
\and D.~Karagiannis\orcid{0000-0002-4927-0816}\inst{\ref{aff115},\ref{aff152}}
\and K.~Kiiveri\inst{\ref{aff74}}
\and C.~C.~Kirkpatrick\inst{\ref{aff74}}
\and S.~Kruk\orcid{0000-0001-8010-8879}\inst{\ref{aff15}}
\and L.~Legrand\orcid{0000-0003-0610-5252}\inst{\ref{aff153},\ref{aff154}}
\and M.~Lembo\orcid{0000-0002-5271-5070}\inst{\ref{aff115},\ref{aff116}}
\and F.~Lepori\orcid{0009-0000-5061-7138}\inst{\ref{aff155}}
\and G.~Leroy\orcid{0009-0004-2523-4425}\inst{\ref{aff156},\ref{aff88}}
\and G.~F.~Lesci\orcid{0000-0002-4607-2830}\inst{\ref{aff87},\ref{aff18}}
\and J.~Lesgourgues\orcid{0000-0001-7627-353X}\inst{\ref{aff40}}
\and L.~Leuzzi\orcid{0009-0006-4479-7017}\inst{\ref{aff87},\ref{aff18}}
\and T.~I.~Liaudat\orcid{0000-0002-9104-314X}\inst{\ref{aff157}}
\and A.~Loureiro\orcid{0000-0002-4371-0876}\inst{\ref{aff158},\ref{aff159}}
\and J.~Macias-Perez\orcid{0000-0002-5385-2763}\inst{\ref{aff160}}
\and G.~Maggio\orcid{0000-0003-4020-4836}\inst{\ref{aff3}}
\and E.~A.~Magnier\orcid{0000-0002-7965-2815}\inst{\ref{aff45}}
\and F.~Mannucci\orcid{0000-0002-4803-2381}\inst{\ref{aff161}}
\and R.~Maoli\orcid{0000-0002-6065-3025}\inst{\ref{aff162},\ref{aff43}}
\and C.~J.~A.~P.~Martins\orcid{0000-0002-4886-9261}\inst{\ref{aff163},\ref{aff30}}
\and L.~Maurin\orcid{0000-0002-8406-0857}\inst{\ref{aff10}}
\and M.~Miluzio\inst{\ref{aff15},\ref{aff164}}
\and P.~Monaco\orcid{0000-0003-2083-7564}\inst{\ref{aff134},\ref{aff3},\ref{aff20},\ref{aff4}}
\and C.~Moretti\orcid{0000-0003-3314-8936}\inst{\ref{aff21},\ref{aff123},\ref{aff3},\ref{aff4},\ref{aff20}}
\and G.~Morgante\inst{\ref{aff18}}
\and K.~Naidoo\orcid{0000-0002-9182-1802}\inst{\ref{aff7}}
\and A.~Navarro-Alsina\orcid{0000-0002-3173-2592}\inst{\ref{aff85}}
\and S.~Nesseris\orcid{0000-0002-0567-0324}\inst{\ref{aff124}}
\and F.~Passalacqua\orcid{0000-0002-8606-4093}\inst{\ref{aff103},\ref{aff104}}
\and K.~Paterson\orcid{0000-0001-8340-3486}\inst{\ref{aff71}}
\and L.~Patrizii\inst{\ref{aff23}}
\and A.~Pisani\orcid{0000-0002-6146-4437}\inst{\ref{aff59},\ref{aff165}}
\and D.~Potter\orcid{0000-0002-0757-5195}\inst{\ref{aff155}}
\and S.~Quai\orcid{0000-0002-0449-8163}\inst{\ref{aff87},\ref{aff18}}
\and M.~Radovich\orcid{0000-0002-3585-866X}\inst{\ref{aff24}}
\and P.-F.~Rocci\inst{\ref{aff10}}
\and G.~Rodighiero\orcid{0000-0002-9415-2296}\inst{\ref{aff103},\ref{aff24}}
\and S.~Sacquegna\orcid{0000-0002-8433-6630}\inst{\ref{aff137},\ref{aff138},\ref{aff139}}
\and M.~Sahl\'en\orcid{0000-0003-0973-4804}\inst{\ref{aff166}}
\and D.~B.~Sanders\orcid{0000-0002-1233-9998}\inst{\ref{aff45}}
\and E.~Sarpa\orcid{0000-0002-1256-655X}\inst{\ref{aff21},\ref{aff123},\ref{aff20}}
\and C.~Scarlata\orcid{0000-0002-9136-8876}\inst{\ref{aff167}}
\and J.~Schaye\orcid{0000-0002-0668-5560}\inst{\ref{aff36}}
\and A.~Schneider\orcid{0000-0001-7055-8104}\inst{\ref{aff155}}
\and M.~Schultheis\inst{\ref{aff89}}
\and D.~Sciotti\orcid{0009-0008-4519-2620}\inst{\ref{aff43},\ref{aff86}}
\and E.~Sellentin\inst{\ref{aff168},\ref{aff36}}
\and L.~C.~Smith\orcid{0000-0002-3259-2771}\inst{\ref{aff169}}
\and S.~A.~Stanford\orcid{0000-0003-0122-0841}\inst{\ref{aff170}}
\and K.~Tanidis\orcid{0000-0001-9843-5130}\inst{\ref{aff118}}
\and G.~Testera\inst{\ref{aff28}}
\and R.~Teyssier\orcid{0000-0001-7689-0933}\inst{\ref{aff165}}
\and S.~Tosi\orcid{0000-0002-7275-9193}\inst{\ref{aff27},\ref{aff28},\ref{aff17}}
\and A.~Troja\orcid{0000-0003-0239-4595}\inst{\ref{aff103},\ref{aff104}}
\and M.~Tucci\inst{\ref{aff58}}
\and C.~Valieri\inst{\ref{aff23}}
\and A.~Venhola\orcid{0000-0001-6071-4564}\inst{\ref{aff171}}
\and D.~Vergani\orcid{0000-0003-0898-2216}\inst{\ref{aff18}}
\and G.~Verza\orcid{0000-0002-1886-8348}\inst{\ref{aff172}}
\and P.~Vielzeuf\orcid{0000-0003-2035-9339}\inst{\ref{aff59}}
\and N.~A.~Walton\orcid{0000-0003-3983-8778}\inst{\ref{aff169}}
\and D.~Scott\orcid{0000-0002-6878-9840}\inst{\ref{aff173}}}
										   
\institute{Universit\'e Paris Cit\'e, CNRS, Astroparticule et Cosmologie, 75013 Paris, France\label{aff1}
\and
CNRS-UCB International Research Laboratory, Centre Pierre Binetruy, IRL2007, CPB-IN2P3, Berkeley, USA\label{aff2}
\and
INAF-Osservatorio Astronomico di Trieste, Via G. B. Tiepolo 11, 34143 Trieste, Italy\label{aff3}
\and
IFPU, Institute for Fundamental Physics of the Universe, via Beirut 2, 34151 Trieste, Italy\label{aff4}
\and
Center for Astronomy and Astrophysics and Department of Physics, Fudan University, Shanghai 200438, People's Republic of China\label{aff5}
\and
School of Physics \& Astronomy, University of Southampton, Highfield Campus, Southampton SO17 1BJ, UK\label{aff6}
\and
Institute of Cosmology and Gravitation, University of Portsmouth, Portsmouth PO1 3FX, UK\label{aff7}
\and
INAF-Istituto di Astrofisica e Planetologia Spaziali, via del Fosso del Cavaliere, 100, 00100 Roma, Italy\label{aff8}
\and
Univ. Lille, CNRS, Centrale Lille, UMR 9189 CRIStAL, 59000 Lille, France\label{aff9}
\and
Universit\'e Paris-Saclay, CNRS, Institut d'astrophysique spatiale, 91405, Orsay, France\label{aff10}
\and
Sterrenkundig Observatorium, Universiteit Gent, Krijgslaan 281 S9, 9000 Gent, Belgium\label{aff11}
\and
Kapteyn Astronomical Institute, University of Groningen, PO Box 800, 9700 AV Groningen, The Netherlands\label{aff12}
\and
School of Physics and Astronomy, Cardiff University, The Parade, Cardiff, CF24 3AA, UK\label{aff13}
\and
INAF-Osservatorio Astronomico di Capodimonte, Via Moiariello 16, 80131 Napoli, Italy\label{aff14}
\and
ESAC/ESA, Camino Bajo del Castillo, s/n., Urb. Villafranca del Castillo, 28692 Villanueva de la Ca\~nada, Madrid, Spain\label{aff15}
\and
School of Mathematics and Physics, University of Surrey, Guildford, Surrey, GU2 7XH, UK\label{aff16}
\and
INAF-Osservatorio Astronomico di Brera, Via Brera 28, 20122 Milano, Italy\label{aff17}
\and
INAF-Osservatorio di Astrofisica e Scienza dello Spazio di Bologna, Via Piero Gobetti 93/3, 40129 Bologna, Italy\label{aff18}
\and
Universit\'e Paris-Saclay, Universit\'e Paris Cit\'e, CEA, CNRS, AIM, 91191, Gif-sur-Yvette, France\label{aff19}
\and
INFN, Sezione di Trieste, Via Valerio 2, 34127 Trieste TS, Italy\label{aff20}
\and
SISSA, International School for Advanced Studies, Via Bonomea 265, 34136 Trieste TS, Italy\label{aff21}
\and
Dipartimento di Fisica e Astronomia, Universit\`a di Bologna, Via Gobetti 93/2, 40129 Bologna, Italy\label{aff22}
\and
INFN-Sezione di Bologna, Viale Berti Pichat 6/2, 40127 Bologna, Italy\label{aff23}
\and
INAF-Osservatorio Astronomico di Padova, Via dell'Osservatorio 5, 35122 Padova, Italy\label{aff24}
\and
Space Science Data Center, Italian Space Agency, via del Politecnico snc, 00133 Roma, Italy\label{aff25}
\and
INAF-Osservatorio Astrofisico di Torino, Via Osservatorio 20, 10025 Pino Torinese (TO), Italy\label{aff26}
\and
Dipartimento di Fisica, Universit\`a di Genova, Via Dodecaneso 33, 16146, Genova, Italy\label{aff27}
\and
INFN-Sezione di Genova, Via Dodecaneso 33, 16146, Genova, Italy\label{aff28}
\and
Department of Physics "E. Pancini", University Federico II, Via Cinthia 6, 80126, Napoli, Italy\label{aff29}
\and
Instituto de Astrof\'isica e Ci\^encias do Espa\c{c}o, Universidade do Porto, CAUP, Rua das Estrelas, PT4150-762 Porto, Portugal\label{aff30}
\and
Faculdade de Ci\^encias da Universidade do Porto, Rua do Campo de Alegre, 4150-007 Porto, Portugal\label{aff31}
\and
Dipartimento di Fisica, Universit\`a degli Studi di Torino, Via P. Giuria 1, 10125 Torino, Italy\label{aff32}
\and
INFN-Sezione di Torino, Via P. Giuria 1, 10125 Torino, Italy\label{aff33}
\and
European Space Agency/ESTEC, Keplerlaan 1, 2201 AZ Noordwijk, The Netherlands\label{aff34}
\and
Institute Lorentz, Leiden University, Niels Bohrweg 2, 2333 CA Leiden, The Netherlands\label{aff35}
\and
Leiden Observatory, Leiden University, Einsteinweg 55, 2333 CC Leiden, The Netherlands\label{aff36}
\and
INAF-IASF Milano, Via Alfonso Corti 12, 20133 Milano, Italy\label{aff37}
\and
Centro de Investigaciones Energ\'eticas, Medioambientales y Tecnol\'ogicas (CIEMAT), Avenida Complutense 40, 28040 Madrid, Spain\label{aff38}
\and
Port d'Informaci\'{o} Cient\'{i}fica, Campus UAB, C. Albareda s/n, 08193 Bellaterra (Barcelona), Spain\label{aff39}
\and
Institute for Theoretical Particle Physics and Cosmology (TTK), RWTH Aachen University, 52056 Aachen, Germany\label{aff40}
\and
Institute of Space Sciences (ICE, CSIC), Campus UAB, Carrer de Can Magrans, s/n, 08193 Barcelona, Spain\label{aff41}
\and
Institut d'Estudis Espacials de Catalunya (IEEC),  Edifici RDIT, Campus UPC, 08860 Castelldefels, Barcelona, Spain\label{aff42}
\and
INAF-Osservatorio Astronomico di Roma, Via Frascati 33, 00078 Monteporzio Catone, Italy\label{aff43}
\and
INFN section of Naples, Via Cinthia 6, 80126, Napoli, Italy\label{aff44}
\and
Institute for Astronomy, University of Hawaii, 2680 Woodlawn Drive, Honolulu, HI 96822, USA\label{aff45}
\and
Dipartimento di Fisica e Astronomia "Augusto Righi" - Alma Mater Studiorum Universit\`a di Bologna, Viale Berti Pichat 6/2, 40127 Bologna, Italy\label{aff46}
\and
Instituto de Astrof\'{\i}sica de Canarias, V\'{\i}a L\'actea, 38205 La Laguna, Tenerife, Spain\label{aff47}
\and
Institute for Astronomy, University of Edinburgh, Royal Observatory, Blackford Hill, Edinburgh EH9 3HJ, UK\label{aff48}
\and
Jodrell Bank Centre for Astrophysics, Department of Physics and Astronomy, University of Manchester, Oxford Road, Manchester M13 9PL, UK\label{aff49}
\and
European Space Agency/ESRIN, Largo Galileo Galilei 1, 00044 Frascati, Roma, Italy\label{aff50}
\and
Universit\'e Claude Bernard Lyon 1, CNRS/IN2P3, IP2I Lyon, UMR 5822, Villeurbanne, F-69100, France\label{aff51}
\and
Institut de Ci\`{e}ncies del Cosmos (ICCUB), Universitat de Barcelona (IEEC-UB), Mart\'{i} i Franqu\`{e}s 1, 08028 Barcelona, Spain\label{aff52}
\and
Instituci\'o Catalana de Recerca i Estudis Avan\c{c}ats (ICREA), Passeig de Llu\'{\i}s Companys 23, 08010 Barcelona, Spain\label{aff53}
\and
UCB Lyon 1, CNRS/IN2P3, IUF, IP2I Lyon, 4 rue Enrico Fermi, 69622 Villeurbanne, France\label{aff54}
\and
Mullard Space Science Laboratory, University College London, Holmbury St Mary, Dorking, Surrey RH5 6NT, UK\label{aff55}
\and
Departamento de F\'isica, Faculdade de Ci\^encias, Universidade de Lisboa, Edif\'icio C8, Campo Grande, PT1749-016 Lisboa, Portugal\label{aff56}
\and
Instituto de Astrof\'isica e Ci\^encias do Espa\c{c}o, Faculdade de Ci\^encias, Universidade de Lisboa, Campo Grande, 1749-016 Lisboa, Portugal\label{aff57}
\and
Department of Astronomy, University of Geneva, ch. d'Ecogia 16, 1290 Versoix, Switzerland\label{aff58}
\and
Aix-Marseille Universit\'e, CNRS/IN2P3, CPPM, Marseille, France\label{aff59}
\and
INFN-Bologna, Via Irnerio 46, 40126 Bologna, Italy\label{aff60}
\and
School of Physics, HH Wills Physics Laboratory, University of Bristol, Tyndall Avenue, Bristol, BS8 1TL, UK\label{aff61}
\and
Universit\"ats-Sternwarte M\"unchen, Fakult\"at f\"ur Physik, Ludwig-Maximilians-Universit\"at M\"unchen, Scheinerstrasse 1, 81679 M\"unchen, Germany\label{aff62}
\and
Max Planck Institute for Extraterrestrial Physics, Giessenbachstr. 1, 85748 Garching, Germany\label{aff63}
\and
NRC Herzberg, 5071 West Saanich Rd, Victoria, BC V9E 2E7, Canada\label{aff64}
\and
Institute of Theoretical Astrophysics, University of Oslo, P.O. Box 1029 Blindern, 0315 Oslo, Norway\label{aff65}
\and
Jet Propulsion Laboratory, California Institute of Technology, 4800 Oak Grove Drive, Pasadena, CA, 91109, USA\label{aff66}
\and
Felix Hormuth Engineering, Goethestr. 17, 69181 Leimen, Germany\label{aff67}
\and
Technical University of Denmark, Elektrovej 327, 2800 Kgs. Lyngby, Denmark\label{aff68}
\and
Cosmic Dawn Center (DAWN), Denmark\label{aff69}
\and
Institut d'Astrophysique de Paris, UMR 7095, CNRS, and Sorbonne Universit\'e, 98 bis boulevard Arago, 75014 Paris, France\label{aff70}
\and
Max-Planck-Institut f\"ur Astronomie, K\"onigstuhl 17, 69117 Heidelberg, Germany\label{aff71}
\and
NASA Goddard Space Flight Center, Greenbelt, MD 20771, USA\label{aff72}
\and
Department of Physics and Astronomy, University College London, Gower Street, London WC1E 6BT, UK\label{aff73}
\and
Department of Physics and Helsinki Institute of Physics, Gustaf H\"allstr\"omin katu 2, 00014 University of Helsinki, Finland\label{aff74}
\and
Universit\'e de Gen\`eve, D\'epartement de Physique Th\'eorique and Centre for Astroparticle Physics, 24 quai Ernest-Ansermet, CH-1211 Gen\`eve 4, Switzerland\label{aff75}
\and
Department of Physics, P.O. Box 64, 00014 University of Helsinki, Finland\label{aff76}
\and
Helsinki Institute of Physics, Gustaf H{\"a}llstr{\"o}min katu 2, University of Helsinki, Helsinki, Finland\label{aff77}
\and
Centre de Calcul de l'IN2P3/CNRS, 21 avenue Pierre de Coubertin 69627 Villeurbanne Cedex, France\label{aff78}
\and
Laboratoire d'etude de l'Univers et des phenomenes eXtremes, Observatoire de Paris, Universit\'e PSL, Sorbonne Universit\'e, CNRS, 92190 Meudon, France\label{aff79}
\and
Aix-Marseille Universit\'e, CNRS, CNES, LAM, Marseille, France\label{aff80}
\and
SKA Observatory, Jodrell Bank, Lower Withington, Macclesfield, Cheshire SK11 9FT, UK\label{aff81}
\and
Dipartimento di Fisica "Aldo Pontremoli", Universit\`a degli Studi di Milano, Via Celoria 16, 20133 Milano, Italy\label{aff82}
\and
INFN-Sezione di Milano, Via Celoria 16, 20133 Milano, Italy\label{aff83}
\and
University of Applied Sciences and Arts of Northwestern Switzerland, School of Computer Science, 5210 Windisch, Switzerland\label{aff84}
\and
Universit\"at Bonn, Argelander-Institut f\"ur Astronomie, Auf dem H\"ugel 71, 53121 Bonn, Germany\label{aff85}
\and
INFN-Sezione di Roma, Piazzale Aldo Moro, 2 - c/o Dipartimento di Fisica, Edificio G. Marconi, 00185 Roma, Italy\label{aff86}
\and
Dipartimento di Fisica e Astronomia "Augusto Righi" - Alma Mater Studiorum Universit\`a di Bologna, via Piero Gobetti 93/2, 40129 Bologna, Italy\label{aff87}
\and
Department of Physics, Institute for Computational Cosmology, Durham University, South Road, Durham, DH1 3LE, UK\label{aff88}
\and
Universit\'e C\^{o}te d'Azur, Observatoire de la C\^{o}te d'Azur, CNRS, Laboratoire Lagrange, Bd de l'Observatoire, CS 34229, 06304 Nice cedex 4, France\label{aff89}
\and
Institut d'Astrophysique de Paris, 98bis Boulevard Arago, 75014, Paris, France\label{aff90}
\and
Institute of Physics, Laboratory of Astrophysics, Ecole Polytechnique F\'ed\'erale de Lausanne (EPFL), Observatoire de Sauverny, 1290 Versoix, Switzerland\label{aff91}
\and
Aurora Technology for European Space Agency (ESA), Camino bajo del Castillo, s/n, Urbanizacion Villafranca del Castillo, Villanueva de la Ca\~nada, 28692 Madrid, Spain\label{aff92}
\and
Institut de F\'{i}sica d'Altes Energies (IFAE), The Barcelona Institute of Science and Technology, Campus UAB, 08193 Bellaterra (Barcelona), Spain\label{aff93}
\and
School of Mathematics, Statistics and Physics, Newcastle University, Herschel Building, Newcastle-upon-Tyne, NE1 7RU, UK\label{aff94}
\and
DARK, Niels Bohr Institute, University of Copenhagen, Jagtvej 155, 2200 Copenhagen, Denmark\label{aff95}
\and
Waterloo Centre for Astrophysics, University of Waterloo, Waterloo, Ontario N2L 3G1, Canada\label{aff96}
\and
Department of Physics and Astronomy, University of Waterloo, Waterloo, Ontario N2L 3G1, Canada\label{aff97}
\and
Perimeter Institute for Theoretical Physics, Waterloo, Ontario N2L 2Y5, Canada\label{aff98}
\and
Centre National d'Etudes Spatiales -- Centre spatial de Toulouse, 18 avenue Edouard Belin, 31401 Toulouse Cedex 9, France\label{aff99}
\and
Institute of Space Science, Str. Atomistilor, nr. 409 M\u{a}gurele, Ilfov, 077125, Romania\label{aff100}
\and
Consejo Superior de Investigaciones Cientificas, Calle Serrano 117, 28006 Madrid, Spain\label{aff101}
\and
Universidad de La Laguna, Departamento de Astrof\'{\i}sica, 38206 La Laguna, Tenerife, Spain\label{aff102}
\and
Dipartimento di Fisica e Astronomia "G. Galilei", Universit\`a di Padova, Via Marzolo 8, 35131 Padova, Italy\label{aff103}
\and
INFN-Padova, Via Marzolo 8, 35131 Padova, Italy\label{aff104}
\and
Institut f\"ur Theoretische Physik, University of Heidelberg, Philosophenweg 16, 69120 Heidelberg, Germany\label{aff105}
\and
Institut de Recherche en Astrophysique et Plan\'etologie (IRAP), Universit\'e de Toulouse, CNRS, UPS, CNES, 14 Av. Edouard Belin, 31400 Toulouse, France\label{aff106}
\and
Universit\'e St Joseph; Faculty of Sciences, Beirut, Lebanon\label{aff107}
\and
Departamento de F\'isica, FCFM, Universidad de Chile, Blanco Encalada 2008, Santiago, Chile\label{aff108}
\and
Satlantis, University Science Park, Sede Bld 48940, Leioa-Bilbao, Spain\label{aff109}
\and
Infrared Processing and Analysis Center, California Institute of Technology, Pasadena, CA 91125, USA\label{aff110}
\and
Instituto de Astrof\'isica e Ci\^encias do Espa\c{c}o, Faculdade de Ci\^encias, Universidade de Lisboa, Tapada da Ajuda, 1349-018 Lisboa, Portugal\label{aff111}
\and
Cosmic Dawn Center (DAWN)\label{aff112}
\and
Niels Bohr Institute, University of Copenhagen, Jagtvej 128, 2200 Copenhagen, Denmark\label{aff113}
\and
Universidad Polit\'ecnica de Cartagena, Departamento de Electr\'onica y Tecnolog\'ia de Computadoras,  Plaza del Hospital 1, 30202 Cartagena, Spain\label{aff114}
\and
Dipartimento di Fisica e Scienze della Terra, Universit\`a degli Studi di Ferrara, Via Giuseppe Saragat 1, 44122 Ferrara, Italy\label{aff115}
\and
Istituto Nazionale di Fisica Nucleare, Sezione di Ferrara, Via Giuseppe Saragat 1, 44122 Ferrara, Italy\label{aff116}
\and
INAF, Istituto di Radioastronomia, Via Piero Gobetti 101, 40129 Bologna, Italy\label{aff117}
\and
Department of Physics, Oxford University, Keble Road, Oxford OX1 3RH, UK\label{aff118}
\and
Zentrum f\"ur Astronomie, Universit\"at Heidelberg, Philosophenweg 12, 69120 Heidelberg, Germany\label{aff119}
\and
INAF - Osservatorio Astronomico di Brera, via Emilio Bianchi 46, 23807 Merate, Italy\label{aff120}
\and
INAF-Osservatorio Astronomico di Brera, Via Brera 28, 20122 Milano, Italy, and INFN-Sezione di Genova, Via Dodecaneso 33, 16146, Genova, Italy\label{aff121}
\and
ICL, Junia, Universit\'e Catholique de Lille, LITL, 59000 Lille, France\label{aff122}
\and
ICSC - Centro Nazionale di Ricerca in High Performance Computing, Big Data e Quantum Computing, Via Magnanelli 2, Bologna, Italy\label{aff123}
\and
Instituto de F\'isica Te\'orica UAM-CSIC, Campus de Cantoblanco, 28049 Madrid, Spain\label{aff124}
\and
CERCA/ISO, Department of Physics, Case Western Reserve University, 10900 Euclid Avenue, Cleveland, OH 44106, USA\label{aff125}
\and
Technical University of Munich, TUM School of Natural Sciences, Physics Department, James-Franck-Str.~1, 85748 Garching, Germany\label{aff126}
\and
Max-Planck-Institut f\"ur Astrophysik, Karl-Schwarzschild-Str.~1, 85748 Garching, Germany\label{aff127}
\and
Laboratoire Univers et Th\'eorie, Observatoire de Paris, Universit\'e PSL, Universit\'e Paris Cit\'e, CNRS, 92190 Meudon, France\label{aff128}
\and
Departamento de F{\'\i}sica Fundamental. Universidad de Salamanca. Plaza de la Merced s/n. 37008 Salamanca, Spain\label{aff129}
\and
Universit\'e de Strasbourg, CNRS, Observatoire astronomique de Strasbourg, UMR 7550, 67000 Strasbourg, France\label{aff130}
\and
Center for Data-Driven Discovery, Kavli IPMU (WPI), UTIAS, The University of Tokyo, Kashiwa, Chiba 277-8583, Japan\label{aff131}
\and
Ludwig-Maximilians-University, Schellingstrasse 4, 80799 Munich, Germany\label{aff132}
\and
Max-Planck-Institut f\"ur Physik, Boltzmannstr. 8, 85748 Garching, Germany\label{aff133}
\and
Dipartimento di Fisica - Sezione di Astronomia, Universit\`a di Trieste, Via Tiepolo 11, 34131 Trieste, Italy\label{aff134}
\and
California Institute of Technology, 1200 E California Blvd, Pasadena, CA 91125, USA\label{aff135}
\and
Department of Physics \& Astronomy, University of California Irvine, Irvine CA 92697, USA\label{aff136}
\and
Department of Mathematics and Physics E. De Giorgi, University of Salento, Via per Arnesano, CP-I93, 73100, Lecce, Italy\label{aff137}
\and
INFN, Sezione di Lecce, Via per Arnesano, CP-193, 73100, Lecce, Italy\label{aff138}
\and
INAF-Sezione di Lecce, c/o Dipartimento Matematica e Fisica, Via per Arnesano, 73100, Lecce, Italy\label{aff139}
\and
Departamento F\'isica Aplicada, Universidad Polit\'ecnica de Cartagena, Campus Muralla del Mar, 30202 Cartagena, Murcia, Spain\label{aff140}
\and
Instituto de Astrof\'isica de Canarias (IAC); Departamento de Astrof\'isica, Universidad de La Laguna (ULL), 38200, La Laguna, Tenerife, Spain\label{aff141}
\and
Instituto de F\'isica de Cantabria, Edificio Juan Jord\'a, Avenida de los Castros, 39005 Santander, Spain\label{aff142}
\and
CEA Saclay, DFR/IRFU, Service d'Astrophysique, Bat. 709, 91191 Gif-sur-Yvette, France\label{aff143}
\and
Department of Astronomy, University of Florida, Bryant Space Science Center, Gainesville, FL 32611, USA\label{aff144}
\and
Department of Computer Science, Aalto University, PO Box 15400, Espoo, FI-00 076, Finland\label{aff145}
\and
Instituto de Astrof\'\i sica de Canarias, c/ Via Lactea s/n, La Laguna 38200, Spain. Departamento de Astrof\'\i sica de la Universidad de La Laguna, Avda. Francisco Sanchez, La Laguna, 38200, Spain\label{aff146}
\and
Ruhr University Bochum, Faculty of Physics and Astronomy, Astronomical Institute (AIRUB), German Centre for Cosmological Lensing (GCCL), 44780 Bochum, Germany\label{aff147}
\and
Department of Physics and Astronomy, Vesilinnantie 5, 20014 University of Turku, Finland\label{aff148}
\and
Serco for European Space Agency (ESA), Camino bajo del Castillo, s/n, Urbanizacion Villafranca del Castillo, Villanueva de la Ca\~nada, 28692 Madrid, Spain\label{aff149}
\and
ARC Centre of Excellence for Dark Matter Particle Physics, Melbourne, Australia\label{aff150}
\and
Centre for Astrophysics \& Supercomputing, Swinburne University of Technology,  Hawthorn, Victoria 3122, Australia\label{aff151}
\and
Department of Physics and Astronomy, University of the Western Cape, Bellville, Cape Town, 7535, South Africa\label{aff152}
\and
DAMTP, Centre for Mathematical Sciences, Wilberforce Road, Cambridge CB3 0WA, UK\label{aff153}
\and
Kavli Institute for Cosmology Cambridge, Madingley Road, Cambridge, CB3 0HA, UK\label{aff154}
\and
Department of Astrophysics, University of Zurich, Winterthurerstrasse 190, 8057 Zurich, Switzerland\label{aff155}
\and
Department of Physics, Centre for Extragalactic Astronomy, Durham University, South Road, Durham, DH1 3LE, UK\label{aff156}
\and
IRFU, CEA, Universit\'e Paris-Saclay 91191 Gif-sur-Yvette Cedex, France\label{aff157}
\and
Oskar Klein Centre for Cosmoparticle Physics, Department of Physics, Stockholm University, Stockholm, SE-106 91, Sweden\label{aff158}
\and
Astrophysics Group, Blackett Laboratory, Imperial College London, London SW7 2AZ, UK\label{aff159}
\and
Univ. Grenoble Alpes, CNRS, Grenoble INP, LPSC-IN2P3, 53, Avenue des Martyrs, 38000, Grenoble, France\label{aff160}
\and
INAF-Osservatorio Astrofisico di Arcetri, Largo E. Fermi 5, 50125, Firenze, Italy\label{aff161}
\and
Dipartimento di Fisica, Sapienza Universit\`a di Roma, Piazzale Aldo Moro 2, 00185 Roma, Italy\label{aff162}
\and
Centro de Astrof\'{\i}sica da Universidade do Porto, Rua das Estrelas, 4150-762 Porto, Portugal\label{aff163}
\and
HE Space for European Space Agency (ESA), Camino bajo del Castillo, s/n, Urbanizacion Villafranca del Castillo, Villanueva de la Ca\~nada, 28692 Madrid, Spain\label{aff164}
\and
Department of Astrophysical Sciences, Peyton Hall, Princeton University, Princeton, NJ 08544, USA\label{aff165}
\and
Theoretical astrophysics, Department of Physics and Astronomy, Uppsala University, Box 515, 751 20 Uppsala, Sweden\label{aff166}
\and
Minnesota Institute for Astrophysics, University of Minnesota, 116 Church St SE, Minneapolis, MN 55455, USA\label{aff167}
\and
Mathematical Institute, University of Leiden, Einsteinweg 55, 2333 CA Leiden, The Netherlands\label{aff168}
\and
Institute of Astronomy, University of Cambridge, Madingley Road, Cambridge CB3 0HA, UK\label{aff169}
\and
Department of Physics and Astronomy, University of California, Davis, CA 95616, USA\label{aff170}
\and
Space physics and astronomy research unit, University of Oulu, Pentti Kaiteran katu 1, FI-90014 Oulu, Finland\label{aff171}
\and
Center for Computational Astrophysics, Flatiron Institute, 162 5th Avenue, 10010, New York, NY, USA\label{aff172}
\and
Department of Physics and Astronomy, University of British Columbia, Vancouver, BC V6T 1Z1, Canada\label{aff173}}    

\abstract{The extent to which the environment affects galaxy evolution has been under scrutiny by researchers for decades. With the first data from \Euclid, we can begin to systematically study a wide range of environments and their effects as a function of redshift, using $63\,\text{deg}^2$ of space-based data. In this paper, we present results from the \Euclid\ Quick Data Release, where we measured the passive-density and morphology-density relations in the redshift range $z=0.25$--$1$. We determined if a galaxy is passive using the specific star formation rate, and we classified the morphologies of galaxies using the S\'ersic index $n$ and the $u-r$ colours. We measured the local environmental density of each galaxy using the $N\text{th}$-nearest neighbour method. We find that at a fixed stellar mass, the quenched fraction (the fraction of galaxies that have ceased star formation) increases with increasing local environmental density {up to $z=1$}. This result is indicative of the separability of the effects from the stellar mass and the environment. Similarly, at all redshifts in this work, the early-type galaxy fraction increases with increasing density at fixed stellar mass, meaning the environment also transforms the morphology of the galaxy independently of stellar mass, up to $M_\ast \lesssim 10^{10.8}\ M_\odot$. For $M_\ast \gtrsim10^{10.8}\ M_\odot$, almost all galaxies are early-types, with minimal impact from the environment. At $z>0.75$, the morphology depends mostly on stellar mass, with only low-mass galaxies being affected by the environment. Given that the morphology classifications use $u-r$ colours, these are correlated to the star formation rate, and as such our morphology results should be taken with caution, yet future morphology classifications should help verify these results. To summarise, we successfully identify the passive-density and morphology-density relations at $0.25<z<1$. Future \Euclid\ data releases are key to confirm these trends at higher redshifts.
}


   \titlerunning{Passive-density and morphology-density relations in Q1 }
   \authorrunning{Euclid Collaboration: Cleland et al.}
   
   \maketitle
    \nolinenumbers
   
\section{\label{sec:Intro}Introduction}

It is clear that the environment plays a significant role in the evolution of galaxies at low redshift. The passive-density relation shows that galaxies in higher-density environments such as groups or clusters are more likely to be quenched than galaxies in the field of the same stellar mass \citep{Peng2010,Vulcani2010,Wetzel2012,Paccagnella2016,Darvish2017,Cleland2021,Corcho-Caballero2023}. Moreover, this effect is more prominent in low-stellar-mass satellite galaxies $(M_\ast\lesssim 10^{10}\,M_\odot)$, while at higher mass, the quenched fractions of satellite galaxies and central galaxies evolve similarly \citep{Cleland2021,Corcho-Caballero2023}, meaning the environment is not preferentially affecting one class of galaxy over another. This phenomenon of quenching at high masses has been described as `mass-quenching' \citep{Peng2010}, and loosely encompasses the quenching mechanisms that occur irrespective of the environment. These quenching mechanisms are also referred to as `internal quenching mechanisms', to distinguish them from `external quenching mechanisms' by the environment. Furthermore, the morphology-density relation is well studied at low $z$: galaxies in high-density environments tend to have more spheroidal morphologies, whereas galaxies in the field are more discy \citep{Dressler1980,Kauffmann2003,Postman2005}. It should be noted that, while the star formation rate (SFR) and morphology of a galaxy are related, they are not the same thing, and the processes that affect them operate in different ways. In other words, many galaxies are passive and early-type, or blue and late-type, but there exist populations of red spirals and blue ellipticals at low redshift \citep{Masters2010,Rowlands2012,Tojeiro2013,McIntosh2014,George2015,George2017}, as well as at high redshift \citep{Mei2006,Mei2015,Fudamoto2022,Mei2023}. The observed differences between these populations are indicative of distinct formation pathways, at low and high redshift. Thus, it is important to disentangle how the star formation and the morphology of galaxies are evolving separately. 

\setlength{\tabcolsep}{10pt} 
\renewcommand{\arraystretch}{1.5} 

These results in the local Universe paint a picture in which high-density environments have a transformative effect on a galaxy's star formation processes and its morphology. These environmental effects include `starvation', a process that prevents the accretion of cool fresh gas onto the galaxy \citep*{Larson1980,Balogh2000}, and `ram-pressure stripping', whereby the gas reservoir of the galaxy is rapidly removed after infall to the cluster \citep{Gunn1972}. Moreover, the structure of the galaxy may be affected by `harassment', whereby increased tidal forces between the galaxy and the cluster, and high-speed encounters with other galaxies, can disrupt the stellar disc \citep{Moore1996}, and by increased rates of mergers in certain environments, which transform the morphology of the galaxy while also accelerating star formation, leading to eventual quenching \citep{Toomre1972}.

This picture is less clear at $z>1$. Much work is ongoing in the search for massive passive galaxies at high redshifts \citep[particularly at $z>2$;][]{Straatman2016,Merlin2018,Merlin2019,Carnall2023,Carnall2024}; however, observing these objects is still a challenge. Quantifying the environments in which they reside is even more difficult. The extent to which the environment affects galaxy properties, and the epoch at which this begins, is under much scrutiny. Notably, the high-density environment looks very different at $z=1$--$2$ compared to that at low $z$. There have been many studies on various observed `proto-clusters' (\citealt*{Chiang2013}; \citealt{Overzier2016,Chiang2017,Lovell2018}), which form the candidate progenitors of galaxy clusters at $z=0$. In the Spiderweb protocluster at $z=2.2$, there is no strong evidence of an environmental effect on the star formation or morphology of the galaxies \citep{Perez-Martinez2023}. However, in a different Spiderweb study, it was found that four out of five massive star-forming galaxies may host an active galactic nucleus (AGN), which may eventually suppress star formation and quench the galaxy \citep{Shimakawa2018b}. In a study of clusters and proto-clusters at $1.3<z<2.8$, there is already evidence of the passive-density relation at $z\sim2$ \citep{Mei2023}. By $z\simeq1.6$, cluster galaxies show strong radial colour gradients compared to field galaxies \citep{Cramer2024}, implying the onset of environmental quenching between $z\approx2$ and $z\approx1$ \citep{Edward2024}. Equally, the redshift at which the morphology-density relation first appears is uncertain. In $z\simeq1$ clusters, there is a clear increase in the fraction of early-type galaxies (ETGs) at higher local densities \citep[$>1000\ \text{gal}\, \text{Mpc}^{-2}$,][]{Postman2005}. Further, as has been demonstrated by \citet{Mei2023}, the morphology-density relation is found to exist at $z\simeq2$; however, more statistics are needed to confirm if the relation is present at $z>2$.

These results suggest the importance of the cluster and proto-cluster environment in building up the star formation of galaxies and then rapidly quenching them \citep{Elbaz2007,Wang2016}. However, identifying large samples of high-density environments over a large redshift range is a challenge. The \Euclid\ mission \citep{EuclidSkyOverview} is well suited to overcome these challenges. The Euclid Wide Survey \citep{Scaramella-EP1} aims to observe up to $14\,000\ \text{deg}^{2}$, providing high-resolution data on billions of galaxies. The size of the \Euclid\ footprint, and the addition of external ground-based data \citep[][]{Q1-TP004, Rhodes2017} will allow for local galaxy density measurements in a wide range of environments, at this key epoch of galaxy evolution, at $z\approx1$. {Despite the fact that this paper focuses on galaxy evolution up to $z=1$, the references above highlight the fact that this epoch serves as a transition between the low-redshift regime and the proto-cluster regime. Future \Euclid releases will help us push even further into cosmic noon. }

In this paper, we use data from the \citet{Q1cite} to investigate the significance of the environment in quenching galaxies and transforming their morphology, as a function of redshift. We calculate the local environmental density for each galaxy, and use that to measure the passive-density and morphology-density relations as functions of redshift. This paper is organised as follows: Sect. \ref{sec:Data} describes the data used, and how we derive the measurements of local environmental density; Sect. \ref{sec:Res} explains our results for the passive-density relation and the morphology-density relation; finally, Sect. \ref{sec:disc} discusses these results in the context of other studies and summarises our results. For cosmological calculations, we use the \textit{Planck} cosmological parameters, \cite{PlanckCollaboration2016}: $\Omega_\text{m}=0.313$, $\Omega_\Lambda=0.687$, $\Omega_\text{b}=0.048$, $H_0=67.31\ \text{km s}^{-1}\ \text{Mpc}^{-1}$, and $n_\text{s}=0.966$.

\section{\label{sec:Data}Data}
\subsection{\label{eucdata}\Euclid\ data}

The \cite{Q1cite} consists of observations of three fields, Euclid Deep Field North (EDF-N), Euclid Deep Field South (EDF-S), and Euclid Deep Field Fornax (EDF-F), consisting of  $22.9\, \deg^2$, $28.1\, \deg^2$, and $12.1\, \deg^2$, respectively \citep{Q1-TP001}. 
This release consists of data products from the OU-MER pipeline \citep{Q1-TP004}, which includes galaxy co-ordinates, photometric data from the VIS \citep{Q1-TP002} and NISP \citep{Q1-TP003} instruments, external data, and also morphology information, such as the S\'ersic index from \texttt{SourceXtractor++} \citep{Bertin2020,Kummel2022}, and other morphological parameters acquired by machine learning \citep{Q1-TP004}. The visible filter, \IE, reaches a $5\,\sigma$ point-source sensitivity of 24.5~mag \citep{EuclidSkyVIS}, and the NIR filters, \YE, \JE, and \HE, each reach 24~mag \citep{EuclidSkyNISP}. For a detailed overview of the \Euclid\ mission, we refer the reader to \citet{EuclidSkyOverview}.

{Photometric redshifts and galaxy physical properties were produced by OU-PHZ \citep{Q1-TP005}. The model \texttt{Nearest-Neighbour Photometric Redshifts} (\texttt{NNPZ}) returns redshifts, stellar masses, SFRs based on a Kroupa inital mass function (IMF) \citep{Kroupa2001}, and galaxy ages, using the nearest neighbours from a calibration sample in the multi-dimensional phase-space \citep{Q1-TP005}. In this work, we utilised the re-run of \texttt{NNPZ} by \citet[][hereafter \citetalias{Q1-SP031}]{Q1-SP031}, which supplements Q1 data with the $3.6\ \mu$m and $4.5\ \mu$m channels from \textit{Spitzer}/IRAC. The addition of these bands significantly improves the accuracy of the estimates of redshifts and galaxy physical properties derived from photometric observations. The re-run also makes slight modifications to some priors, which ultimately prevents the need to remove unphysical data\footnote{The original Q1 data exhibited, for example, an unrealistic population of galaxies with unphysically young ages near the prior boundary, which previously had to be removed using a cut.}. We refer the reader to \citetalias{Q1-SP031} for a thorough explanation on this \texttt{NNPZ} re-run.} 

\begin{table}[]
    \centering
    \caption{Median uncertainty ($\sigma$) on the redshift for each redshift bin.}
    \begin{tabular}{c c}
    \hline\hline
        Redshift bin $(z)$& Median uncertainty $(\text{d}z)$\\
        \hline
        $0.25$--$0.50$ & $0.007$ \\
        $0.50$--$0.75$ & $0.018$ \\
        $0.75$--$1.00$ & $0.025$\\
        \hline
    \end{tabular}
    
    \label{tab:z_err}
\end{table}

A delayed exponential star formation history model was used to generate the model spectral energy distributions, and two dust extinction laws were used, \citet{Calzetti2000} and \citet{Prevot1984}; full details can be found in \citet{Q1-TP005}. For the redshifts, stellar masses, SFRs, and galaxy ages, we used the medians of the posterior distributions provided by \texttt{NNPZ}. We then measured the specific star formation rate (sSFR) by dividing the SFR by the stellar mass, i.e. $\text{sSFR}=\left(\text{SFR} /M_\ast\right)$.

\begin{figure}
    \centering
            \includegraphics[width=\columnwidth]{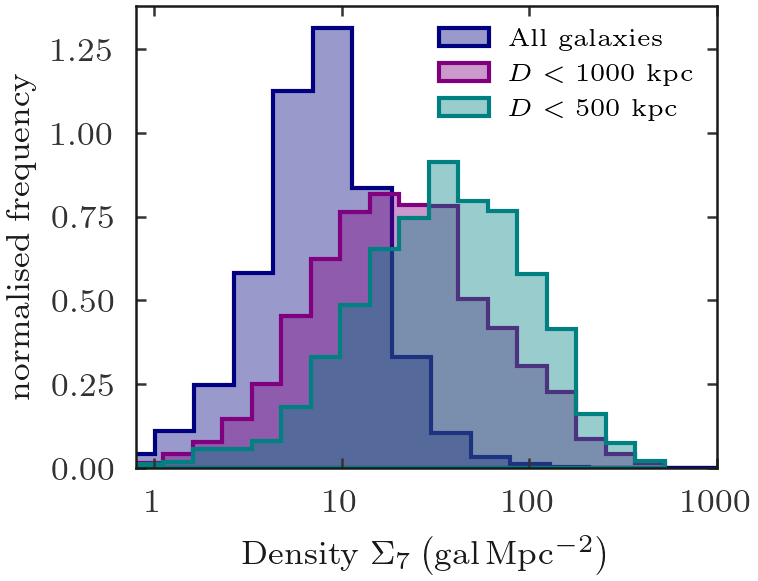}
    \caption{{Distribution of local density for all Q1 galaxies, and for galaxies where the projected distance, $D$, to a known cluster is $<1000\ \text{kpc}$ and $<500\ \text{kpc}$. We see that the Q1 galaxies at closer distances to known clusters have larger values of local density.}}
    \label{fig:close_gals}
\end{figure}

{The catalogue provided by \citetalias{Q1-SP031} already has some quality cuts made. These cuts remove spurious, saturated, or masked data, as well as stars. We describe the cuts below:}

\begin{itemize}
    \item \texttt{SPURIOUS$\_$FLAG} $= 0$;
    \item \texttt{DET\_QUALITY\_FLAG} $< 4$;
    \item \texttt{MUMAX\_MINUS\_MAG} $> -2.6$.
\end{itemize}

{Furthermore, we imposed a magnitude limit of \HE $<24$, and only included galaxies above the 85\% stellar mass completeness limit for passive galaxies at $z=1$, which we calculated to be $M_\ast>10^{9.5}\ M_\odot$, following \citet{Pozzetti2010}. We verified that this limit in stellar mass corresponds well to the magnitude cut, such that when we cut in magnitude we are not further biasing our results against faint, passive galaxies that would otherwise meet the threshold in stellar mass. Additionally, we checked how this magnitude cut affects our quenched fractions. We found there is a negligible difference, of about 2 percentage points, in the quenched fractions in all redshift bins, at $M_\ast<10^{11.5}\ \msun$. Finally, for the purposes of this paper we restricted our analysis to $0.25<z<1$, due to the fact that at $z>1$ the median uncertainty on the redshift increases. This results in a final sample of \num{1155139} galaxies.}

\subsection{\label{sec:dens}Density measurements}
The local environmental density was calculated for each galaxy using the $N$th-nearest neighbour method \citep{Postman2005,Mei2023}. The density was calculated as $\Sigma_N={N}\, /\left({\pi\, D^2_N}\right)$, where $N$ is the number of galaxy neighbours and $D_N$ is defined as the on-sky distance to the $N$th-nearest neighbour. Results are stable in the range $N=5$--$10$, and we used $N=7$ to be consistent with previous galaxy projected surface density estimates \citep{Postman2005,Mei2023}. 

For every galaxy in the final sample, we calculated the proper distance to the seventh-nearest neighbour, within a redshift slice, $\Delta z$, centred on the galaxy. This redshift slice corresponds to three times the median uncertainty (the difference between the redshift and the average of the $68\,\%$ upper and lower limits) on the redshift in each photometric redshift bin, i.e. $3\sigma$ (listed in Table \ref{tab:z_err}). Note that in the highest redshift bin, $z=0.75$--$1$, the density calculation includes galaxies at higher redshifts within the redshift slice. This ensures that we do not miss any nearby galaxies that are outside the redshift bin. We note that the large uncertainties on the photometric redshifts result in a large redshift range in which to search for overdensities, which may ultimately smooth out any environmental effects at high $z$. We divided the number of galaxies (in this case, this is seven) by the on-sky distance to the seventh-nearest neighbour in the redshift slice to obtain the surface density of galaxies, $\Sigma_7$. This corresponds to the number of galaxies within an on-sky area with radius $1\ \text{Mpc}$. We calculated uncertainties on $\Sigma_7$ by propagating the uncertainties on the photo-$z$s in the cosmological distance measurement, and the Poissonian uncertainties in counting the number of nearby galaxy neighbours, i.e. $\sqrt{7}$. Hereafter, we refer to $\Sigma_7$ as density. 

\subsection{Comparison of density method with known clusters in Q1}

{In order to check the performance of our density calculation method, we used the list of known clusters in Q1, compiled and presented in \citet{Q1-SP050}. Taking each cluster in turn, we selected Q1 galaxies that have redshift $0.9 z_{cl}<z<1.1 z_{cl}$, where $z_{cl}$ is the redshift of the known cluster. There are 337 clusters that have Q1 galaxies within this redshift range. In fact, these 337 clusters all have at least \num{12464} galaxies within this redshift range, with the distribution peaking at $\gtrsim$\num{28000} galaxies. We then computed the on-sky projected proper distance to the Q1 galaxies within this range. We found that, of the 337 known clusters that have Q1 galaxies within redshift range, 11 of them have no galaxy within $1000\ \text{kpc}$ and 25 of them have no galaxy within $500\ \text{kpc}$. From visual inspection, it appears that many of those clusters have clear regions of high-density nearby, and so this may simply be due to an offset in the cluster detection location and the location of the Q1 galaxies. We plot a histogram of the values of $\Sigma_7$ for the entire Q1 sample, and for galaxies within $500\ \text{kpc}$ and $1000\ \text{kpc}$ of a known cluster in Fig. \ref{fig:close_gals}. We can see that, overall, the distribution in local density skews towards larger values for galaxies closer to known clusters. This shows that our method of estimating local densities is reliable for identifying high-density environments.}

\begin{figure}
    \centering
    \includegraphics[width=\columnwidth]{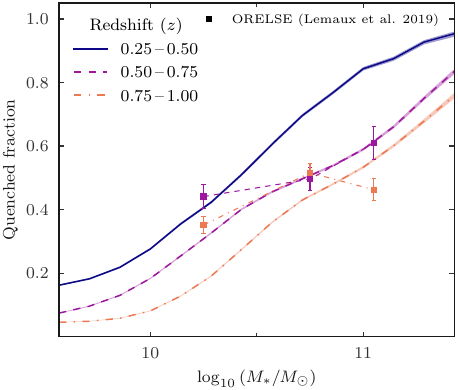}
    \caption{Fraction of galaxy classes as quenched as a function of stellar mass, binned by redshift. Shaded regions show $68\,\%$ confidence intervals. Only bins where the total number of galaxies is greater than ten are plotted. ORELSE quenched fractions \citep{Lemaux2019} are plotted in coloured squares. Their fractions are higher on average, likely because they purposely probe areas with higher galaxy densities than the Q1 fields.}
    \label{fig:qf_redshift}
\end{figure}

\begin{figure*}
    \centering
            \includegraphics[width=0.9\hsize]{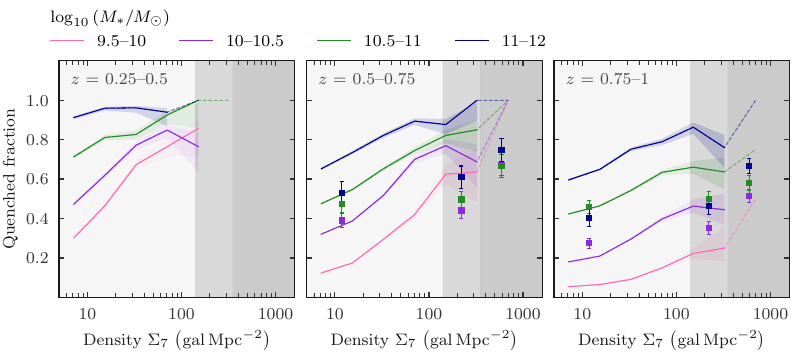}
    \caption{Fraction of galaxies classed as quenched as a function of galaxy density, binned by stellar mass and redshift. The stellar mass bins listed in the legend are in units of $\logten (M_\ast/M_\odot)$. The shaded grey regions in the background are indicative of the three density bins used in \citet{Lemaux2019}; see text for details. The shaded coloured regions show $68\,\%$ confidence intervals. Dashed lines indicate density bins where the total number of galaxies is <5.}
    \label{fig:qf_dens}
\end{figure*}

\section{\label{sec:Res}Results}
\subsection{\label{subsec:qf}Quenched fraction}

First, we considered quenched galaxies, which are galaxies in which star formation was suppressed.
We determined if a galaxy is quenched following the criterion in \citet{Franx2008}; that is, if $\text{sSFR}<0.3\, t_{\text{H},z}^{-1}$, where $t_{\text{H},z}$ is the Hubble time at a given redshift, which gives a timescale for the expansion of the Universe at that redshift. This means that the threshold in sSFR for a galaxy to be considered quenched changes with redshift: at $z=0$ this threshold is $-10.68$ and at $z=1$ it is $-10.43$. This is due to the fact that the average galaxy sSFR on the main sequence also increases with redshift at a given mass \citep{Speagle2014}.

We find that the number of passive galaxies in the data drops significantly at $z>1$, likely due to the depth of the optical surveys in the Q1 area that causes a lack of high-quality photometric redshifts at high $z$. Thus, for the current work, we limit our analysis to $z\leq1$.
In Fig. \ref{fig:qf_redshift}, we plot the quenched fraction as a function of stellar mass, binned by redshift. We see that the quenched fraction increases with increasing stellar mass, and decreasing redshift. {Qualitatively, the behaviour in Fig. \ref{fig:qf_redshift} compares well with results described in two other Q1 papers. \citet{Q1-SP044} models the values of sSFR$_\tau$ (the average sSFR over a given time period, $\tau$) in order to characterise the star formation history of each galaxy. They find the fraction of slowly-quenched galaxies\footnote{This paper uses `Ageing', `Quenching', and `Retired' categorisations of galaxies, referring to the star formation history to get a sense of how quickly the galaxy quenched or is quenching. Our classification of passive galaxies corresponds roughly to their Quenching and Retired galaxies.} is approximately $60\%$ at $0.6<z<0.8$ at $M_\ast>10^{11}\,M_\odot$. On the other hand, \citet{Q1-SP031} use NUV$-r^+$ and $r^+-J$ colours derived from \texttt{NNPZ} to determine passive status of Q1 galaxies.} They report a decrease in the fraction of passive galaxies with increasing redshift. They find, for all galaxies above their $95\%$ completeness limit ($\approx10^{9}\, M_\odot$ at $z\sim1$), a quenched fraction of $23\%$ at $0.2<z<0.5$, $15\%$ at $0.5<z<0.8$, and $8\%$ at $0.8<z<1.5$. When computing the quenched fraction for all galaxies in our sample with $10^{9.5}<M_\ast\,/\,M_\odot<10^{12}$, we find fractions of $39\%$ at $0.25<z<0.5$, $28\%$ at $0.5<z<0.8$, and $19\%$ at $0.8<z<1$. Note that we do not have exactly the same redshift bins. The relative increase in the quenched fraction in our sample compared to theirs in the last redshift bin can be explained by the fact that their bin will have even more star-forming galaxies at $z>1$, diluting their quenched fraction at that redshift. 

{It should be noted that due to differences in quality cuts chosen, and in passive galaxy selection, it is difficult to make a direct quantitative comparison between our results and these two Q1 papers, and slight variance in the fractions of quenched galaxies is expected. That said, we find our results qualitatively match these two papers well despite using different methods to classify quenched galaxies. However, we cannot exclude biases in the galaxy photometry and from the large uncertainties in photometric redshift that could potentially impact these analyses. The effect of uncertainties on our results is explored in Sect. \ref{sec:unc}.}

\begin{figure*}
    \centering
    \includegraphics[width=0.9\hsize]{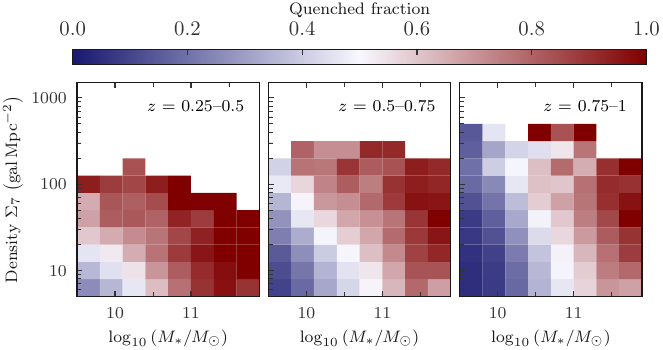}
    \caption{Distribution of the quenched fraction as a function of both stellar mass and galaxy density, in bins of increasing redshift from left to right. Only bins where the total number of galaxies is >5 are shown. The colour bar indicates the quenched fraction in each bin. The separability of the impact of stellar mass and the environment on the quenched fractions is clearly visible up to $z=1$.}
    \label{fig:qf_dens_2d}
\end{figure*}

We also compared our results to the quenched fractions from the Observations of Redshift Evolution in Large-Scale Environments \citep[ORELSE,][]{Lemaux2019} survey. This survey studied quenched fractions in a large range of environments, in the redshift range $0.55<z<1.4$. Quenched galaxies were selected using $\text{NUV}$, $r$, and $J$ colours. Their passive selection corresponds to galaxies with $\text{sSFR}<10^{-11}\,\text{yr}^{-1}$, which is a slightly more conservative cut than ours. In Fig. \ref{fig:qf_redshift}, we plot the ORELSE quenched fractions in the mass and redshift bins provided in their Table 3. {Since this survey targeted overdense environments, the underlying distribution of galaxy densities is different to ours, which did not target any environment in particular. Quenched galaxies are more prolific in high-density environments, so this difference in distribution will result in a higher number of quenched galaxies across the whole sample. Therefore, we weighted the ORELSE quenched fractions according to the ratio between the relative abundances of galaxies in their sample and in ours, for each density bin. The corrected fractions, shown in Fig. \ref{fig:qf_redshift}, are approximately 0.1 less than the original quenched fractions in the two lowest-mass bins; in the highest-mass bin the corrected fractions are approximately 0.03 higher.} We see a steady increase in the quenched fraction from $10^{10}\ M_\odot$ to $>10^{11.5}\ M_\odot$, reaching $60$--$80\,\%$ at the high-mass end. 

Despite our method of applying weights to the quenched fractions, there is still a discrepancy between our quenched fractions and the ORELSE quenched fractions, of at most about $10$--$20\,\%$, although we note that the sample size for ORELSE, $\num{5000}$ spectroscopically confirmed galaxies, is much smaller than ours. However, even taking error bars into account, our quenched fractions are lower than the ORELSE fractions at the low-mass end. However, we also note that our fractions have not been corrected for the quenched and star-forming galaxies selection functions, since they are not yet available for Q1. Furthermore, it is found in Sect. \ref{sec:unc} that the uncertainties in the Q1 dataset may result in the underestimation of the quenched fractions. Therefore, these results should be confirmed with further \Euclid\ photometry characterisation.

We plot the quenched fraction as a function of environmental density in Fig. \ref{fig:qf_dens}. Galaxies are binned by stellar mass and by redshift. Again, we plot the ORELSE quenched fractions. They calculated their density $\logten(1+\delta)$ using a Voronoi Monte-Carlo method \citep[see][]{Lubin2009,Tomczak2017}. They presented their results in three environmental density bins: low density $\logten(1+\delta)<0.3$, intermediate density $0.3<\logten(1+\delta)<0.7$, and high density $\logten(1+\delta)>0.7$. These density bins may be converted to bins in $\Sigma_7$, and the result is $\Sigma_7<140\ \text{gal}\ \text{Mpc}^{-2}$, $140<\Sigma_7\,/\,\text{gal}\ \text{Mpc}^{-2}<350$, and $\Sigma_7>350\ \text{gal}\ \text{Mpc}^{-2}$ for low-, intermediate-, and high-density bins, respectively \citep{Mei2023}. It has been shown by \citet{Darvish2015} that the conversion between the nearest-neighbour method and the Voronoi Monte-Carlo method works well, and at the same depth as this work. The ORELSE density bins are indicated on Fig. \ref{fig:qf_dens} by the shaded grey regions. It is clear that the quenched fraction increases with local environmental density at all stellar masses. This effect from the environment is present even at $z=1$, although it is not as strong as at lower redshifts. We plot the quenched fractions from the ORELSE survey \citep{Lemaux2019}, in the same stellar mass and redshift bins. 

Quantitatively, we find that the differential change in the fraction as a function of density, which corresponds to the overall slope of each coloured line in Fig. \ref{fig:qf_dens}, indeed decreases with increasing redshift. Here, the slope is an indicator of how strongly the environment affects the quenched fraction, such that a higher slope means a stronger impact. Notably, the slope decreases over the range $z=0.25$ to $z=0.8$, and by $z=1$, the slope is $\lesssim0.2$ at all stellar masses. We also find that the overall decrease in the slope (i.e. the change in the impact of the environment with increasing redshift) is larger for low-mass galaxies with $M_\ast<10^{10.5}\,M_\odot$, whereas the decrease in the slope for high-mass galaxies is shallower. 

In order to further verify that the quenching effect does not simply trace the stellar mass, we plot the two-dimensional distribution of the quenched fraction with respect to the stellar mass and density in Fig. \ref{fig:qf_dens_2d}. Here, galaxies are binned by stellar mass and local density, and panels from left to right show increasing redshift bins. {Only bins where the total number of galaxies is >5 are shown.} The colour bar shows the quenched fraction in each bin, such that redder bins contain a higher number of quenched galaxies compared to bluer bins, which are mostly star forming. The size of the bins on the $x$ axis and $y$ axis were determined by the typical uncertainties in the stellar mass and local density, respectively. In the figure, we see that at fixed stellar mass, the quenched fraction increases with increasing local environmental density. This effect is visible even at $z=1$. At $z<0.75$, the effect from the environment is strongest for low-mass galaxies, and gets weaker with increasing stellar mass. This behaviour illustrates how the environment of a galaxy can quench its star formation, at a fixed stellar mass, reflective of the result in \citet{Peng2010}, which is for the red fraction of SDSS galaxies at $z<0.2$. However, there are differences between these results. Following the line in Fig. \ref{fig:qf_dens_2d} where the quenched fraction is equal to $0.5$, the line shows little to no curve. The line where the red fraction is equal to $0.5$ in Fig. 6 of \citet{Peng2010} is more curved, implying that there is a density threshold (mass threshold) after which quenching can happen at a fixed stellar mass (density). On the other hand, our results imply a more continuous change with increasing stellar mass or density, at least at $z\lesssim0.75$. At $z>0.75$, we observe a weaker effect from the environment; only at $M_\ast\gtrsim10^{10.5}\ M_\odot$ do galaxies begin to quench. 

{In Figs. \ref{fig:qf_dens} and \ref{fig:qf_dens_2d}, we note that in Q1 there is a lack of massive galaxies in high-density environments at $0.25<z<0.5$. We investigate here if this deficit is due to a limitation of the telescope, or if it is expected. By calculating the volume of Q1 in this redshift bin, we can estimate the number of expected galaxies of a certain stellar mass. We find the volume of Q1 in this redshift bin to be approximately $3.8\times10^7\ \text{Mpc}^3$. By following \citet{Conroy2009}, we estimate the number of massive ($>10^{11}\ \msun$) galaxies in this volume to be $\approx$ \num{14500}. In our sample, we find $\approx$ \num{9800}, which is within 1-1.5$\sigma$ of cosmic variance at this redshift \citep{Moster2011}. Thus, we are sure that Q1 is capable of detecting massive galaxies at this redshift. Therefore, the deficit of galaxies must be due to the lack of high-density environments. We remind the reader that Q1 consists of non-contiguous regions that sum to $60\ \text{deg}^2$, and that these regions were deliberately chosen to be in low-density regions away from known clusters. Thus, high-density regions will be rare in these small volumes. In fact, in this redshift bin, we find only 49 galaxies with $\Sigma_7>100\ \text{gal}\ \text{Mpc}^{-2}$. The lack of massive galaxies in high-density environments is therefore not necessarily a limitation of the instrument, but rather a consequence of probing a small volume in a low-density region. }

\begin{figure}
    \centering
    \includegraphics[width=\hsize]{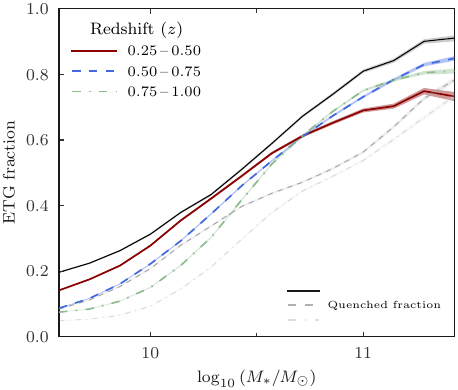}
    \caption{Fractions of galaxies classed as ETGs as a function of stellar mass, binned by redshift. Shaded regions show $68\,\%$ confidence intervals. Only bins where the total number of galaxies is greater than 5 are plotted. The quenched fractions from Fig. \ref{fig:qf_redshift} are plotted for comparison in the same redshift bins, in grey. The ETG fraction increases with stellar mass at all redshifts.}
    \label{fig:etg_redshift}
\end{figure}

\begin{figure*}
    \centering
    \includegraphics[width=0.9\hsize]{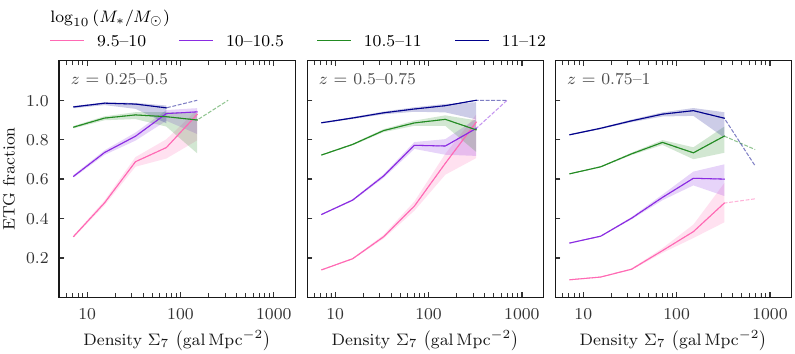}
    \caption{Fraction of galaxies classed as ETGs as a function of galaxy density, binned by stellar mass and redshift. Shaded regions show $68\,\%$ confidence intervals. Dashed lines indicate density bins where the total number of galaxies is <5. At all redshifts, there is a significant increase in the ETG fraction of galaxies from low-density environments to high-density environments, for low-mass galaxies. However, more massive galaxies are already mostly ETGs. At $z>0.75$, the environmental effect on the ETG fraction is slightly weaker.}
    \label{fig:etg_dens}
\end{figure*}

\subsection{\label{subsec:etg}Morphology}

Usually, galaxy morphological classification is based on galaxy specific visual features, such as the presence of a bulge or a disc and their respective predominance in the rest-frame $B$ band (for an example at $z\sim1$, see \citealp{Postman2005}). However, for our sample the rest-frame $B$ band corresponds to $i$ and $z$ apparent magnitudes, which are not observed by \Euclid.
Therefore, we used the morphological classification from \citet[][hereafter \citetalias{Q1-SP040}]{Q1-SP040} to identify early-type galaxies (ETGs) and late-type ones (LTGs). This classification is based on the separation of these two galaxy morphological types in the S\'ersic index, $n$, versus the $u-r$ colour plane. We note that, since this classification uses colour, this is not a classification exclusively based on visual features. Our ETG sample will be biased towards red galaxies. This means that while the ETG sample contains a mix of passive and star-forming galaxies, almost every passive galaxy is an ETG ($90\%$). This is an unavoidable degeneracy at this stage, and as such results should be interpreted cautiously. However machine-learning based classifications should be available for \Euclid\ data in the future \citep{DominguezSanchez2022}.

To distinguish between ETGs and LTGs, we use a demarcation line (Eq. 4 in \citetalias{Q1-SP040}), such that ETGs are located above this line and LTGs are below it. The equation for this line is in Eq. \ref{eq:line} below
\begin{equation}\label{eq:line}
    (u-r)_{\text{lim},n} = 2.32-1.32\log_{10}(n).
\end{equation}

We plot the ETG fraction as a function of stellar mass, in bins of redshift, in Fig. \ref{fig:etg_redshift}. We also plot the quenched fraction from Fig. \ref{fig:qf_redshift} for comparison. Here, we see a rise in the ETG fraction up to $90\,\%$ in the most massive galaxies. The ETG fraction also increases with decreasing redshift. We do not plot ETG fractions at $z>1$ because, as with the passive galaxies, the \citetalias{Q1-SP040} morphological classification is less reliable at these high redshifts.

\begin{figure}
    \centering
    \includegraphics[width=0.9\hsize]{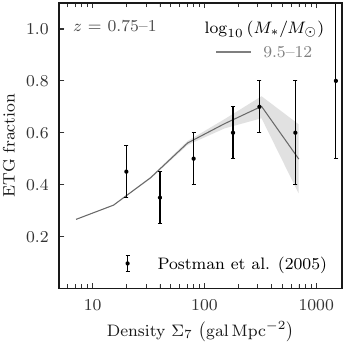}
    \caption{Fraction of ETGs as a function of galaxy density, in the highest-redshift bin, $0.75<z<1$. The stellar mass range is shown in the top right corner. The ETG fractions from the entire cluster sample of \citet{Postman2005} are plotted as black dots. The two samples are comparable, as they are selected at roughly the same depth; however, the morphological classifications were not done in the same way, which may possibly lead to differences in the ETG fractions. Despite these slight differences, the fractions are in good agreement except for the fact that we do not reach $\Sigma_7\approx1000\ \text{gal}\ \text{Mpc}^{-2}$. }
    \label{fig:etg_postman}
\end{figure}

\begin{figure*}
    \centering
    \includegraphics[width=0.9\hsize]{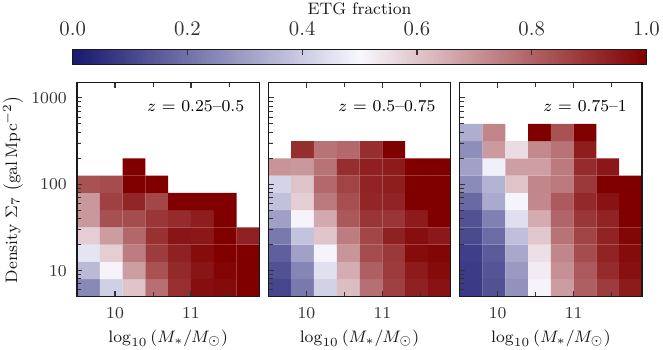}
    \caption{Distribution of the ETG fraction as a function of both stellar mass and galaxy density, in increasing redshift bins from left to right. Only bins where the total number of galaxies is >5 are shown. The colour bar indicates the mean ETG fraction in each bin. Similar to in Fig. \ref{fig:qf_dens_2d}, the separability of the stellar mass and environment in producing the ETG fraction is clear up to $z=0.75$. However, it is most evident only up to $M_\ast\approx10^{10.8}\ M_\odot$, after which almost all galaxies are ETGs, regardless of the environment.}
    \label{fig:etg_dens_2d}
\end{figure*}

Similar to Fig. \ref{fig:qf_dens}, we plot the ETG fraction as a function of local density in Fig. \ref{fig:etg_dens}. At $0.25<z<0.75$, we see a significant increase in the ETG fraction of galaxies with  $M_\ast<10^{10.5}\ M_\odot$ from low-density environments to high-density ones. More massive galaxies are already mostly ETGs, as is seen in Fig. \ref{fig:etg_redshift}. At $z>0.75$ the effect of the environment is still present, but weaker. The effect is most prominent for low-mass galaxies, which show an increase in the ETG fraction from low density to high density of $\gtrsim20\,\%$.

\begin{figure*}
    \centering
    \includegraphics[width=0.9\hsize]{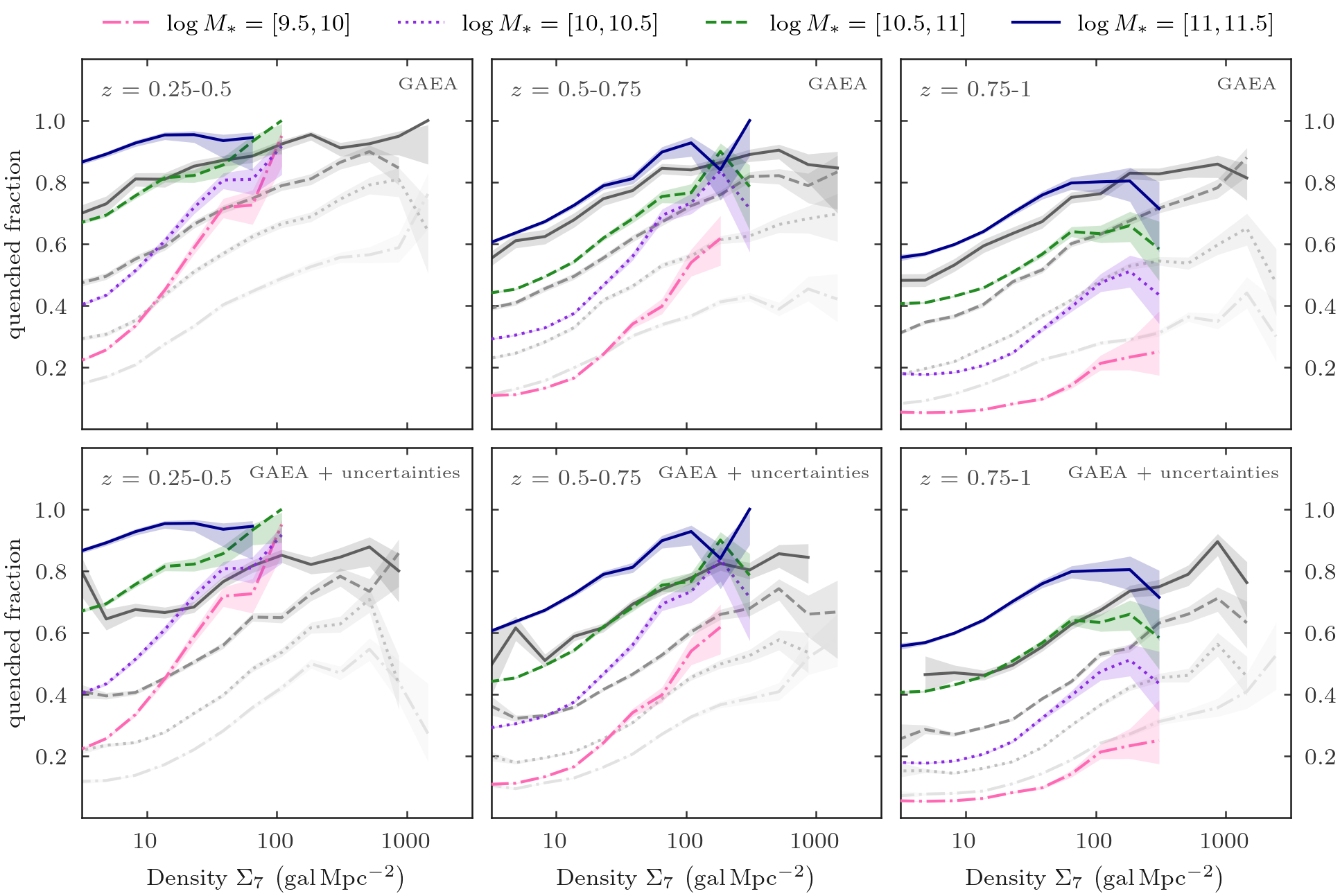}
    \caption{{Top: Quenched fractions as in Fig. \ref{fig:qf_dens} in coloured lines, compared to quenched fractions from the GAEA semi-analytic model \citep{DeLucia2024} in grey lines. Bottom: As above, but with \Euclid-like uncertainties applied to the GAEA data. We see good agreement between the Q1 data and the GAEA data, particularly at $z>0.5$, even with observational uncertainties applied.}}
    \label{fig:QF-dens-GAEA}
\end{figure*}

In the last redshift bin, we can compare our results with the ETG fractions from seven clusters observed with the \textit{Hubble} Space Telescope at $0.8<z<1.3$ from \citet{Postman2005} in Fig.~\ref{fig:etg_postman}. We note that these fractions are not separated by stellar mass, and so we plot our ETG fraction for all stellar masses. The galaxies in \citet{Postman2005} were selected at roughly the same depth as ours. Here, we see that the \citet{Postman2005} ETG fractions are generally in good agreement with ours, within error bars. Given that galaxy densities were calculated at the same depth and with the same method, any discrepancies might be due to the different morphological classification, which was performed based on visual morphology in \citet{Postman2005}, compared to the S\'ersic index-colour plane in \citetalias{Q1-SP040}. Another aspect to consider is that we do not identify as many high-density regions ($\gtrsim1000\ \text{gal}\ \text{Mpc}^{-2}$), such as those probed in \citet{Postman2005}, due to the small volume probed in Q1. In fact, we only have $<100$ galaxies with $\Sigma_7>500\ \text{gal}\ \text{Mpc}^{-2} $ in our entire sample.

We further illustrate the interplay between the stellar mass and the local density in Fig. \ref{fig:etg_dens_2d}, where most of the high-mass galaxies are already ETGs even at high redshift. The transformative effect of the environment can be seen, and it is much stronger at lower redshifts compared to higher redshifts, and in lower stellar-mass galaxies.

We quantify these trends described in Sec. \ref{subsec:qf} and \ref{subsec:etg} in Appendix \ref{sec:quant}, by plotting the differential change in the quenched fraction and the ETG fraction due to the environment, in fixed stellar mass and redshift bins. We confirm that the differential change in the quenched fraction or ETG fraction decreases at all stellar masses at $z>0.7$, and that the environmental effect is strongest for the lowest mass galaxies at low redshifts.

Additionally, we ran a Monte Carlo test 1000 times, using the uncertainties in the local density to verify if these results persist. These figures for the quenched fraction and the ETG fraction are shown in Appendix \ref{sec:mc}. The trends with stellar mass and density do not change; however, we observe that the fraction of quenched galaxies drops at  higher stellar masses and higher densities.

\section{\label{sec:disc}Discussion}
\subsection{\label{sec:unc}The impact of uncertainties on our results as estimated from simulations}

{Despite the improvements in redshift and physical property estimation brought by utilising the \citetalias{Q1-SP031} catalogue, the \Euclid\ data products still present uncertainties.} Specifically, there are significant uncertainties in the redshift, stellar mass, and SFR estimates from \texttt{NNPZ}. Here, we examine the impact of those uncertainties on our results by using simulations. 

For this, we used a lightcone from the Galaxy Evolution and Assembly semi-analytic model \citep[GAEA,][]{DeLucia2007,Hirschmann2016,Xie2017,DeLucia2024}. This model originally aimed to study the evolution of the brightest cluster galaxies (BCGs), uncovering the hierarchical nature of these objects. Since the original formulation of \citet{DeLucia2007}, there have been several additions, including improved treatment of stellar feedback, an updated AGN feedback model, and treatment of satellite galaxies \citep{Hirschmann2016,Fontanot2020,Xie2020,DeLucia2024}. In this model, galaxies are assigned an instantaneous sSFR, and are then quenched according to a threshold in sSFR, such that $\text{sSFR}<0.3\, t_{\text{H},z}^{-1}$. Note that this is the same way that we selected passive galaxies in this work. We selected the same sample of galaxies, using the same redshift, stellar mass, and magnitude constraints. 

There are no uncertainties reported in the GAEA dataset. In order to understand the impact of the uncertainties in \Euclid\ data on our results, we must assign uncertainties to the GAEA data. Ideally we would run GAEA (or some other simulated dataset) through \texttt{NNPZ} so as to recover `observed' values for simulated data in the same way as in Q1. This is not available at the time of doing this work, but will be available in the future. In the meantime, we achieved this in an approximate way. We began with the distribution of uncertainties in photometric redshift, stellar mass, and SFR from \Euclid, in narrow bins.{ It should be pointed out that, rather than drawing from a multi-variate distribution for all quantities at once (as is the case for \texttt{NNPZ}), we pulled from each distribution separately. Thus, this method is an approximation.} For the stellar mass and the SFR, these uncertainties are on the logarithmic quantities. We then binned the GAEA values according to the same bins as the \Euclid\ data. The median uncertainty and standard deviation of the \Euclid\ uncertainties in each bin were used to construct a Gaussian distribution, from which a value was randomly chosen as the uncertainty. This value was either added or subtracted to the GAEA quantity, resulting in a new value that was representative of a real value with observational uncertainty. From this we selected passive galaxies in the same way as before, except now the new value for sSFR contained observational-like uncertainties in the stellar mass and the SFR. We also obtained measurements for the local density using these observation-like quantities.

{In Fig. \ref{fig:QF-dens-GAEA}, we plot these results. The upper panels show the observed Q1 quenched fraction in colour lines, the same as in Fig. \ref{fig:qf_dens}, and the quenched fraction as a function of local density in GAEA in grey lines. The bottom panels show the same Q1 quenched fraction, but now with the quenched fraction from GAEA with \Euclid\ uncertainties added. We can see that the inclusion of uncertainties brings about a reduction in the quenched fraction of about 0.1, particularly in the low-density regions. In fact, we see that the quenched fractions converge at higher values of $\Sigma_7$, including at $z\sim1$. From this, we can be confident that uncertainties in \Euclid's Q1 are not overly biasing our results, although we may be slightly underestimating quenched fractions in low-density regions by about 0.1.}

\subsection{Comparing to other results}

In this paper, we have showcased the early capabilities of the \Euclid\ project, in terms of galaxy evolution and identifying a wide range of environments. Here, we identify and analyse the passive-density and morphology-density relations from $z=0.25$ to $z=1$. Many previous studies have examined quenched fractions and ETG fractions as functions of stellar mass and environment; however, not all of them probe the effect of the environment at fixed stellar mass, and as a function of redshift. Our approach allows us to quantify the effect of the environment for each stellar mass and redshift bin. This approach is similar to that in \citet{Peng2010}, which demonstrated the separability of the effects of stellar mass and environment at $z<0.2$. This paper also found that these effects are separable to $z\sim1$; however, only in density quartiles. We have now shown on a continuous scale of density and stellar mass how the differential effects change with redshift. Similarly, while previous studies such as \citet{Postman2005,Lemaux2019,Mei2023} showed these relations at $z\sim1$, we have expanded on these results by including a wider range of environments, as opposed to targeting overdense regions. 

In Fig. \ref{fig:qf_dens}, we find the quenched fraction of galaxies increases with increasing local density at all stellar masses up to $z=1$, although the effect is slightly weaker at this redshift. This is consistent with the results of many studies \citep{Cucciati2017,Kawinwanichakij2017,Strazzullo2019,Chartab2020,vanderBurg2020,Baxter2023,Shi2024,Taamoli2024,Trudeau2024}; however, quantitatively, the values of the quenched fractions differ slightly. As has already been shown in Fig. \ref{fig:qf_redshift}, the ORELSE survey \citep{Lemaux2019} is in decent agreement with Q1 in terms of quenched fractions. Yet, in Fig. \ref{fig:qf_dens}, when we plot these fractions as a function of density, the ORELSE quenched fractions evolve very little across the density bins and the redshift bins. In these redshift bins the ORELSE quenched fractions show little difference between the stellar mass bins, whereas ours depend on the stellar mass as well as the local density. Furthermore, other papers with results from this redshift range show how the quenched fraction of galaxies depends strongly on the stellar mass and the environment \citep[e.g.,][]{Davidzon2016,Weaver2023}.

On the other hand, quenched fractions presented in the CARLA study \citep{Mei2023} reach $>60\,\%$ in the highest-density environments, which is also in good agreement with our results. However, it must be noted that these results are for $1.3<z<1.9$, and so caution must be taken when directly comparing these results. Nonetheless, our quenched fractions at $z\sim1$ are slightly underestimated than those in \citet{Mei2023} in the same mass bins used; only galaxies at $M_\ast>10^{10.5}\ M_\odot$ reach quenched fractions of $60\,\%$. It should also be noted that the CARLA results show little dependence on stellar mass, possibly because of the high uncertainty on their mass measurements, where our results and other studies show significant mass dependence at fixed local density. For instance, it has been shown that the environmental quenching efficiency increases with decreasing redshift, and this effect is strongest at low stellar masses \citep{Cucciati2010,Cucciati2017,Lemaux2019}, which is consistent with our results in Fig. \ref{fig:qf_dens}.

For ETGs, we find that massive galaxies $\left(M_\ast>10^{11}\ M_\odot\right)$ are $>80\,\%$ ETGs at $z<1$. For low-mass galaxies, there is a considerable effect from the environment on their morphology at up to $z=1$. There are many studies that show that the morphology-density relation is already in place by $z\sim1$ \citep{Postman2005,Shi2024}, or even at $z\sim2$ \citep{Sazonova2020,Mei2023}. In \citet{Sazonova2020} the effect is stronger at low $M_\ast$, which is also seen in this work. Additionally, our ETG fractions plotted in Fig. \ref{fig:etg_postman} are in good agreement with \citet{Postman2005} at $z\sim1$, within the error bars. It is difficult to say anything conclusively about the densest environments due to the very small number of galaxies with $\Sigma_7\gtrsim500\ \text{gal}\ \text{Mpc}^{-2}$. Qualitatively speaking, there is clear evidence of a morphology-density relation at $0.75<z<1$ across all stellar masses, and more prominently at $M_\ast<10^{10.5}\ M_\odot$.

\section{Summary}\label{sec:summ}
{
In this paper we showcase the early capabilities of the \Euclid\ survey, with analysis from the first Q1 results. We computed the quenched fractions and ETG fractions of a mass complete sample, from $0.25<z<1$, and found a significant stellar mass and redshift evolution of these quantities. Additionally, we calculated the local environmental density on each galaxy in our sample using the $N$th-nearest neighbour method. We then plotted the quenched fractions and ETG fractions as functions of local density. We find that the strength of the environmental effect depends strongly on stellar mass and redshift, with low-mass galaxies at low $z$ being the most susceptible to the transformative environmental effects. We find broad qualitative agreement with previous observational results \citep[e.g.,][]{Postman2005,Lemaux2019}, and other Q1 results \citep{Q1-SP031,Q1-SP044}; however, we caution that these results are preliminary and subject to uncertainties. That said, by performing Monte Carlo tests taking these uncertainties into account, we find that the overall trends remain, although the environmental effect is weaker at $z\sim1$.}

{Our main results can be summarised as follows:}

\begin{itemize}
    \item At $M_*>10^{11}\ M_\odot$, we find that the quenched fraction of galaxies is at least $80\ \%$ at $0.25<z<0.5$, and $\gtrsim50\ \%$ at $0.75<z<1$. 
    \item There is a clear environmental effect, causing galaxies at the same stellar mass to quench in higher-density environments. In particular, low-mass galaxies are strongly affected, with the quenched fraction increasing from $30\ \%$ to $80\ \%$ at low $z$. 
    \item At $z\sim1$, the environmental effect is weaker but still present, with $M_*>10^{11}\ M_\odot$ galaxies reaching $80\ \%$ quenched in high-density regions.
    \item We find that the fraction of ETGs increases monotonically with stellar mass, with $\gtrsim 80\ \%$ of high-mass galaxies having early-type morphology at all redshifts.
    \item Similarly, we find a positive correlation between the ETG fraction and the local density, with low-mass galaxies being the most affected by the environment. Here, the ETG fraction of galaxies with $10^{9.5}<M_*/M_\odot<10^{10}$ increases from $20\ \%$ to $\gtrsim80\ \%$ at $z<0.5$.
    \item Using the GAEA semi-analytic model, we applied \Euclid-like uncertainties to simulated data to test the effect of these uncertainties on our results. We find our density measurements to be reliable, and that the uncertainties may lead to an underestimation of the quenched fractions of about 0.1, particularly in low-density regions. This result highlights the importance of future data releases for making our results more robust.
\end{itemize}
In a future paper, we plan to extend this work to $z>1$, and include results from other high-redshift studies to compare.

The first \Euclid\ results provide observations, redshifts, and physical properties of at least one million galaxies in a wide variety of environments. With the upcoming public release of DR1, the number of observed sources and the number of high-density environments in \Euclid\ data will increase by orders of magnitude.

\begin{acknowledgements}
We thank the anonymous reviewer for their insightful comments, which helped improve this paper. This work was supported by CNES, focused on the \Euclid\ space mission.
\AckQone 
\AckEC 
The authors acknowledge support from the ELSA project. "ELSA: Euclid Legacy Science Advanced analysis tools" (Grant Agreement no. 101135203) is funded by the European Union. Views and opinions expressed are however those of the author(s) only and do not necessarily reflect those of the European Union or Innovate UK. Neither the European Union nor the granting authority can be held responsible for them. UK participation is funded through the UK Horizon guarantee scheme under Innovate UK grant 10093177.
This research was supported in part by grant NSF PHY-2309135 to the Kavli Institute for Theoretical Physics (KITP). 
\end{acknowledgements}

\bibliography{bib}

@ARTICLE{DeLucia2024,
       author = {{De Lucia}, Gabriella and {Fontanot}, Fabio and {Xie}, Lizhi and {Hirschmann}, Michaela},
        title = "{Tracing the quenching journey across cosmic time}",
      journal = {\aap},
         year = 2024,
        month = jul,
       volume = {687},
          eid = {A68},
        pages = {A68},
          doi = {10.1051/0004-6361/202349045},
archivePrefix = {arXiv},
       eprint = {2401.06211},
 primaryClass = {astro-ph.GA},
       adsurl = {https://ui.adsabs.harvard.edu/abs/2024A&A...687A..68D}
}

@ARTICLE{PlanckCollaboration2016,
       author = {{Planck Collaboration} and {Ade}, P.~A.~R. and {Aghanim}, N. and {Arnaud}, M. and {Ashdown}, M. and {Aumont}, J. and {Baccigalupi}, C. and {Banday}, A.~J. and {Barreiro}, R.~B. and {Bartlett}, J.~G. and et al.},
        title = "{Planck 2015 results. XIII. Cosmological parameters}",
      journal = {\aap},
         year = 2016,
        month = sep,
       volume = {594},
          eid = {A13},
        pages = {A13},
          doi = {10.1051/0004-6361/201525830},
archivePrefix = {arXiv},
       eprint = {1502.01589},
 primaryClass = {astro-ph.CO},
       adsurl = {https://ui.adsabs.harvard.edu/abs/2016A&A...594A..13P}
}

@ARTICLE{Wang2016,
       author = {{Wang}, Tao and {Elbaz}, David and {Daddi}, Emanuele and {Finoguenov}, Alexis and {Liu}, Daizhong and {Schreiber}, Corentin and {Mart{\'\i}n}, Sergio and {Strazzullo}, Veronica and {Valentino}, Francesco and {van der Burg}, Remco and {Zanella}, Anita and {Ciesla}, Laure and {Gobat}, Raphael and {Le Brun}, Amandine and {Pannella}, Maurilio and {Sargent}, Mark and {Shu}, Xinwen and {Tan}, Qinghua and {Cappelluti}, Nico and {Li}, Yanxia},
        title = "{Discovery of a Galaxy Cluster with a Violently Starbursting Core at z = 2.506}",
      journal = {\apj},
         year = 2016,
        month = sep,
       volume = {828},
       number = {1},
          eid = {56},
        pages = {56},
          doi = {10.3847/0004-637X/828/1/56},
archivePrefix = {arXiv},
       eprint = {1604.07404},
 primaryClass = {astro-ph.GA},
       adsurl = {https://ui.adsabs.harvard.edu/abs/2016ApJ...828...56W}
}

@ARTICLE{Carnall2024,
       author = {{Carnall}, A.~C. and {Cullen}, F. and {McLure}, R.~J. and {McLeod}, D.~J. and {Begley}, R. and {Donnan}, C.~T. and {Dunlop}, J.~S. and {Shapley}, A.~E. and {Rowlands}, K. and {Almaini}, O. and {Arellano-C{\'o}rdova}, K.~Z. and {Barrufet}, L. and {Cimatti}, A. and {Ellis}, R.~S. and {Grogin}, N.~A. and {Hamadouche}, M.~L. and {Illingworth}, G.~D. and {Koekemoer}, A.~M. and {Leung}, H.-H. and {Lovell}, C.~C. and {P{\'e}rez-Gonz{\'a}lez}, P.~G. and {Santini}, P. and {Stanton}, T.~M. and {Wild}, V.},
        title = "{The JWST EXCELS survey: too much, too young, too fast? Ultra-massive quiescent galaxies at 3 < z < 5}",
      journal = {\mnras},
         year = 2024,
        month = oct,
       volume = {534},
       number = {1},
        pages = {325-348},
          doi = {10.1093/mnras/stae2092},
archivePrefix = {arXiv},
       eprint = {2405.02242},
 primaryClass = {astro-ph.GA},
       adsurl = {https://ui.adsabs.harvard.edu/abs/2024MNRAS.534..325C}
}

@ARTICLE{Elbaz2007,
       author = {{Elbaz}, D. and {Daddi}, E. and {Le Borgne}, D. and {Dickinson}, M. and {Alexander}, D.~M. and {Chary}, R. -R. and {Starck}, J. -L. and {Brandt}, W.~N. and {Kitzbichler}, M. and {MacDonald}, E. and {Nonino}, M. and {Popesso}, P. and {Stern}, D. and {Vanzella}, E.},
        title = "{The reversal of the star formation-density relation in the distant universe}",
      journal = {\aap},
         year = 2007,
        month = jun,
       volume = {468},
       number = {1},
        pages = {33-48},
          doi = {10.1051/0004-6361:20077525},
archivePrefix = {arXiv},
       eprint = {astro-ph/0703653},
 primaryClass = {astro-ph},
       adsurl = {https://ui.adsabs.harvard.edu/abs/2007A&A...468...33E}
}

@ARTICLE{Franx2008,
       author = {{Franx}, Marijn and {van Dokkum}, Pieter G. and {F{\"o}rster Schreiber}, Natascha M. and {Wuyts}, Stijn and {Labb{\'e}}, Ivo and {Toft}, Sune},
        title = "{Structure and Star Formation in Galaxies out to z = 3: Evidence for Surface Density Dependent Evolution and Upsizing}",
      journal = {\apj},
         year = 2008,
        month = dec,
       volume = {688},
       number = {2},
        pages = {770-788},
          doi = {10.1086/592431},
archivePrefix = {arXiv},
       eprint = {0808.2642},
 primaryClass = {astro-ph},
       adsurl = {https://ui.adsabs.harvard.edu/abs/2008ApJ...688..770F}
}

@ARTICLE{Cramer2024,
       author = {{Cramer}, W.~J. and {Noble}, A.~G. and {Rudnick}, G. and {Pigarelli}, A. and {Wilson}, G. and {Bah{\'e}}, Y.~M. and {Cooper}, M.~C. and {Demarco}, R. and {Matharu}, J. and {Miller}, T.~B. and {Muzzin}, A. and {Nantais}, J. and {Sportsman}, W. and {van Kampen}, E. and {Webb}, T.~M.~A. and {Yee}, H.~K.~C.},
        title = "{Resolved UV and Optical Color Gradients Reveal Environmental Influence on Galaxy Evolution at Redshift z {\ensuremath{\sim}} 1.6}",
      journal = {\apj},
         year = 2024,
        month = nov,
       volume = {975},
       number = {1},
          eid = {144},
        pages = {144},
          doi = {10.3847/1538-4357/ad7798},
archivePrefix = {arXiv},
       eprint = {2404.07355},
 primaryClass = {astro-ph.GA},
       adsurl = {https://ui.adsabs.harvard.edu/abs/2024ApJ...975..144C}
}

@ARTICLE{Edward2024,
       author = {{Edward}, Adit H. and {Balogh}, Michael L. and {Bah{\'e}}, Yannick M. and {Cooper}, M.~C. and {Hatch}, Nina A. and {Marchioni}, Justin and {Muzzin}, Adam and {Noble}, Allison and {Rudnick}, Gregory H. and {Vulcani}, Benedetta and {Wilson}, Gillian and {De Lucia}, Gabriella and {Demarco}, Ricardo and {Forrest}, Ben and {Hirschmann}, Michaela and {Castignani}, Gianluca and {Cerulo}, Pierluigi and {Finn}, Rose A. and {Hewitt}, Guillaume and {Jablonka}, Pascale and {Kodama}, Tadayuki and {Maurogordato}, Sophie and {Nantais}, Julie and {Xie}, Lizhi},
        title = "{The stellar mass function of quiescent galaxies in 2 < z < 2.5 protoclusters}",
      journal = {\mnras},
         year = 2024,
        month = jan,
       volume = {527},
       number = {3},
        pages = {8598-8617},
          doi = {10.1093/mnras/stad3751},
archivePrefix = {arXiv},
       eprint = {2312.12380},
 primaryClass = {astro-ph.GA},
       adsurl = {https://ui.adsabs.harvard.edu/abs/2024MNRAS.527.8598E}
}

@ARTICLE{Carnall2023,
       author = {{Carnall}, A.~C. and {McLeod}, D.~J. and {McLure}, R.~J. and {Dunlop}, J.~S. and {Begley}, R. and {Cullen}, F. and {Donnan}, C.~T. and {Hamadouche}, M.~L. and {Jewell}, S.~M. and {Jones}, E.~W. and {Pollock}, C.~L. and {Wild}, V.},
        title = "{A surprising abundance of massive quiescent galaxies at 3 < z < 5 in the first data from JWST CEERS}",
      journal = {\mnras},
         year = 2023,
        month = apr,
       volume = {520},
       number = {3},
        pages = {3974-3985},
          doi = {10.1093/mnras/stad369},
archivePrefix = {arXiv},
       eprint = {2208.00986},
 primaryClass = {astro-ph.GA},
       adsurl = {https://ui.adsabs.harvard.edu/abs/2023MNRAS.520.3974C}
}

@ARTICLE{Mei2023,
       author = {{Mei}, Simona and {Hatch}, Nina A. and {Amodeo}, Stefania and {Afanasiev}, Anton V. and {De Breuck}, Carlos and {Stern}, Daniel and {Cooke}, Elizabeth A. and {Gonzalez}, Anthony H. and {Noirot}, Ga{\"e}l and {Rettura}, Alessandro and {Seymour}, Nick and {Stanford}, Spencer A. and {Vernet}, Jo{\"e}l and {Wylezalek}, Dominika},
        title = "{Morphology-density relation, quenching, and mergers in CARLA clusters and protoclusters at 1.4 < z < 2.8}",
      journal = {\aap},
         year = 2023,
        month = feb,
       volume = {670},
          eid = {A58},
        pages = {A58},
          doi = {10.1051/0004-6361/202243551},
archivePrefix = {arXiv},
       eprint = {2209.02078},
 primaryClass = {astro-ph.GA},
       adsurl = {https://ui.adsabs.harvard.edu/abs/2023A&A...670A..58M}
}

@ARTICLE{Perez-Martinez2023,
       author = {{P{\'e}rez-Mart{\'\i}nez}, J.~M. and {Dannerbauer}, H. and {Kodama}, T. and {Koyama}, Y. and {Shimakawa}, R. and {Suzuki}, T.~L. and {Calvi}, R. and {Chen}, Z. and {Daikuhara}, K. and {Hatch}, N.~A. and {Laza-Ramos}, A. and {Sobral}, D. and {Stott}, J.~P. and {Tanaka}, I.},
        title = "{Signs of environmental effects on star-forming galaxies in the Spiderweb protocluster at z = 2.16}",
      journal = {\mnras},
         year = 2023,
        month = jan,
       volume = {518},
       number = {2},
        pages = {1707-1734},
          doi = {10.1093/mnras/stac2784},
archivePrefix = {arXiv},
       eprint = {2209.13069},
 primaryClass = {astro-ph.GA},
       adsurl = {https://ui.adsabs.harvard.edu/abs/2023MNRAS.518.1707P}
}

@ARTICLE{Xie2020,
       author = {{Xie}, Lizhi and {De Lucia}, Gabriella and {Hirschmann}, Michaela and {Fontanot}, Fabio},
        title = "{The influence of environment on satellite galaxies in the GAEA semi-analytic model}",
      journal = {\mnras},
         year = 2020,
        month = nov,
       volume = {498},
       number = {3},
        pages = {4327-4344},
          doi = {10.1093/mnras/staa2370},
archivePrefix = {arXiv},
       eprint = {2003.12757},
 primaryClass = {astro-ph.GA},
       adsurl = {https://ui.adsabs.harvard.edu/abs/2020MNRAS.498.4327X}
}

@ARTICLE{Fontanot2020,
       author = {{Fontanot}, Fabio and {De Lucia}, Gabriella and {Hirschmann}, Michaela and {Xie}, Lizhi and {Monaco}, Pierluigi and {Menci}, Nicola and {Fiore}, Fabrizio and {Feruglio}, Chiara and {Cristiani}, Stefano and {Shankar}, Francesco},
        title = "{The rise of active galactic nuclei in the galaxy evolution and assembly semi-analytic model}",
      journal = {\mnras},
         year = 2020,
        month = aug,
       volume = {496},
       number = {3},
        pages = {3943-3960},
          doi = {10.1093/mnras/staa1716},
archivePrefix = {arXiv},
       eprint = {2002.10576},
 primaryClass = {astro-ph.CO},
       adsurl = {https://ui.adsabs.harvard.edu/abs/2020MNRAS.496.3943F}
}

@ARTICLE{Shimakawa2018b,
       author = {{Shimakawa}, Rhythm and {Koyama}, Yusei and {R{\"o}ttgering}, Huub J.~A. and {Kodama}, Tadayuki and {Hayashi}, Masao and {Hatch}, Nina A. and {Dannerbauer}, Helmut and {Tanaka}, Ichi and {Tadaki}, Ken-ichi and {Suzuki}, Tomoko L. and {Fukagawa}, Nao and {Cai}, Zheng and {Kurk}, Jaron D.},
        title = "{MAHALO Deep Cluster Survey II. Characterizing massive forming galaxies in the Spiderweb protocluster at z = 2.2}",
      journal = {\mnras},
         year = 2018,
        month = dec,
       volume = {481},
       number = {4},
        pages = {5630-5650},
          doi = {10.1093/mnras/sty2618},
archivePrefix = {arXiv},
       eprint = {1809.08755},
 primaryClass = {astro-ph.GA},
       adsurl = {https://ui.adsabs.harvard.edu/abs/2018MNRAS.481.5630S}
}

@ARTICLE{Xie2017,
       author = {{Xie}, Lizhi and {De Lucia}, Gabriella and {Hirschmann}, Michaela and {Fontanot}, Fabio and {Zoldan}, Anna},
        title = "{H$_{2}$-based star formation laws in hierarchical models of galaxy formation}",
      journal = {\mnras},
         year = 2017,
        month = jul,
       volume = {469},
       number = {1},
        pages = {968-993},
          doi = {10.1093/mnras/stx889},
archivePrefix = {arXiv},
       eprint = {1611.09372},
 primaryClass = {astro-ph.GA},
       adsurl = {https://ui.adsabs.harvard.edu/abs/2017MNRAS.469..968X}
}

@ARTICLE{Rhodes2017,
       author = {{Rhodes}, Jason and {Nichol}, Robert C. and {Aubourg}, {\'E}ric and {Bean}, Rachel and {Boutigny}, Dominique and {Bremer}, Malcolm N. and {Capak}, Peter and {Cardone}, Vincenzo and {Carry}, Beno{\^\i}t and {Conselice}, Christopher J. and {Connolly}, Andrew J. and {Cuillandre}, Jean-Charles and {Hatch}, N.~A. and {Helou}, George and {Hemmati}, Shoubaneh and {Hildebrandt}, Hendrik and {Hlo{\v{z}}ek}, Ren{\'e}e and {Jones}, Lynne and {Kahn}, Steven and {Kiessling}, Alina and {Kitching}, Thomas and {Lupton}, Robert and {Mandelbaum}, Rachel and {Markovic}, Katarina and {Marshall}, Phil and {Massey}, Richard and {Maughan}, Ben J. and {Melchior}, Peter and {Mellier}, Yannick and {Newman}, Jeffrey A. and {Robertson}, Brant and {Sauvage}, Marc and {Schrabback}, Tim and {Smith}, Graham P. and {Strauss}, Michael A. and {Taylor}, Andy and {Von Der Linden}, Anja},
        title = "{Scientific Synergy between LSST and Euclid}",
      journal = {\apjs},
         year = 2017,
        month = dec,
       volume = {233},
       number = {2},
          eid = {21},
        pages = {21},
          doi = {10.3847/1538-4365/aa96b0},
archivePrefix = {arXiv},
       eprint = {1710.08489},
 primaryClass = {astro-ph.IM},
       adsurl = {https://ui.adsabs.harvard.edu/abs/2017ApJS..233...21R}
}

@ARTICLE{Hirschmann2016,
       author = {{Hirschmann}, Michaela and {De Lucia}, Gabriella and {Fontanot}, Fabio},
        title = "{Galaxy assembly, stellar feedback and metal enrichment: the view from the GAEA model}",
      journal = {\mnras},
         year = 2016,
        month = sep,
       volume = {461},
       number = {2},
        pages = {1760-1785},
          doi = {10.1093/mnras/stw1318},
archivePrefix = {arXiv},
       eprint = {1512.04531},
 primaryClass = {astro-ph.GA},
       adsurl = {https://ui.adsabs.harvard.edu/abs/2016MNRAS.461.1760H}
}

@ARTICLE{Chiang2013,
       author = {{Chiang}, Yi-Kuan and {Overzier}, Roderik and {Gebhardt}, Karl},
        title = "{Ancient Light from Young Cosmic Cities: Physical and Observational Signatures of Galaxy Proto-clusters}",
      journal = {\apj},
         year = 2013,
        month = dec,
       volume = {779},
       number = {2},
          eid = {127},
        pages = {127},
          doi = {10.1088/0004-637X/779/2/127},
archivePrefix = {arXiv},
       eprint = {1310.2938},
 primaryClass = {astro-ph.CO},
       adsurl = {https://ui.adsabs.harvard.edu/abs/2013ApJ...779..127C}
}

@ARTICLE{Wetzel2012,
       author = {{Wetzel}, Andrew R. and {Tinker}, Jeremy L. and {Conroy}, Charlie},
        title = "{Galaxy evolution in groups and clusters: star formation rates, red sequence fractions and the persistent bimodality}",
      journal = {\mnras},
         year = 2012,
        month = jul,
       volume = {424},
       number = {1},
        pages = {232-243},
          doi = {10.1111/j.1365-2966.2012.21188.x},
archivePrefix = {arXiv},
       eprint = {1107.5311},
 primaryClass = {astro-ph.CO},
       adsurl = {https://ui.adsabs.harvard.edu/abs/2012MNRAS.424..232W}
}

@ARTICLE{Peng2010,
       author = {{Peng}, Ying-jie and {Lilly}, Simon J. and {Kova{\v{c}}}, Katarina and {Bolzonella}, Micol and {Pozzetti}, Lucia and {Renzini}, Alvio and {Zamorani}, Gianni and {Ilbert}, Olivier and {Knobel}, Christian and {Iovino}, Angela and {Maier}, Christian and {Cucciati}, Olga and {Tasca}, Lidia and {Carollo}, C. Marcella and {Silverman}, John and {Kampczyk}, Pawel and {de Ravel}, Loic and {Sanders}, David and {Scoville}, Nicholas and {Contini}, Thierry and {Mainieri}, Vincenzo and {Scodeggio}, Marco and {Kneib}, Jean-Paul and {Le F{\`e}vre}, Olivier and {Bardelli}, Sandro and {Bongiorno}, Angela and {Caputi}, Karina and {Coppa}, Graziano and {de la Torre}, Sylvain and {Franzetti}, Paolo and {Garilli}, Bianca and {Lamareille}, Fabrice and {Le Borgne}, Jean-Francois and {Le Brun}, Vincent and {Mignoli}, Marco and {Perez Montero}, Enrique and {Pello}, Roser and {Ricciardelli}, Elena and {Tanaka}, Masayuki and {Tresse}, Laurence and {Vergani}, Daniela and {Welikala}, Niraj and {Zucca}, Elena and {Oesch}, Pascal and {Abbas}, Ummi and {Barnes}, Luke and {Bordoloi}, Rongmon and {Bottini}, Dario and {Cappi}, Alberto and {Cassata}, Paolo and {Cimatti}, Andrea and {Fumana}, Marco and {Hasinger}, Gunther and {Koekemoer}, Anton and {Leauthaud}, Alexei and {Maccagni}, Dario and {Marinoni}, Christian and {McCracken}, Henry and {Memeo}, Pierdomenico and {Meneux}, Baptiste and {Nair}, Preethi and {Porciani}, Cristiano and {Presotto}, Valentina and {Scaramella}, Roberto},
        title = "{Mass and Environment as Drivers of Galaxy Evolution in SDSS and zCOSMOS and the Origin of the Schechter Function}",
      journal = {\apj},
         year = 2010,
        month = sep,
       volume = {721},
       number = {1},
        pages = {193-221},
          doi = {10.1088/0004-637X/721/1/193},
archivePrefix = {arXiv},
       eprint = {1003.4747},
 primaryClass = {astro-ph.CO},
       adsurl = {https://ui.adsabs.harvard.edu/abs/2010ApJ...721..193P}
}

@ARTICLE{Lubin2009,
       author = {{Lubin}, L.~M. and {Gal}, R.~R. and {Lemaux}, B.~C. and {Kocevski}, D.~D. and {Squires}, G.~K.},
        title = "{The Observations of Redshift Evolution in Large-Scale Environments (ORELSE) Survey. I. The Survey Design and First Results on CL 0023+0423 at z = 0.84 and RX J1821.6+6827 at z = 0.82}",
      journal = {\aj},
         year = 2009,
        month = jun,
       volume = {137},
       number = {6},
        pages = {4867-4883},
          doi = {10.1088/0004-6256/137/6/4867},
archivePrefix = {arXiv},
       eprint = {0809.2092},
 primaryClass = {astro-ph},
       adsurl = {https://ui.adsabs.harvard.edu/abs/2009AJ....137.4867L}
}

@ARTICLE{DeLucia2007,
       author = {{De Lucia}, Gabriella and {Blaizot}, J{\'e}r{\'e}my},
        title = "{The hierarchical formation of the brightest cluster galaxies}",
      journal = {\mnras},
         year = 2007,
        month = feb,
       volume = {375},
       number = {1},
        pages = {2-14},
          doi = {10.1111/j.1365-2966.2006.11287.x},
archivePrefix = {arXiv},
       eprint = {astro-ph/0606519},
 primaryClass = {astro-ph},
       adsurl = {https://ui.adsabs.harvard.edu/abs/2007MNRAS.375....2D}
}

@ARTICLE{Postman2005,
       author = {{Postman}, M. and {Franx}, M. and {Cross}, N.~J.~G. and {Holden}, B. and {Ford}, H.~C. and {Illingworth}, G.~D. and {Goto}, T. and {Demarco}, R. and {Rosati}, P. and {Blakeslee}, J.~P. and {Tran}, K. -V. and {Ben{\'\i}tez}, N. and {Clampin}, M. and {Hartig}, G.~F. and {Homeier}, N. and {Ardila}, D.~R. and {Bartko}, F. and {Bouwens}, R.~J. and {Bradley}, L.~D. and {Broadhurst}, T.~J. and {Brown}, R.~A. and {Burrows}, C.~J. and {Cheng}, E.~S. and {Feldman}, P.~D. and {Golimowski}, D.~A. and {Gronwall}, C. and {Infante}, L. and {Kimble}, R.~A. and {Krist}, J.~E. and {Lesser}, M.~P. and {Martel}, A.~R. and {Mei}, S. and {Menanteau}, F. and {Meurer}, G.~R. and {Miley}, G.~K. and {Motta}, V. and {Sirianni}, M. and {Sparks}, W.~B. and {Tran}, H.~D. and {Tsvetanov}, Z.~I. and {White}, R.~L. and {Zheng}, W.},
        title = "{The Morphology-Density Relation in z \raisebox{-0.5ex}\textasciitilde 1 Clusters}",
      journal = {\apj},
         year = 2005,
        month = apr,
       volume = {623},
       number = {2},
        pages = {721-741},
          doi = {10.1086/428881},
archivePrefix = {arXiv},
       eprint = {astro-ph/0501224},
 primaryClass = {astro-ph},
       adsurl = {https://ui.adsabs.harvard.edu/abs/2005ApJ...623..721P}
}

@ARTICLE{Chiang2017,
       author = {{Chiang}, Yi-Kuan and {Overzier}, Roderik A. and {Gebhardt}, Karl and {Henriques}, Bruno},
        title = "{Galaxy Protoclusters as Drivers of Cosmic Star Formation History in the First 2 Gyr}",
      journal = {\apjl},
         year = 2017,
        month = aug,
       volume = {844},
       number = {2},
          eid = {L23},
        pages = {L23},
          doi = {10.3847/2041-8213/aa7e7b},
archivePrefix = {arXiv},
       eprint = {1705.01634},
 primaryClass = {astro-ph.GA},
       adsurl = {https://ui.adsabs.harvard.edu/abs/2017ApJ...844L..23C}
}

@ARTICLE{Dressler1980,
       author = {{Dressler}, A.},
        title = "{Galaxy morphology in rich clusters: implications for the formation and evolution of galaxies.}",
      journal = {\apj},
         year = 1980,
        month = mar,
       volume = {236},
        pages = {351-365},
          doi = {10.1086/157753},
       adsurl = {https://ui.adsabs.harvard.edu/abs/1980ApJ...236..351D}
}

@ARTICLE{Pozzetti2010,
       author = {{Pozzetti}, L. and {Bolzonella}, M. and {Zucca}, E. and {Zamorani}, G. and {Lilly}, S. and {Renzini}, A. and {Moresco}, M. and {Mignoli}, M. and {Cassata}, P. and {Tasca}, L. and {Lamareille}, F. and {Maier}, C. and {Meneux}, B. and {Halliday}, C. and {Oesch}, P. and {Vergani}, D. and {Caputi}, K. and {Kova{\v{c}}}, K. and {Cimatti}, A. and {Cucciati}, O. and {Iovino}, A. and {Peng}, Y. and {Carollo}, M. and {Contini}, T. and {Kneib}, J. -P. and {Le F{\'e}vre}, O. and {Mainieri}, V. and {Scodeggio}, M. and {Bardelli}, S. and {Bongiorno}, A. and {Coppa}, G. and {de la Torre}, S. and {de Ravel}, L. and {Franzetti}, P. and {Garilli}, B. and {Kampczyk}, P. and {Knobel}, C. and {Le Borgne}, J. -F. and {Le Brun}, V. and {Pell{\`o}}, R. and {Perez Montero}, E. and {Ricciardelli}, E. and {Silverman}, J.~D. and {Tanaka}, M. and {Tresse}, L. and {Abbas}, U. and {Bottini}, D. and {Cappi}, A. and {Guzzo}, L. and {Koekemoer}, A.~M. and {Leauthaud}, A. and {Maccagni}, D. and {Marinoni}, C. and {McCracken}, H.~J. and {Memeo}, P. and {Porciani}, C. and {Scaramella}, R. and {Scarlata}, C. and {Scoville}, N.},
        title = "{zCOSMOS - 10k-bright spectroscopic sample. The bimodality in the galaxy stellar mass function: exploring its evolution with redshift}",
      journal = {\aap},
         year = 2010,
        month = nov,
       volume = {523},
          eid = {A13},
        pages = {A13},
          doi = {10.1051/0004-6361/200913020},
archivePrefix = {arXiv},
       eprint = {0907.5416},
 primaryClass = {astro-ph.CO},
       adsurl = {https://ui.adsabs.harvard.edu/abs/2010A&A...523A..13P}
}

@ARTICLE{Lemaux2019,
       author = {{Lemaux}, B.~C. and {Tomczak}, A.~R. and {Lubin}, L.~M. and {Gal}, R.~R. and {Shen}, L. and {Pelliccia}, D. and {Wu}, P. -F. and {Hung}, D. and {Mei}, S. and {Le F{\`e}vre}, O. and {Rumbaugh}, N. and {Kocevski}, D.~D. and {Squires}, G.~K.},
        title = "{Persistence of the colour-density relation and efficient environmental quenching to z {\ensuremath{\sim}} 1.4}",
      journal = {\mnras},
         year = 2019,
        month = nov,
       volume = {490},
       number = {1},
        pages = {1231-1254},
          doi = {10.1093/mnras/stz2661},
archivePrefix = {arXiv},
       eprint = {1812.04624},
 primaryClass = {astro-ph.GA},
       adsurl = {https://ui.adsabs.harvard.edu/abs/2019MNRAS.490.1231L}
}

@INPROCEEDINGS{Bertin2020,
       author = {{Bertin}, E. and {Schefer}, M. and {Apostolakos}, N. and {{\'A}lvarez-Ayll{\'o}n}, A. and {Dubath}, P. and {K{\"u}mmel}, M.},
        title = "{The SourceXtractor++ Software}",
    booktitle = {Astronomical Data Analysis Software and Systems XXIX},
         year = 2020,
       editor = {{Pizzo}, R. and {Deul}, E.~R. and {Mol}, J.~D. and {de Plaa}, J. and {Verkouter}, H.},
       series = {Astronomical Society of the Pacific Conference Series},
       volume = {527},
        month = jan,
        pages = {461},
       adsurl = {https://ui.adsabs.harvard.edu/abs/2020ASPC..527..461B}
}

@ARTICLE{Tomczak2017,
       author = {{Tomczak}, Adam R. and {Lemaux}, Brian C. and {Lubin}, Lori M. and {Gal}, Roy R. and {Wu}, Po-Feng and {Holden}, Bradford and {Kocevski}, Dale D. and {Mei}, Simona and {Pelliccia}, Debora and {Rumbaugh}, Nicholas and {Shen}, Lu},
        title = "{Glimpsing the imprint of local environment on the galaxy stellar mass function}",
      journal = {\mnras},
         year = 2017,
        month = dec,
       volume = {472},
       number = {3},
        pages = {3512-3531},
          doi = {10.1093/mnras/stx2245},
archivePrefix = {arXiv},
       eprint = {1709.00011},
 primaryClass = {astro-ph.GA},
       adsurl = {https://ui.adsabs.harvard.edu/abs/2017MNRAS.472.3512T}
}

@ARTICLE{Cucciati2010,
       author = {{Cucciati}, O. and {Iovino}, A. and {Kova{\v{c}}}, K. and {Scodeggio}, M. and {Lilly}, S.~J. and {Bolzonella}, M. and {Bardelli}, S. and {Vergani}, D. and {Tasca}, L.~A.~M. and {Zucca}, E. and {Zamorani}, G. and {Pozzetti}, L. and {Knobel}, C. and {Oesch}, P. and {Lamareille}, F. and {Caputi}, K. and {Kampczyk}, P. and {Tresse}, L. and {Maier}, C. and {Carollo}, C.~M. and {Contini}, T. and {Kneib}, J. -P. and {Le F{\`e}vre}, O. and {Mainieri}, V. and {Renzini}, A. and {Bongiorno}, A. and {Coppa}, G. and {de la Torre}, S. and {de Ravel}, L. and {Franzetti}, P. and {Garilli}, B. and {Le Borgne}, J. -F. and {Le Brun}, V. and {Mignoli}, M. and {Pell{\`o}}, R. and {Peng}, Y. and {Perez-Montero}, E. and {Ricciardelli}, E. and {Silverman}, J.~D. and {Tanaka}, M. and {Koekemoer}, A.~M. and {Scoville}, N. and {Abbas}, U. and {Bottini}, D. and {Cappi}, A. and {Cassata}, P. and {Cimatti}, A. and {Guzzo}, L. and {Leauthaud}, A. and {Maccagni}, D. and {Marinoni}, C. and {McCracken}, H.~J. and {Memeo}, P. and {Meneux}, B. and {Porciani}, C. and {Scaramella}, R.},
        title = "{The zCOSMOS 10k-sample: the role of galaxy stellar mass in the colour-density relation up to z \raisebox{-0.5ex}\textasciitilde 1}",
      journal = {\aap},
         year = 2010,
        month = dec,
       volume = {524},
          eid = {A2},
        pages = {A2},
          doi = {10.1051/0004-6361/200912585},
archivePrefix = {arXiv},
       eprint = {1007.3841},
 primaryClass = {astro-ph.CO},
       adsurl = {https://ui.adsabs.harvard.edu/abs/2010A&A...524A...2C}
}

@ARTICLE{Cucciati2017,
       author = {{Cucciati}, O. and {Davidzon}, I. and {Bolzonella}, M. and {Granett}, B.~R. and {De Lucia}, G. and {Branchini}, E. and {Zamorani}, G. and {Iovino}, A. and {Garilli}, B. and {Guzzo}, L. and {Scodeggio}, M. and {de la Torre}, S. and {Abbas}, U. and {Adami}, C. and {Arnouts}, S. and {Bottini}, D. and {Cappi}, A. and {Franzetti}, P. and {Fritz}, A. and {Krywult}, J. and {Le Brun}, V. and {Le F{\`e}vre}, O. and {Maccagni}, D. and {Ma{\l}ek}, K. and {Marulli}, F. and {Moutard}, T. and {Polletta}, M. and {Pollo}, A. and {Tasca}, L.~A.~M. and {Tojeiro}, R. and {Vergani}, D. and {Zanichelli}, A. and {Bel}, J. and {Blaizot}, J. and {Coupon}, J. and {Hawken}, A. and {Ilbert}, O. and {Moscardini}, L. and {Peacock}, J.~A. and {Gargiulo}, A.},
        title = "{The VIMOS Public Extragalactic Redshift Survey (VIPERS). The decline of cosmic star formation: quenching, mass, and environment connections}",
      journal = {\aap},
         year = 2017,
        month = jun,
       volume = {602},
          eid = {A15},
        pages = {A15},
          doi = {10.1051/0004-6361/201630113},
archivePrefix = {arXiv},
       eprint = {1611.07049},
 primaryClass = {astro-ph.GA},
       adsurl = {https://ui.adsabs.harvard.edu/abs/2017A&A...602A..15C}
}

@ARTICLE{Trudeau2024,
       author = {{Trudeau}, A. and {Gonzalez}, Anthony H. and {Thongkham}, K. and {Lee}, Kyoung-Soo and {Alberts}, Stacey and {Brodwin}, M. and {Connor}, Thomas and {Eisenhardt}, Peter R.~M. and {Moravec}, Emily and {Puvvada}, Eshwar and {Stanford}, S.~A.},
        title = "{The Massive and Distant Clusters of WISE Survey 2: A Stacking Analysis Investigating the Evolution of Star Formation Rates and Stellar Masses in Groups and Clusters}",
      journal = {\apj},
         year = 2024,
        month = sep,
       volume = {972},
       number = {1},
          eid = {27},
        pages = {27},
          doi = {10.3847/1538-4357/ad5545},
archivePrefix = {arXiv},
       eprint = {2406.03633},
 primaryClass = {astro-ph.GA},
       adsurl = {https://ui.adsabs.harvard.edu/abs/2024ApJ...972...27T}
}

@ARTICLE{Taamoli2024,
       author = {{Taamoli}, Sina and {Mobasher}, Bahram and {Chartab}, Nima and {Darvish}, Behnam and {Weaver}, John R. and {Hemmati}, Shoubaneh and {Casey}, Caitlin M. and {Sattari}, Zahra and {Brammer}, Gabriel and {Capak}, Peter L. and {Ilbert}, Olivier and {Kartaltepe}, Jeyhan S. and {McCracken}, Henry J. and {Moneti}, Andrea and {Sanders}, David B. and {Scoville}, Nicholas and {Steinhardt}, Charles L. and {Toft}, Sune},
        title = "{Large-scale Structures in COSMOS2020: Evolution of Star Formation Activity in Different Environments at 0.4 < z < 4}",
      journal = {\apj},
         year = 2024,
        month = may,
       volume = {966},
       number = {1},
          eid = {18},
        pages = {18},
          doi = {10.3847/1538-4357/ad32c5},
archivePrefix = {arXiv},
       eprint = {2312.10222},
 primaryClass = {astro-ph.GA},
       adsurl = {https://ui.adsabs.harvard.edu/abs/2024ApJ...966...18T}
}

@ARTICLE{Baxter2023,
       author = {{Baxter}, Devontae C. and {Cooper}, M.~C. and {Balogh}, Michael L. and {Rudnick}, Gregory H. and {De Lucia}, Gabriella and {Demarco}, Ricardo and {Finoguenov}, Alexis and {Forrest}, Ben and {Muzzin}, Adam and {Reeves}, Andrew M.~M. and {Sarron}, Florian and {Vulcani}, Benedetta and {Wilson}, Gillian and {Zaritsky}, Dennis},
        title = "{When the well runs dry: modelling environmental quenching of high-mass satellites in massive clusters at z {\ensuremath{\gtrsim}} 1}",
      journal = {\mnras},
         year = 2023,
        month = dec,
       volume = {526},
       number = {3},
        pages = {3716-3729},
          doi = {10.1093/mnras/stad2995},
archivePrefix = {arXiv},
       eprint = {2306.09404},
 primaryClass = {astro-ph.GA},
       adsurl = {https://ui.adsabs.harvard.edu/abs/2023MNRAS.526.3716B}
}

@ARTICLE{Sazonova2020,
       author = {{Sazonova}, Elizaveta and {Alatalo}, Katherine and {Lotz}, Jennifer and {Rowlands}, Kate and {Snyder}, Gregory F. and {Boone}, Kyle and {Brodwin}, Mark and {Hayden}, Brian and {Lanz}, Lauranne and {Perlmutter}, Saul and {Rodriguez-Gomez}, Vicente},
        title = "{The Morphology-Density Relationship in 1 < z < 2 Clusters}",
      journal = {\apj},
         year = 2020,
        month = aug,
       volume = {899},
       number = {1},
          eid = {85},
        pages = {85},
          doi = {10.3847/1538-4357/aba42f},
archivePrefix = {arXiv},
       eprint = {2007.03698},
 primaryClass = {astro-ph.GA},
       adsurl = {https://ui.adsabs.harvard.edu/abs/2020ApJ...899...85S}
}

@ARTICLE{Shi2024,
       author = {{Shi}, Ke and {Malavasi}, Nicola and {Toshikawa}, Jun and {Zheng}, Xianzhong},
        title = "{Nature versus Nurture: Revisiting the Environmental Impact on Star Formation Activities of Galaxies}",
      journal = {\apj},
         year = 2024,
        month = jan,
       volume = {961},
       number = {1},
          eid = {39},
        pages = {39},
          doi = {10.3847/1538-4357/ad11d7},
archivePrefix = {arXiv},
       eprint = {2311.18427},
 primaryClass = {astro-ph.GA},
       adsurl = {https://ui.adsabs.harvard.edu/abs/2024ApJ...961...39S}
}

@ARTICLE{vanderBurg2020,
       author = {{van der Burg}, Remco F.~J. and {Rudnick}, Gregory and {Balogh}, Michael L. and {Muzzin}, Adam and {Lidman}, Chris and {Old}, Lyndsay J. and {Shipley}, Heath and {Gilbank}, David and {McGee}, Sean and {Biviano}, Andrea and {Cerulo}, Pierluigi and {Chan}, Jeffrey C.~C. and {Cooper}, Michael and {De Lucia}, Gabriella and {Demarco}, Ricardo and {Forrest}, Ben and {Gwyn}, Stephen and {Jablonka}, Pascale and {Kukstas}, Egidijus and {Marchesini}, Danilo and {Nantais}, Julie and {Noble}, Allison and {Pintos-Castro}, Irene and {Poggianti}, Bianca and {Reeves}, Andrew M.~M. and {Stefanon}, Mauro and {Vulcani}, Benedetta and {Webb}, Kristi and {Wilson}, Gillian and {Yee}, Howard and {Zaritsky}, Dennis},
        title = "{The GOGREEN Survey: A deep stellar mass function of cluster galaxies at 1.0 < z < 1.4 and the complex nature of satellite quenching}",
      journal = {\aap},
         year = 2020,
        month = jun,
       volume = {638},
          eid = {A112},
        pages = {A112},
          doi = {10.1051/0004-6361/202037754},
archivePrefix = {arXiv},
       eprint = {2004.10757},
 primaryClass = {astro-ph.GA},
       adsurl = {https://ui.adsabs.harvard.edu/abs/2020A&A...638A.112V}
}

@ARTICLE{Strazzullo2019,
       author = {{Strazzullo}, V. and {Pannella}, M. and {Mohr}, J.~J. and {Saro}, A. and {Ashby}, M.~L.~N. and {Bayliss}, M.~B. and {Bocquet}, S. and {Bulbul}, E. and {Khullar}, G. and {Mantz}, A.~B. and {Stanford}, S.~A. and {Benson}, B.~A. and {Bleem}, L.~E. and {Brodwin}, M. and {Canning}, R.~E.~A. and {Capasso}, R. and {Chiu}, I. and {Gonzalez}, A.~H. and {Gupta}, N. and {Hlavacek-Larrondo}, J. and {Klein}, M. and {McDonald}, M. and {Noordeh}, E. and {Rapetti}, D. and {Reichardt}, C.~L. and {Schrabback}, T. and {Sharon}, K. and {Stalder}, B.},
        title = "{Galaxy populations in the most distant SPT-SZ clusters. I. Environmental quenching in massive clusters at 1.4 {\ensuremath{\lesssim}} z {\ensuremath{\lesssim}} 1.7}",
      journal = {\aap},
         year = 2019,
        month = feb,
       volume = {622},
          eid = {A117},
        pages = {A117},
          doi = {10.1051/0004-6361/201833944},
archivePrefix = {arXiv},
       eprint = {1807.09768},
 primaryClass = {astro-ph.GA},
       adsurl = {https://ui.adsabs.harvard.edu/abs/2019A&A...622A.117S}
}

@article{Vulcani2010,
doi = {10.1088/2041-8205/710/1/L1},
url = {https://dx.doi.org/10.1088/2041-8205/710/1/L1},
year = {2010},
month = {jan},
publisher = {The American Astronomical Society},
volume = {710},
number = {1},
pages = {L1},
author = {Vulcani, Benedetta and Poggianti, Bianca M. and Finn, Rose A. and Rudnick, Gregory and Desai, Vandana and Bamford, Steven},
title = {COMPARING THE RELATION BETWEEN STAR FORMATION AND GALAXY MASS IN DIFFERENT ENVIRONMENTS},
journal = {\apj}
}

@ARTICLE{Paccagnella2016,
       author = {{Paccagnella}, A. and {Vulcani}, B. and {Poggianti}, B.~M. and {Moretti}, A. and {Fritz}, J. and {Gullieuszik}, M. and {Couch}, W. and {Bettoni}, D. and {Cava}, A. and {D'Onofrio}, M. and {Fasano}, G.},
        title = "{Slow Quenching of Star Formation in OMEGAWINGS Clusters: Galaxies in Transition in the Local Universe}",
      journal = {\apjl},
         year = 2016,
        month = jan,
       volume = {816},
       number = {2},
          eid = {L25},
        pages = {L25},
          doi = {10.3847/2041-8205/816/2/L25},
archivePrefix = {arXiv},
       eprint = {1512.04549},
 primaryClass = {astro-ph.GA},
       adsurl = {https://ui.adsabs.harvard.edu/abs/2016ApJ...816L..25P}
}

@ARTICLE{Darvish2017,
       author = {{Darvish}, Behnam and {Mobasher}, Bahram and {Martin}, D. Christopher and {Sobral}, David and {Scoville}, Nick and {Stroe}, Andra and {Hemmati}, Shoubaneh and {Kartaltepe}, Jeyhan},
        title = "{Cosmic Web of Galaxies in the COSMOS Field: Public Catalog and Different Quenching for Centrals and Satellites}",
      journal = {\apj},
         year = 2017,
        month = mar,
       volume = {837},
       number = {1},
          eid = {16},
        pages = {16},
          doi = {10.3847/1538-4357/837/1/16},
archivePrefix = {arXiv},
       eprint = {1611.05451},
 primaryClass = {astro-ph.GA},
       adsurl = {https://ui.adsabs.harvard.edu/abs/2017ApJ...837...16D}
}

@ARTICLE{Kauffmann2003,
       author = {{Kauffmann}, Guinevere and {Heckman}, Timothy M. and {White}, Simon D.~M. and {Charlot}, St{\'e}phane and {Tremonti}, Christy and {Peng}, Eric W. and {Seibert}, Mark and {Brinkmann}, Jon and {Nichol}, Robert C. and {SubbaRao}, Mark and {York}, Don},
        title = "{The dependence of star formation history and internal structure on stellar mass for {}10$^{5}$ low-redshift galaxies}",
      journal = {\mnras},
         year = 2003,
        month = may,
       volume = {341},
       number = {1},
        pages = {54-69},
          doi = {10.1046/j.1365-8711.2003.06292.x},
archivePrefix = {arXiv},
       eprint = {astro-ph/0205070},
 primaryClass = {astro-ph},
       adsurl = {https://ui.adsabs.harvard.edu/abs/2003MNRAS.341...54K}
}

@ARTICLE{Balogh2000,
       author = {{Balogh}, Michael L. and {Navarro}, Julio F. and {Morris}, Simon L.},
        title = "{The Origin of Star Formation Gradients in Rich Galaxy Clusters}",
      journal = {\apj},
         year = 2000,
        month = sep,
       volume = {540},
       number = {1},
        pages = {113-121},
          doi = {10.1086/309323},
archivePrefix = {arXiv},
       eprint = {astro-ph/0004078},
 primaryClass = {astro-ph},
       adsurl = {https://ui.adsabs.harvard.edu/abs/2000ApJ...540..113B}
}

@ARTICLE{Gunn1972,
       author = {{Gunn}, James E. and {Gott}, III, J. Richard},
        title = "{On the Infall of Matter Into Clusters of Galaxies and Some Effects on Their Evolution}",
      journal = {\apj},
         year = 1972,
        month = aug,
       volume = {176},
        pages = {1},
          doi = {10.1086/151605},
       adsurl = {https://ui.adsabs.harvard.edu/abs/1972ApJ...176....1G}
}

@ARTICLE{Moore1996,
       author = {{Moore}, Ben and {Katz}, Neal and {Lake}, George and {Dressler}, Alan and {Oemler}, Augustus},
        title = "{Galaxy harassment and the evolution of clusters of galaxies}",
      journal = {\nat},
         year = 1996,
        month = feb,
       volume = {379},
       number = {6566},
        pages = {613-616},
          doi = {10.1038/379613a0},
archivePrefix = {arXiv},
       eprint = {astro-ph/9510034},
 primaryClass = {astro-ph},
       adsurl = {https://ui.adsabs.harvard.edu/abs/1996Natur.379..613M}
}

@ARTICLE{Toomre1972,
       author = {{Toomre}, Alar and {Toomre}, Juri},
        title = "{Galactic Bridges and Tails}",
      journal = {\apj},
         year = 1972,
        month = dec,
       volume = {178},
        pages = {623-666},
          doi = {10.1086/151823},
       adsurl = {https://ui.adsabs.harvard.edu/abs/1972ApJ...178..623T}
}

@ARTICLE{Overzier2016,
       author = {{Overzier}, Roderik A.},
        title = "{The realm of the galaxy protoclusters. A review}",
      journal = {\aapr},
         year = 2016,
        month = nov,
       volume = {24},
       number = {1},
          eid = {14},
        pages = {14},
          doi = {10.1007/s00159-016-0100-3},
archivePrefix = {arXiv},
       eprint = {1610.05201},
 primaryClass = {astro-ph.GA},
       adsurl = {https://ui.adsabs.harvard.edu/abs/2016A&ARv..24...14O}
}

@ARTICLE{Lovell2018,
       author = {{Lovell}, Christopher C. and {Thomas}, Peter A. and {Wilkins}, Stephen M.},
        title = "{Characterising and identifying galaxy protoclusters}",
      journal = {\mnras},
         year = 2018,
        month = mar,
       volume = {474},
       number = {4},
        pages = {4612-4628},
          doi = {10.1093/mnras/stx3090},
archivePrefix = {arXiv},
       eprint = {1710.02148},
 primaryClass = {astro-ph.GA},
       adsurl = {https://ui.adsabs.harvard.edu/abs/2018MNRAS.474.4612L}
}

@ARTICLE{Merlin2018,
       author = {{Merlin}, E. and {Fontana}, A. and {Castellano}, M. and {Santini}, P. and {Torelli}, M. and {Boutsia}, K. and {Wang}, T. and {Grazian}, A. and {Pentericci}, L. and {Schreiber}, C. and {Ciesla}, L. and {McLure}, R. and {Derriere}, S. and {Dunlop}, J.~S. and {Elbaz}, D.},
        title = "{Chasing passive galaxies in the early Universe: a critical analysis in CANDELS GOODS-South}",
      journal = {\mnras},
         year = 2018,
        month = jan,
       volume = {473},
       number = {2},
        pages = {2098-2123},
          doi = {10.1093/mnras/stx2385},
archivePrefix = {arXiv},
       eprint = {1709.00429},
 primaryClass = {astro-ph.GA},
       adsurl = {https://ui.adsabs.harvard.edu/abs/2018MNRAS.473.2098M}
}

@ARTICLE{Merlin2019,
       author = {{Merlin}, E. and {Fortuni}, F. and {Torelli}, M. and {Santini}, P. and {Castellano}, M. and {Fontana}, A. and {Grazian}, A. and {Pentericci}, L. and {Pilo}, S. and {Schmidt}, K.~B.},
        title = "{Red and dead CANDELS: massive passive galaxies at the dawn of the Universe}",
      journal = {\mnras},
         year = 2019,
        month = dec,
       volume = {490},
       number = {3},
        pages = {3309-3328},
          doi = {10.1093/mnras/stz2615},
archivePrefix = {arXiv},
       eprint = {1909.07996},
 primaryClass = {astro-ph.GA},
       adsurl = {https://ui.adsabs.harvard.edu/abs/2019MNRAS.490.3309M}
}

@ARTICLE{Speagle2014,
       author = {{Speagle}, J.~S. and {Steinhardt}, C.~L. and {Capak}, P.~L. and {Silverman}, J.~D.},
        title = "{A Highly Consistent Framework for the Evolution of the Star-Forming ``Main Sequence'' from z \raisebox{-0.5ex}\textasciitilde 0-6}",
      journal = {\apjs},
         year = 2014,
        month = oct,
       volume = {214},
       number = {2},
          eid = {15},
        pages = {15},
          doi = {10.1088/0067-0049/214/2/15},
archivePrefix = {arXiv},
       eprint = {1405.2041},
 primaryClass = {astro-ph.GA},
       adsurl = {https://ui.adsabs.harvard.edu/abs/2014ApJS..214...15S}
}

@ARTICLE{Larson1980,
       author = {{Larson}, R.~B. and {Tinsley}, B.~M. and {Caldwell}, C.~N.},
        title = "{The evolution of disk galaxies and the origin of S0 galaxies}",
      journal = {\apj},
         year = 1980,
        month = may,
       volume = {237},
        pages = {692-707},
          doi = {10.1086/157917},
       adsurl = {https://ui.adsabs.harvard.edu/abs/1980ApJ...237..692L}
}

@ARTICLE{Corcho-Caballero2023,
       author = {{Corcho-Caballero}, Pablo and {Ascasibar}, Yago and {Cortese}, Luca and {S{\'a}nchez}, Sebasti{\'a}n F. and {L{\'o}pez-S{\'a}nchez}, {\'A}ngel R. and {Fraser-McKelvie}, Amelia and {Zafar}, Tayyaba},
        title = "{Ageing and quenching through the Ageing Diagram - II. Physical characterization of galaxies}",
      journal = {\mnras},
         year = 2023,
        month = sep,
       volume = {524},
       number = {3},
        pages = {3692-3704},
          doi = {10.1093/mnras/stad2096},
archivePrefix = {arXiv},
       eprint = {2307.02024},
 primaryClass = {astro-ph.GA},
       adsurl = {https://ui.adsabs.harvard.edu/abs/2023MNRAS.524.3692C}
}

@ARTICLE{Cleland2021,
       author = {{Cleland}, Cressida and {McGee}, Sean L.},
        title = "{The environmental dependence of rapidly quenching and rejuvenating galaxies}",
      journal = {\mnras},
         year = 2021,
        month = jan,
       volume = {500},
       number = {1},
        pages = {590-602},
          doi = {10.1093/mnras/staa3267},
archivePrefix = {arXiv},
       eprint = {2006.16307},
 primaryClass = {astro-ph.GA},
       adsurl = {https://ui.adsabs.harvard.edu/abs/2021MNRAS.500..590C}
}

@ARTICLE{DominguezSanchez2022,
       author = {{Dom{\'\i}nguez S{\'a}nchez}, H. and {Margalef}, B. and {Bernardi}, M. and {Huertas-Company}, M.},
        title = "{SDSS-IV DR17: final release of MaNGA PyMorph photometric and deep-learning morphological catalogues}",
      journal = {\mnras},
         year = 2022,
        month = jan,
       volume = {509},
       number = {3},
        pages = {4024-4036},
          doi = {10.1093/mnras/stab3089},
archivePrefix = {arXiv},
       eprint = {2110.10694},
 primaryClass = {astro-ph.GA},
       adsurl = {https://ui.adsabs.harvard.edu/abs/2022MNRAS.509.4024D}
}

@ARTICLE{Straatman2016,
       author = {{Straatman}, Caroline M.~S. and {Spitler}, Lee R. and {Quadri}, Ryan F. and {Labb{\'e}}, Ivo and {Glazebrook}, Karl and {Persson}, S. Eric and {Papovich}, Casey and {Tran}, Kim-Vy H. and {Brammer}, Gabriel B. and {Cowley}, Michael and {Tomczak}, Adam and {Nanayakkara}, Themiya and {Alcorn}, Leo and {Allen}, Rebecca and {Broussard}, Adam and {van Dokkum}, Pieter and {Forrest}, Ben and {van Houdt}, Josha and {Kacprzak}, Glenn G. and {Kawinwanichakij}, Lalitwadee and {Kelson}, Daniel D. and {Lee}, Janice and {McCarthy}, Patrick J. and {Mehrtens}, Nicola and {Monson}, Andrew and {Murphy}, David and {Rees}, Glen and {Tilvi}, Vithal and {Whitaker}, Katherine E.},
        title = "{The FourStar Galaxy Evolution Survey (ZFOURGE): Ultraviolet to Far-infrared Catalogs, Medium-bandwidth Photometric Redshifts with Improved Accuracy, Stellar Masses, and Confirmation of Quiescent Galaxies to z {\ensuremath{\sim}} 3.5}",
      journal = {\apj},
         year = 2016,
        month = oct,
       volume = {830},
       number = {1},
          eid = {51},
        pages = {51},
          doi = {10.3847/0004-637X/830/1/51},
archivePrefix = {arXiv},
       eprint = {1608.07579},
 primaryClass = {astro-ph.GA},
       adsurl = {https://ui.adsabs.harvard.edu/abs/2016ApJ...830...51S}
}

@ARTICLE{Kummel2022,
       author = {{K{\"u}mmel}, M. and {{\'A}lvarez-Ayll{\'o}n}, A. and {Bertin}, E. and {Dubath}, P. and {Gavazzi}, R. and {Hartley}, W. and {Schefer}, M.},
        title = "{Using the SourceXtractor++ package for data reduction}",
      journal = {arXiv e-prints},
         year = 2022,
        month = dec,
          eid = {arXiv:2212.02428},
        pages = {arXiv:2212.02428},
          doi = {10.48550/arXiv.2212.02428},
archivePrefix = {arXiv},
       eprint = {2212.02428},
 primaryClass = {astro-ph.IM},
       adsurl = {https://ui.adsabs.harvard.edu/abs/2022arXiv221202428K}
}

@ARTICLE{Mei2006,
       author = {{Mei}, S. and {Blakeslee}, J.~P. and {Stanford}, S.~A. and {Holden}, B.~P. and {Rosati}, P. and {Strazzullo}, V. and {Homeier}, N. and {Postman}, M. and {Franx}, M. and {Rettura}, A. and {Ford}, H. and {Illingworth}, G.~D. and {Ettori}, S. and {Bouwens}, R.~J. and {Demarco}, R. and {Martel}, A.~R. and {Clampin}, M. and {Hartig}, G.~F. and {Eisenhardt}, P. and {Ardila}, D.~R. and {Bartko}, F. and {Ben{\'\i}tez}, N. and {Bradley}, L.~D. and {Broadhurst}, T.~J. and {Brown}, R.~A. and {Burrows}, C.~J. and {Cheng}, E.~S. and {Cross}, N.~J.~G. and {Feldman}, P.~D. and {Golimowski}, D.~A. and {Goto}, T. and {Gronwall}, C. and {Infante}, L. and {Kimble}, R.~A. and {Krist}, J.~E. and {Lesser}, M.~P. and {Menanteau}, F. and {Meurer}, G.~R. and {Miley}, G.~K. and {Motta}, V. and {Sirianni}, M. and {Sparks}, W.~B. and {Tran}, H.~D. and {Tsvetanov}, Z.~I. and {White}, R.~L. and {Zheng}, W.},
        title = "{Evolution of the Color-Magnitude Relation in High-Redshift Clusters: Blue Early-Type Galaxies and Red Pairs in RDCS J0910+5422}",
      journal = {\apj},
         year = 2006,
        month = mar,
       volume = {639},
       number = {1},
        pages = {81-94},
          doi = {10.1086/499259},
archivePrefix = {arXiv},
       eprint = {astro-ph/0601327},
 primaryClass = {astro-ph},
       adsurl = {https://ui.adsabs.harvard.edu/abs/2006ApJ...639...81M}
}

@ARTICLE{Mei2015,
       author = {{Mei}, Simona and {Scarlata}, Claudia and {Pentericci}, Laura and {Newman}, Jeffrey A. and {Weiner}, Benjamin J. and {Ashby}, Matthew L.~N. and {Castellano}, Marco and {Conselice}, Chistopher J. and {Finkelstein}, Steven L. and {Galametz}, Audrey and {Grogin}, Norman A. and {Koekemoer}, Anton M. and {Huertas-Company}, Marc and {Lani}, Caterina and {Lucas}, Ray A. and {Papovich}, Casey and {Rafelski}, Marc and {Teplitz}, Harry I.},
        title = "{Star-forming Blue ETGs in Two Newly Discovered Galaxy Overdensities in the HUDF at z=1.84 and 1.9: Unveiling the Progenitors of Passive ETGs in Cluster Cores}",
      journal = {\apj},
         year = 2015,
        month = may,
       volume = {804},
       number = {2},
          eid = {117},
        pages = {117},
          doi = {10.1088/0004-637X/804/2/117},
archivePrefix = {arXiv},
       eprint = {1403.7524},
 primaryClass = {astro-ph.GA},
       adsurl = {https://ui.adsabs.harvard.edu/abs/2015ApJ...804..117M}
}

@ARTICLE{McIntosh2014,
       author = {{McIntosh}, Daniel H. and {Wagner}, Cory and {Cooper}, Andrew and {Bell}, Eric F. and {Kere{\v{s}}}, Du{\v{s}}an and {Bosch}, Frank C. van den and {Gallazzi}, Anna and {Haines}, Tim and {Mann}, Justin and {Pasquali}, Anna and {Christian}, Allison M.},
        title = "{A new population of recently quenched elliptical galaxies in the SDSS}",
      journal = {\mnras},
         year = 2014,
        month = jul,
       volume = {442},
       number = {1},
        pages = {533-557},
          doi = {10.1093/mnras/stu808},
archivePrefix = {arXiv},
       eprint = {1308.0054},
 primaryClass = {astro-ph.GA},
       adsurl = {https://ui.adsabs.harvard.edu/abs/2014MNRAS.442..533M}
}

@ARTICLE{Rowlands2012,
       author = {{Rowlands}, K. and {Dunne}, L. and {Maddox}, S. and {Bourne}, N. and {Gomez}, H.~L. and {Kaviraj}, S. and {Bamford}, S.~P. and {Brough}, S. and {Charlot}, S. and {da Cunha}, E. and {Driver}, S.~P. and {Eales}, S.~A. and {Hopkins}, A.~M. and {Kelvin}, L. and {Nichol}, R.~C. and {Sansom}, A.~E. and {Sharp}, R. and {Smith}, D.~J.~B. and {Temi}, P. and {van der Werf}, P. and {Baes}, M. and {Cava}, A. and {Cooray}, A. and {Croom}, S.~M. and {Dariush}, A. and {de Zotti}, G. and {Dye}, S. and {Fritz}, J. and {Hopwood}, R. and {Ibar}, E. and {Ivison}, R.~J. and {Liske}, J. and {Loveday}, J. and {Madore}, B. and {Norberg}, P. and {Popescu}, C.~C. and {Rigby}, E.~E. and {Robotham}, A. and {Rodighiero}, G. and {Seibert}, M. and {Tuffs}, R.~J.},
        title = "{Herschel-ATLAS/GAMA: dusty early-type galaxies and passive spirals}",
      journal = {\mnras},
         year = 2012,
        month = jan,
       volume = {419},
       number = {3},
        pages = {2545-2578},
          doi = {10.1111/j.1365-2966.2011.19905.x},
archivePrefix = {arXiv},
       eprint = {1109.6274},
 primaryClass = {astro-ph.CO},
       adsurl = {https://ui.adsabs.harvard.edu/abs/2012MNRAS.419.2545R}
}

@ARTICLE{Fudamoto2022,
       author = {{Fudamoto}, Yoshinobu and {Inoue}, Akio K. and {Sugahara}, Yuma},
        title = "{Red Spiral Galaxies at Cosmic Noon Unveiled in the First JWST Image}",
      journal = {\apjl},
         year = 2022,
        month = oct,
       volume = {938},
       number = {2},
          eid = {L24},
        pages = {L24},
          doi = {10.3847/2041-8213/ac982b},
archivePrefix = {arXiv},
       eprint = {2208.00132},
 primaryClass = {astro-ph.GA},
       adsurl = {https://ui.adsabs.harvard.edu/abs/2022ApJ...938L..24F}
}

@ARTICLE{Tojeiro2013,
       author = {{Tojeiro}, Rita and {Masters}, Karen L. and {Richards}, Joshua and {Percival}, Will J. and {Bamford}, Steven P. and {Maraston}, Claudia and {Nichol}, Robert C. and {Skibba}, Ramin and {Thomas}, Daniel},
        title = "{The different star formation histories of blue and red spiral and elliptical galaxies}",
      journal = {\mnras},
         year = 2013,
        month = jun,
       volume = {432},
       number = {1},
        pages = {359-373},
          doi = {10.1093/mnras/stt484},
archivePrefix = {arXiv},
       eprint = {1303.3551},
 primaryClass = {astro-ph.CO},
       adsurl = {https://ui.adsabs.harvard.edu/abs/2013MNRAS.432..359T}
}

@ARTICLE{Masters2010,
       author = {{Masters}, Karen L. and {Mosleh}, Moein and {Romer}, A. Kathy and {Nichol}, Robert C. and {Bamford}, Steven P. and {Schawinski}, Kevin and {Lintott}, Chris J. and {Andreescu}, Dan and {Campbell}, Heather C. and {Crowcroft}, Ben and {Doyle}, Isabelle and {Edmondson}, Edward M. and {Murray}, Phil and {Raddick}, M. Jordan and {Slosar}, An{\v{z}}e and {Szalay}, Alexander S. and {Vandenberg}, Jan},
        title = "{Galaxy Zoo: passive red spirals}",
      journal = {\mnras},
         year = 2010,
        month = jun,
       volume = {405},
       number = {2},
        pages = {783-799},
          doi = {10.1111/j.1365-2966.2010.16503.x},
archivePrefix = {arXiv},
       eprint = {0910.4113},
 primaryClass = {astro-ph.CO},
       adsurl = {https://ui.adsabs.harvard.edu/abs/2010MNRAS.405..783M}
}

@ARTICLE{George2017,
       author = {{George}, Koshy},
        title = "{Structural analysis of star-forming blue early-type galaxies. Merger-driven star formation in elliptical galaxies}",
      journal = {\aap},
         year = 2017,
        month = feb,
       volume = {598},
          eid = {A45},
        pages = {A45},
          doi = {10.1051/0004-6361/201629667},
archivePrefix = {arXiv},
       eprint = {1612.04000},
 primaryClass = {astro-ph.GA},
       adsurl = {https://ui.adsabs.harvard.edu/abs/2017A&A...598A..45G}
}

@ARTICLE{George2015,
       author = {{George}, Koshy and {Zingade}, Kshama},
        title = "{Revealing the nature of star forming blue early-type galaxies at low redshift}",
      journal = {\aap},
         year = 2015,
        month = nov,
       volume = {583},
          eid = {A103},
        pages = {A103},
          doi = {10.1051/0004-6361/201424826},
archivePrefix = {arXiv},
       eprint = {1510.01931},
 primaryClass = {astro-ph.GA},
       adsurl = {https://ui.adsabs.harvard.edu/abs/2015A&A...583A.103G}
}

@ARTICLE{Darvish2015,
       author = {{Darvish}, Behnam and {Mobasher}, Bahram and {Sobral}, David and {Scoville}, Nicholas and {Aragon-Calvo}, Miguel},
        title = "{A Comparative Study of Density Field Estimation for Galaxies: New Insights into the Evolution of Galaxies with Environment in COSMOS out to z{\ensuremath{\sim}}3}",
      journal = {\apj},
         year = 2015,
        month = jun,
       volume = {805},
       number = {2},
          eid = {121},
        pages = {121},
          doi = {10.1088/0004-637X/805/2/121},
archivePrefix = {arXiv},
       eprint = {1503.07879},
 primaryClass = {astro-ph.GA},
       adsurl = {https://ui.adsabs.harvard.edu/abs/2015ApJ...805..121D}
}

@ARTICLE{Calzetti2000,
       author = {{Calzetti}, Daniela and {Armus}, Lee and {Bohlin}, Ralph C. and {Kinney}, Anne L. and {Koornneef}, Jan and {Storchi-Bergmann}, Thaisa},
        title = "{The Dust Content and Opacity of Actively Star-forming Galaxies}",
      journal = {\apj},
         year = 2000,
        month = apr,
       volume = {533},
       number = {2},
        pages = {682-695},
          doi = {10.1086/308692},
archivePrefix = {arXiv},
       eprint = {astro-ph/9911459},
 primaryClass = {astro-ph},
       adsurl = {https://ui.adsabs.harvard.edu/abs/2000ApJ...533..682C}
}

@ARTICLE{Prevot1984,
       author = {{Prevot}, M.~L. and {Lequeux}, J. and {Maurice}, E. and {Prevot}, L. and {Rocca-Volmerange}, B.},
        title = "{The typical interstellar extinction in the Small Magellanic Cloud.}",
      journal = {\aap},
         year = 1984,
        month = mar,
       volume = {132},
        pages = {389-392},
       adsurl = {https://ui.adsabs.harvard.edu/abs/1984A&A...132..389P}
}

@ARTICLE{Weaver2023,
       author = {{Weaver}, J.~R. and {Davidzon}, I. and {Toft}, S. and {Ilbert}, O. and {McCracken}, H.~J. and {Gould}, K.~M.~L. and {Jespersen}, C.~K. and {Steinhardt}, C. and {Lagos}, C.~D.~P. and {Capak}, P.~L. and {Casey}, C.~M. and {Chartab}, N. and {Faisst}, A.~L. and {Hayward}, C.~C. and {Kartaltepe}, J.~S. and {Kauffmann}, O.~B. and {Koekemoer}, A.~M. and {Kokorev}, V. and {Laigle}, C. and {Liu}, D. and {Long}, A. and {Magdis}, G.~E. and {McPartland}, C.~J.~R. and {Milvang-Jensen}, B. and {Mobasher}, B. and {Moneti}, A. and {Peng}, Y. and {Sanders}, D.~B. and {Shuntov}, M. and {Sneppen}, A. and {Valentino}, F. and {Zalesky}, L. and {Zamorani}, G.},
        title = "{COSMOS2020: The galaxy stellar mass function. The assembly and star formation cessation of galaxies at 0.2< z {\ensuremath{\leq}} 7.5}",
      journal = {\aap},
         year = 2023,
        month = sep,
       volume = {677},
          eid = {A184},
        pages = {A184},
          doi = {10.1051/0004-6361/202245581},
archivePrefix = {arXiv},
       eprint = {2212.02512},
 primaryClass = {astro-ph.GA},
       adsurl = {https://ui.adsabs.harvard.edu/abs/2023A&A...677A.184W}
}

@ARTICLE{Davidzon2016,
       author = {{Davidzon}, I. and {Cucciati}, O. and {Bolzonella}, M. and {De Lucia}, G. and {Zamorani}, G. and {Arnouts}, S. and {Moutard}, T. and {Ilbert}, O. and {Garilli}, B. and {Scodeggio}, M. and {Guzzo}, L. and {Abbas}, U. and {Adami}, C. and {Bel}, J. and {Bottini}, D. and {Branchini}, E. and {Cappi}, A. and {Coupon}, J. and {de la Torre}, S. and {Di Porto}, C. and {Fritz}, A. and {Franzetti}, P. and {Fumana}, M. and {Granett}, B.~R. and {Guennou}, L. and {Iovino}, A. and {Krywult}, J. and {Le Brun}, V. and {Le F{\`e}vre}, O. and {Maccagni}, D. and {Ma{\l}ek}, K. and {Marulli}, F. and {McCracken}, H.~J. and {Mellier}, Y. and {Moscardini}, L. and {Polletta}, M. and {Pollo}, A. and {Tasca}, L.~A.~M. and {Tojeiro}, R. and {Vergani}, D. and {Zanichelli}, A.},
        title = "{The VIMOS Public Extragalactic Redshift Survey (VIPERS). Environmental effects shaping the galaxy stellar mass function}",
      journal = {\aap},
         year = 2016,
        month = feb,
       volume = {586},
          eid = {A23},
        pages = {A23},
          doi = {10.1051/0004-6361/201527129},
archivePrefix = {arXiv},
       eprint = {1511.01145},
 primaryClass = {astro-ph.GA},
       adsurl = {https://ui.adsabs.harvard.edu/abs/2016A&A...586A..23D}
}

@ARTICLE{Kroupa2001,
       author = {{Kroupa}, Pavel},
        title = "{On the variation of the initial mass function}",
      journal = {\mnras},
         year = 2001,
        month = apr,
       volume = {322},
       number = {2},
        pages = {231-246},
          doi = {10.1046/j.1365-8711.2001.04022.x},
archivePrefix = {arXiv},
       eprint = {astro-ph/0009005},
 primaryClass = {astro-ph},
       adsurl = {https://ui.adsabs.harvard.edu/abs/2001MNRAS.322..231K}
}

@ARTICLE{Conroy2009,
       author = {{Conroy}, Charlie and {Wechsler}, Risa H.},
        title = "{Connecting Galaxies, Halos, and Star Formation Rates Across Cosmic Time}",
      journal = {\apj},
         year = 2009,
        month = may,
       volume = {696},
       number = {1},
        pages = {620-635},
          doi = {10.1088/0004-637X/696/1/620},
archivePrefix = {arXiv},
       eprint = {0805.3346},
 primaryClass = {astro-ph},
       adsurl = {https://ui.adsabs.harvard.edu/abs/2009ApJ...696..620C}
}

@ARTICLE{Moster2011,
       author = {{Moster}, Benjamin P. and {Somerville}, Rachel S. and {Newman}, Jeffrey A. and {Rix}, Hans-Walter},
        title = "{A Cosmic Variance Cookbook}",
      journal = {\apj},
         year = 2011,
        month = apr,
       volume = {731},
       number = {2},
          eid = {113},
        pages = {113},
          doi = {10.1088/0004-637X/731/2/113},
archivePrefix = {arXiv},
       eprint = {1001.1737},
 primaryClass = {astro-ph.CO},
       adsurl = {https://ui.adsabs.harvard.edu/abs/2011ApJ...731..113M}
}

@ARTICLE{Kawinwanichakij2017,
       author = {{Kawinwanichakij}, Lalitwadee and {Papovich}, Casey and {Quadri}, Ryan F. and {Glazebrook}, Karl and {Kacprzak}, Glenn G. and {Allen}, Rebecca J. and {Bell}, Eric F. and {Croton}, Darren J. and {Dekel}, Avishai and {Ferguson}, Henry C. and {Forrest}, Ben and {Grogin}, Norman A. and {Guo}, Yicheng and {Kocevski}, Dale D. and {Koekemoer}, Anton M. and {Labb{\'e}}, Ivo and {Lucas}, Ray A. and {Nanayakkara}, Themiya and {Spitler}, Lee R. and {Straatman}, Caroline M.~S. and {Tran}, Kim-Vy H. and {Tomczak}, Adam and {van Dokkum}, Pieter},
        title = "{Effect of Local Environment and Stellar Mass on Galaxy Quenching and Morphology at 0.5 < z < 2.0}",
      journal = {\apj},
         year = 2017,
        month = oct,
       volume = {847},
       number = {2},
          eid = {134},
        pages = {134},
          doi = {10.3847/1538-4357/aa8b75},
archivePrefix = {arXiv},
       eprint = {1706.03780},
 primaryClass = {astro-ph.GA},
       adsurl = {https://ui.adsabs.harvard.edu/abs/2017ApJ...847..134K}
}

@ARTICLE{Chartab2020,
       author = {{Chartab}, Nima and {Mobasher}, Bahram and {Darvish}, Behnam and {Finkelstein}, Steve and {Guo}, Yicheng and {Kodra}, Dritan and {Lee}, Kyoung-Soo and {Newman}, Jeffrey A. and {Pacifici}, Camilla and {Papovich}, Casey and {Sattari}, Zahra and {Shahidi}, Abtin and {Dickinson}, Mark E. and {Faber}, Sandra M. and {Ferguson}, Henry C. and {Giavalisco}, Mauro and {Jafariyazani}, Marziye},
        title = "{Large-scale Structures in the CANDELS Fields: The Role of the Environment in Star Formation Activity}",
      journal = {\apj},
         year = 2020,
        month = feb,
       volume = {890},
       number = {1},
          eid = {7},
        pages = {7},
          doi = {10.3847/1538-4357/ab61fd},
archivePrefix = {arXiv},
       eprint = {1912.04890},
 primaryClass = {astro-ph.GA},
       adsurl = {https://ui.adsabs.harvard.edu/abs/2020ApJ...890....7C}
}

@ARTICLE{Q1-TP001,
       author = {{Euclid Collaboration: Aussel}, H. and {Tereno}, I. and {Schirmer}, M. and others},
        title = "{Euclid Quick Data Release (Q1) - Data release overview}",
      journal = {A\&A, accepted (Euclid Q1 SI)},
         year = 2025,
        month = mar,
          eid = {arXiv:2503.15302},
        pages = {arXiv:2503.15302},
archivePrefix = {arXiv},
       eprint = {2503.15302},
 primaryClass = {astro-ph.GA},
       adsurl = {https://ui.adsabs.harvard.edu/abs/2025arXiv250315302E}
}

@ARTICLE{Q1-TP002,
       author = {{Euclid Collaboration: McCracken}, H.~J. and {Benson}, K. and {Dolding}, C. and others},
        title = "{Euclid Quick Data Release (Q1): VIS processing and data products}",
      journal = {A\&A, in press (Euclid Q1 SI), \url{https://doi.org/10.1051/0004-6361/202554594}},
         year = 2025,
        month = mar,
          eid = {arXiv:2503.15303},
        pages = {arXiv:2503.15303},
archivePrefix = {arXiv},
       eprint = {2503.15303},
 primaryClass = {astro-ph.IM},
       adsurl = {https://ui.adsabs.harvard.edu/abs/2025arXiv250315303E}
}

@ARTICLE{Q1-TP003,
       author = {{Euclid Collaboration: Polenta}, G. and {Frailis}, M. and {Alavi}, A. and others},
        title = "{Euclid Quick Data Release (Q1). NIR processing and data products}",
      journal = {A\&A, in press (Euclid Q1 SI), \url{https://doi.org/10.1051/0004-6361/202554657}},
         year = 2025,
        month = mar,
          eid = {arXiv:2503.15304},
        pages = {arXiv:2503.15304},
archivePrefix = {arXiv},
       eprint = {2503.15304},
 primaryClass = {astro-ph.CO},
       adsurl = {https://ui.adsabs.harvard.edu/abs/2025arXiv250315304E}
}

@ARTICLE{Q1-TP004,
       author = {{Euclid Collaboration: Romelli}, E. and {K\"ummel}, M. and {Dole}, H. and others},
        title = "{Euclid Quick Data Release (Q1). From images to multiwavelength catalogues: the Euclid MERge Processing Function}",
      journal = {A\&A, in press (Euclid Q1 SI), \url{https://doi.org/10.1051/0004-6361/202554586}},
         year = 2025,
        month = mar,
          eid = {arXiv:2503.15305},
        pages = {arXiv:2503.15305},
archivePrefix = {arXiv},
       eprint = {2503.15305},
 primaryClass = {astro-ph.CO},
       adsurl = {https://ui.adsabs.harvard.edu/abs/2025arXiv250315305E}
}

@ARTICLE{Q1-TP005,
       author = {{Euclid Collaboration: Tucci}, M. and {Paltani}, S. and {Hartley}, W.~G. and others},
        title = "{Euclid Quick Data Release (Q1). Photometric redshifts and physical properties of galaxies through the PHZ processing function}",
      journal = {A\&A, in press (Euclid Q1 SI), \url{https://doi.org/10.1051/0004-6361/202554588}},
         year = 2025,
        month = mar,
          eid = {arXiv:2503.15306},
        pages = {arXiv:2503.15306},
archivePrefix = {arXiv},
       eprint = {2503.15306},
 primaryClass = {astro-ph.GA},
       adsurl = {https://ui.adsabs.harvard.edu/abs/2025arXiv250315306E}
}

@ARTICLE{Q1-SP040,
       author = {{Euclid Collaboration: Quilley}, L. and {Damjanov}, I. and {de Lapparent}, V. and others},
        title = "{Euclid Quick Data Release (Q1). Exploring galaxy morphology across cosmic time through Sersic fits}",
      journal = {A\&A, in press (Euclid Q1 SI), \url{https://doi.org/10.1051/0004-6361/202554585}},
         year = 2025,
        month = mar,
          eid = {arXiv:2503.15309},
        pages = {arXiv:2503.15309},
archivePrefix = {arXiv},
       eprint = {2503.15309},
 primaryClass = {astro-ph.GA},
       adsurl = {https://ui.adsabs.harvard.edu/abs/2025arXiv250315309E}
}

@ARTICLE{Q1-SP031,
       author = {{Euclid Collaboration: Enia}, A. and {Pozzetti}, L. and {Bolzonella}, M. and others},
       title = "{Euclid Quick Data Release (Q1). A first view of the star-forming main sequence in the Euclid Deep Fields}",
      journal = {A\&A, in press (Euclid Q1 SI), \url{https://doi.org/10.1051/0004-6361/202554576}},
         year = 2025,
        month = mar,
          eid = {arXiv:2503.15314},
        pages = {arXiv:2503.15314},
archivePrefix = {arXiv},
       eprint = {2503.15314},
 primaryClass = {astro-ph.GA},
       adsurl = {https://ui.adsabs.harvard.edu/abs/2025arXiv250315314E}
}

@ARTICLE{Q1-SP044,
       author = {{Euclid Collaboration: Corcho-Caballero}, P. and {Ascasibar}, Y. and {Verdoes Kleijn}, G. and others},
       title = "{Euclid Quick Data Release (Q1). A probabilistic classification of quenched galaxies}",
      journal = {A\&A, in press (Euclid Q1 SI), \url{https://doi.org/10.1051/0004-6361/202554582}},
         year = 2025,
        month = mar,
          eid = {arXiv:2503.15315},
        pages = {arXiv:2503.15315},
archivePrefix = {arXiv},
       eprint = {2503.15315},
 primaryClass = {astro-ph.GA},
       adsurl = {https://ui.adsabs.harvard.edu/abs/2025arXiv250315315E}
}

@ARTICLE{Q1-SP050,
       author = {{Euclid Collaboration: Bhargava}, S. and {Benoist}, C. and {Gonzalez}, A.~H. and others},
       title = "{Euclid Quick Data Release (Q1). First detections from the galaxy cluster workflow}",
      journal = {A\&A, in press (Euclid Q1 SI), \url{ https://doi.org/10.1051/0004-6361/202554937}},
         year = 2025,
        month = mar,
          eid = {arXiv:2503.19196},
        pages = {arXiv:2503.19196},
          doi = {10.48550/arXiv.2503.19196},
archivePrefix = {arXiv},
       eprint = {2503.19196},
 primaryClass = {astro-ph.CO},
       adsurl = {https://ui.adsabs.harvard.edu/abs/2025arXiv250319196E}
}

@ARTICLE{EuclidSkyOverview,
author = {{Euclid Collaboration: Mellier}, Y. and {Abdurro'uf} and {Acevedo~Barroso}, J.A. and others},
	title = {Euclid - I. Overview of the Euclid mission},
	DOI= "10.1051/0004-6361/202450810",
	url= "https://doi.org/10.1051/0004-6361/202450810",
	journal = {A\&A},
	year = 2025,
	volume = 697,
	pages = "A1",
}

@ARTICLE{EuclidSkyVIS,
author = {{Euclid Collaboration: Cropper}, M. and {Al-Bahlawan}, A. and {Amiaux}, J. and others},
	title = {Euclid - II. The VIS instrument},
	DOI= "10.1051/0004-6361/202450996",
	url= "https://doi.org/10.1051/0004-6361/202450996",
	journal = {A\&A},
	year = 2025,
	volume = 697,
	pages = "A2",
}

@ARTICLE{EuclidSkyNISP,
author = {{Euclid Collaboration: Jahnke}, K. and {Gillard}, W. and {Schirmer}, M. and others},
	title = {Euclid - III. The NISP Instrument},
	DOI= "10.1051/0004-6361/202450786",
	url= "https://doi.org/10.1051/0004-6361/202450786",
	journal = {A\&A},
	year = 2025,
	volume = 697,
	pages = "A3",
}

@misc{Q1cite,
author = "{Euclid Quick Release Q1}",
howpublished = "\url{https://doi.org/10.57780/esa-2853f3b}",
year = 2025
}

@ARTICLE{Scaramella-EP1,
       author = {{Euclid Collaboration: Scaramella}, R. and {Amiaux}, J. and {Mellier}, Y. and others},
        title = "{Euclid preparation. I. The Euclid Wide Survey}",
      journal = {\aap},
         year = 2022,
        month = jun,
       volume = {662},
          eid = {A112},
        pages = {A112},
          doi = {10.1051/0004-6361/202141938},
archivePrefix = {arXiv},
       eprint = {2108.01201},
 primaryClass = {astro-ph.CO},
       adsurl = {https://ui.adsabs.harvard.edu/abs/2022A&A...662A.112E}
}

\begin{appendix}
\nolinenumbers
\onecolumn
\section{Quantifying the effect of the environment}\label{sec:quant}

\begin{figure}[h]
    \centering
    \includegraphics[width=0.7\hsize]{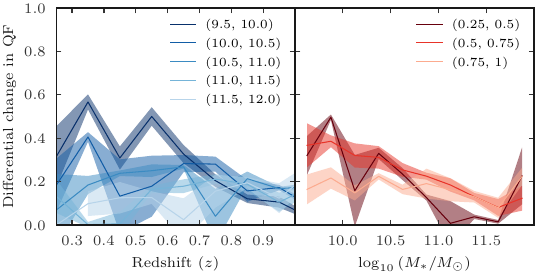}
    \caption{Left: Differential change in the quenched fraction because of the environment, as a function of redshift in fixed mass bins. The environment has the strongest effect at low-mass, and this effect is lessened with increasing redshift. Right: Same as the left panel, but as a function of stellar mass in fixed redshift bins. At low redshift, low-mass galaxies are most affected, and this effect weakens at high mass, while at high redshift, the effect increases slightly with increasing stellar mass. Shaded regions show the $1\ \sigma$ uncertainty on the slope}
    \label{fig:qf-quant}
\end{figure}

Here we quantify the differential effect of the environment in fixed stellar mass or redshift bins. The differential effect here is defined as the slope of the fitted line of the quenched fraction as a function of increasing density in stellar mass or redshift bins. This allows us to quantify the effect of the environment, and identify overall trends between low and high mass, and between low and high redshift. The results of this analysis are plotted in Fig. \ref{fig:qf-quant}.

In the left panel Fig. \ref{fig:qf-quant}, we see the differential change in the quenched fraction as a function of redshift, in fixed stellar mass bins. We see that at low redshift, the effect of the environment becomes stronger with decreasing stellar mass. This effect drops off at $z\sim0.7$. In the right panel of Fig. \ref{fig:qf-quant}, we plot the same thing but as a function of stellar mass, in fixed redshift bins. These are the same redshift bins used in the main text, however when increasing the number of bins, the main results remain the same. Again we see that at $z<0.75$, the environmental effect is strongest at low mass, and decreases with increasing mass. At $0.75<z<1$, the differential environmental effect is on average lower than in the other two redshift bins, however there is evidence that the environmental effect actually increases at high mass here. That said, the uncertainty on the slope is large.

\begin{figure}[h]
    \centering
    \includegraphics[width=0.7\hsize]{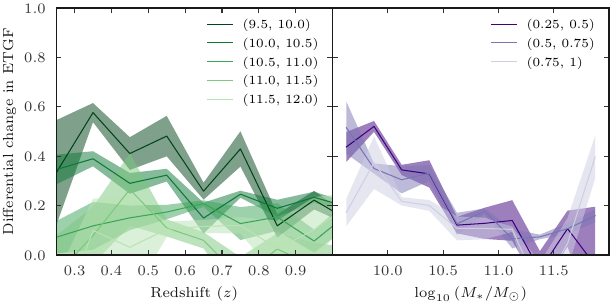}
    \caption{Same as Fig. \ref{fig:qf-quant} but for the ETG fraction.}
    \label{fig:etg-quant}
\end{figure}

We plot the same thing but for the ETG fraction in Fig. \ref{fig:etg-quant}. The main difference here is that the differential change on the ETG fraction due to the environment is very low at $M_\ast>10^{10.5}\ M_\odot$. Again, we see that, at low redshift, the environmental effect is strongest for low-mass galaxies.

\section{Monte Carlo tests}\label{sec:mc}
We present the results of our Monte Carlo experiments, after 1000 runs. For each realisation, we draw a value from a Gaussian distribution, centered on $\Sigma_7$ and with the uncertainty on $\Sigma_7$, as described in Sect. \ref{sec:dens}, as the standard deviation of the distribution. This results in a perturbed value for the local density. We use this perturbed value to create the same binned plots in Figs. \ref{fig:qf_dens_2d} and \ref{fig:etg_dens_2d}. After 1000 realisations, we take the mean value in each bin to obtain a `smoothed' version of the results in the main text. This procedure allows us to identify if the trends persist after taking into account the uncertainties on the density measurements.

\begin{figure}[h]
    \centering
    \includegraphics[width=0.9\hsize]{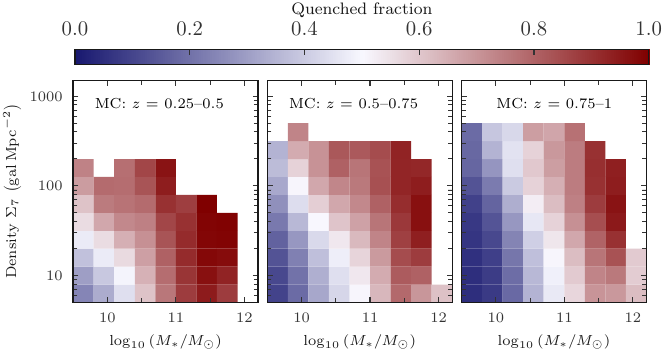}
    \caption{Same as Fig. \ref{fig:qf_dens_2d}, but after 1000 Monte Carlo tests run to show the dependence of the quenched fraction on environment and stellar mass.}
    \label{fig:qf-mc}
\end{figure}

In Fig. \ref{fig:qf-mc}, we see that for the quenched fraction, the dependence on both stellar mass and environment at $z<0.75$ remains, although the quenched fraction is reduced in regions of the plot where there were very few galaxies (high stellar mass and high local density). At $z>0.75$, the stellar mass is the dominant factor in determining the mean quenched fraction, with only a very slight dependence on environment.

\begin{figure}[h]
    \centering
    \includegraphics[width=0.9\hsize]{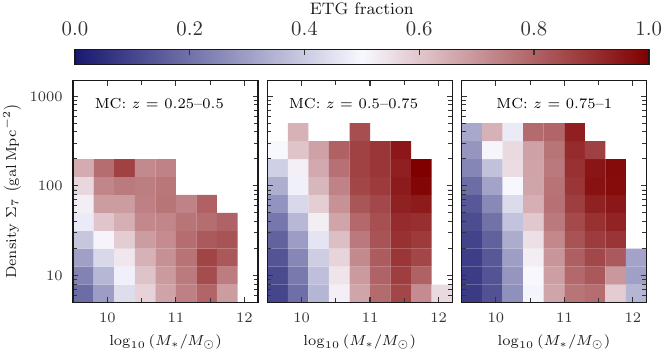}
    \caption{Same as Fig. \ref{fig:etg_dens_2d}, but after 1000 Monte Carlo tests run to show the dependence of the ETG fraction on environment and stellar mass.}
    \label{fig:etg-mc}
\end{figure}

For the ETG fraction, in Fig. \ref{fig:etg-mc}, the result that the environment has the biggest effect on low-mass galaxies at $z<0.5$ is still visible. The ETG fraction depends strongly on the stellar mass in all redshift bins. At $0.5<z<0.75$, the ETG fraction depends less on the environment, but the effect is still there at $M_\ast<10^{10.5}\ M_\odot$. At $z>0.75$, there is very little evidence that the ETG fraction depends on environment.

\end{appendix}

\end{document}